\documentclass[10pt,twocolumn,showpacs,amsmath,amssymb, aps, prb]{revtex4-1}

\usepackage{graphicx}
\usepackage{dcolumn}
\usepackage{bm}
\usepackage{hyperref}
\usepackage[cp1250]{inputenc}
\usepackage[usenames,dvipsnames]{color}
\usepackage{tabularx}
\usepackage[normalem]{ulem}
\usepackage{cancel}
\usepackage{color}

\def\Xint#1{\mathchoice
   {\XXint\displaystyle\textstyle{#1}}%
   {\XXint\textstyle\scriptstyle{#1}}%
   {\XXint\scriptstyle\scriptscriptstyle{#1}}%
   {\XXint\scriptscriptstyle\scriptscriptstyle{#1}}%
   \!\int}
\def\XXint#1#2#3{{\setbox0=\hbox{$#1{#2#3}{\int}$}
     \vcenter{\hbox{$#2#3$}}\kern-.5\wd0}}

\def\dashint{\Xint-}

\newcommand{\vecl}{{\rm\bf l}}
\newcommand{\vecr}{{\rm\bf r}}
\newcommand{\arctanh}{{\rm arctanh}}
\newcommand{\arccsch}{{\rm arccsch}}
\newcommand{\Cot}{{\rm Cot}}
\newcommand{\Ci}{{\rm Ci}}
\newcommand{\Si}{{\rm Si}}
\newcommand{\sgn}{{\rm sgn}}
\newcommand{\rmi}{{\rm i}}
\newcommand{\rmd}{{\rm d}}
\newcommand{\veck}{{\rm\bf k}}

\newcommand{\arccot}{{\rm arccot}}
\newcommand{\sech}{{\rm sech}}
\newcommand{\Ei}{{\rm Ei}}
\newcommand{\Erfi}{{\rm Erfi}}

\renewcommand{\today}{April 23, 2019; Phys.\ Rev.\ B~{\bf 99}, 165130 (2019)}

\begin{document}

\title{Ground-state properties of the symmetric single-impurity Anderson model 
on a ring\\
from Density-Matrix Renormalization Group, Hartree-Fock, and Gutzwiller theory}

\author{Gergely Barcza$^1$}
\author{Florian Gebhard$^2$}
\email{florian.gebhard@physik.uni-marburg.de}
\author{Thorben Linneweber$^2$}
\author{\"Ors Legeza$^1$}
\affiliation{$^1$Strongly Correlated Systems Lend\"ulet Research Group, 
Institute for Solid State Physics and Optics, MTA Wigner Research Centre for
Physics, P.O.\ Box 49, H-1525 Budapest, Hungary}
\affiliation{$^2$Fachbereich Physik, Philipps-Universit\"at Marburg,
D-35032 Marburg, Germany}

\date{\today}

\begin{abstract}%
We analyze the ground-state energy, magnetization,
magnetic susceptibility,
and Kondo screening cloud of the symmetric single-impurity
Anderson model (SIAM) that is characterized by the band width~$W$,
the impurity interaction strength~$U$, and the local hybridization~$V$.
We compare Gutzwiller variational and 
magnetic Hartree-Fock results in the thermodynamic limit
with numerically exact data
from the Density-Matrix Renormalization Group (DMRG) method
on large rings. To improve the DMRG performance, we use a canonical
transformation to map the SIAM onto a chain with half the system size 
and open boundary conditions. We compare to Bethe-Ansatz results
for the ground-state energy, magnetization, and spin susceptibility
that become exact in the wide-band limit. 
Our detailed comparison shows that the field-theoretical description 
is applicable to the SIAM on a ring for a broad parameter range.
Hartree-Fock theory gives an excellent ground-state energy
and local moment for intermediate and strong interactions.
However, it lacks spin fluctuations and thus cannot screen the impurity spin.
The Gutzwiller variational energy bound becomes very poor
for large interactions because it does not describe properly
the charge fluctuations.
Nevertheless, the Gutzwiller approach provides a qualitatively correct 
description of the zero-field susceptibility and the Kondo screening cloud. 
The DMRG provides excellent data for the ground-state energy and the 
magnetization for finite external fields. At strong interactions, 
finite-size effects make it extremely difficult to recover the exponentially large 
zero-field susceptibility and the mesoscopically large Kondo screening cloud.
\end{abstract}



\maketitle

\section{Introduction}

The single-impurity Anderson model (SIAM) describes 
an impurity where electrons interact locally 
in a metallic host.~\cite{PhysRev.124.41,RevModPhys.50.191}
It poses one of the best studied and understood 
fundamental many-body problems; for a review, see
Ref.~[\onlinecite{Hewson}]. 
Therefore, it still serves as a benchmark test for the development of 
advanced analytical many-body techniques, e.g., the functional 
renormalization group 
technique.~\cite{PhysRevB.81.195109,PhysRevB.87.035111,PhysRevB.93.195160}
For the symmetric SIAM,
the low-energy physics is similar to that of the single-impurity $s$-$d$ or 
Kondo model~\cite{PhysRev.81.440,KondoPTHeorPhys}
where an impurity spin couples to the host electrons' spin degrees of freedom:
at zero temperature, the impurity local moment is screened by the host electrons
which gives rise to a narrow Abrikosov-Suhl or Kondo resonance
in the impurity spectral function at the Fermi level.~\cite{Hewson} 
The resonance can be resolved using the Numerical Renormalization Group (NRG)
technique; for a review, see Ref.~[\onlinecite{RevModPhys.80.395}].
At higher energies, Hubbard satellites appear in the impurity spectral function
that describe the local charge fluctuations.
Both the Kondo resonance and the Hubbard satellite 
are accessible from the analytic Local-Moment 
Approach.~\cite{Logan1998,Logan2000,Logan2009}

More recently, the real-space features of the screening were studied 
for the Kondo model using 
NRG,~\cite{PhysRevB.75.041307,PhysRevB.84.115120,PhysRevB.90.045117}
and the analytical 
coherent-state expansion.~\cite{PhysRevB.92.195106}
For the non-interacting SIAM (resonant-level model) in the wide-band limit,
the screening cloud was analyzed analytically,~\cite{1742-5468-2014-4-P04011}
and the magnetic properties of the interacting SIAM were studied numerically using
the Density Matrix Renormalization Group 
(DMRG) method.~\cite{PhysRevB.80.205114}
The various methods show that the screening cloud extends very far into the
host metal. In the Kondo regime, an algebraic decay sets in only beyond
a characteristic (Kondo) length scale that is proportional to the inverse of the Kondo
temperature.

Less attention was dedicated to ground-state properties 
of the SIAM and Kondo models because they are solvable by 
Bethe Ansatz.~\cite{PhysRevLett.45.379,0022-3719-14-10-014,%
0022-3719-16-12-017,0022-3719-16-12-018} Therefore, important
quantities such as the ground-state energy, magnetization, 
and magnetic susceptibility at zero field are known explicitly.
The Bethe Ansatz is based on the wide-band limit, $W\to \infty$, so that the dispersion
relation of the host electrons can be linearized around the Fermi energy.
However, the implicit assumption that the Hubbard interaction~$U$ is small
compared to the bandwidth, $U\ll W$, impedes a comparison with methods that
treat the SIAM on a lattice such as the DMRG method,
and the Hartree-Fock~\cite{PhysRev.124.41}
and Gutzwiller wave functions.~\cite{Schoenhammer1990,PhysRevB.41.9452}

As one of the best studied many-body problems, the SIAM
is particularly suitable to test existing, and conceivable future,
many-body methods. 
Since these are often customized for the treatment of lattice Hamiltonians,
it is one of the purposes of this work to provide tangible results for 
a ring geometry; for other recent numerical
treatments of finite structures, 
see Refs.~\onlinecite{PhysRevB.88.245113,PhysRevB.92.155104}.
Given the high accuracy of the DMRG data
for large system sizes, an extrapolation of most ground-state properties
to the thermodynamic limit is unproblematic.
As we shall see, the wide-band limit remains applicable
for fairly large interaction strengths even for a substantial hybridization
which justifies the application of the wide-band limit even
for sizable Coulomb parameters.

In this work, we use the DMRG to calculate numerically exactly
the ground-state energy, 
the local magnetic moment, the zero-field susceptibility, 
and the screening cloud of the single-impurity Anderson model on large rings.
The Gutzwiller and Hartree-Fock approaches provide complementary
insights. The Hartree-Fock variational estimate
of the ground-state energy is very satisfactory for moderate to large Hubbard 
interactions whereas the Gutzwiller estimate is acceptable only for small~$U$.
On the other hand, the Gutzwiller approach provides a qualitatively correct
description of the magnetic properties whereas Hartree-Fock theory 
fails to screen the impurity spin even at infinitely large distances.
Since the Gutzwiller approach is heavily based 
on the exact results for the non-interacting SIAM, 
we compile the results for the resonant-level model in the
appendix.

Our work is structured as follows.
In Sect.~\ref{sec:Hamiltonian},
we introduce the one-dimensional SIAM on a ring
with local hybridization at particle-hole and spin symmetry.
We map the model onto a 
two-chain problem~\cite{PhysRevB.74.195112,PhysRevB.84.115403} 
where the two chains separate 
in the thermodynamic limit. The reduced model provides
the basis for our numerical DMRG investigations.
In Sect.~\ref{sec:nonint} we discuss the ground-state energy, magnetization,
and spin correlation function between the impurity and the bath sites
for the non-interacting SIAM for small hybridizations. 
The derivation of the formulae is deferred
to the appendices. In Sect.~\ref{sec:Gutzwiller},
we evaluate the Gutzwiller variational wave function
for the SIAM and determine an analytical variational upper bound
for the ground-state energy. Moreover, we calculate the variational
magnetization and spin correlation function. 
In Sect.~\ref{sec:DMRG} we compare our results for the ground-state
energy, magnetization, and the spin correlation function
with numerically exact DMRG data for large system sizes.
We include the results from Bethe Ansatz and a magnetic Hartree-Fock 
calculation, see the appendix for their derivation.
Short conclusions, Sect.~\ref{sec:conclusions}, close our presentation.

\section{Symmetric single-impurity 
Anderson model on a ring}
\label{sec:Hamiltonian}

We study the particle-hole
and spin symmetric SIAM on a ring.~\cite{PhysRev.124.41,RevModPhys.50.191} 
For strong interactions, this model maps onto
the one-dimensional Kondo impurity model.~\cite{PhysRev.149.491}

\subsection{Hamiltonian}
\label{subsec:Hamilt}

The Hamilton operator
for the one-dimensional single-impurity Anderson model 
reads~\cite{PhysRev.124.41,Hewson}
\begin{eqnarray}
\hat{H}&=& \hat{H}_0+ \hat{H}_{\rm int} \; , \nonumber \\
\hat{H}_{\rm int} &=& U\left(\hat{n}_{d,\uparrow}-1/2\right)
\left(\hat{n}_{d,\downarrow}-1/2\right)  \;,
\label{eq:SIAMH}
\end{eqnarray}
where $\hat{n}_{d,\sigma}=\hat{d}_{\sigma}^{+} \hat{d}_{\sigma}^{\vphantom{+}}$
counts the number of $\sigma$-electrons 
on the impurity site ($\sigma=\uparrow,\downarrow$).
Only the electrons on the impurity site repel each other with
strengths~$U>0$.
The non-interacting Hamiltonian,
\begin{equation}
\hat{H}_0= \hat{T}+\hat{B}+\hat{V} +\hat{P}\; ,
\label{eq:defHzero}
\end{equation}
describes bath electrons that move between neighboring sites on a ring 
with $L$~sites,
\begin{equation}
\hat{T}= -\frac{W}{4}\sum_{n=0,\sigma}^{L-1} 
\left(\hat{c}_{n,\sigma}^{+}
\hat{c}_{n+1,\sigma}^{\vphantom{+}} +
\hat{c}_{n+1,\sigma}^{+}
\hat{c}_{n,\sigma}^{\vphantom{+}}\right) \; ,
\label{eq:defT}
\end{equation}
where the band width provides our unit of energy, $W\equiv 1$.

In the presence of an external magnetic field~${\cal H}_{\text{bath}}$
we may include the magnetic term
\begin{equation}
\hat{B}= -B_{\text{bath}}\sum_{n=0}^{L-1} 
\left(\hat{c}_{n,\uparrow}^{+}\hat{c}_{n,\uparrow}^{\vphantom{+}} -
\hat{c}_{n,\downarrow}^{+}
\hat{c}_{n,\downarrow}^{\vphantom{+}}\right) \; .
\label{eq:defB}
\end{equation}
Here, we abbreviated $B_{\text{bath}}=g\mu_{\rm B}{\cal H}_{\text{bath}}/2$
where $\mu_{\rm B}$ is the Bohr magneton and $g\approx 2$ is the electrons'
gyromagnetic factor.

The bath electrons hybridize at the origin, $n=0$, with the impurity electrons
with strength $V>0$,
\begin{equation}
\hat{V}= V \sum_{\sigma} \left( \hat{d}_{\sigma}^{+}
\hat{c}_{0,\sigma}^{\vphantom{+}} 
+ \hat{c}_{0,\sigma}^{+} \hat{d}_{\sigma}^{\vphantom{+}} \right) \; .
\label{eq:defV}
\end{equation}
The system is half filled, i.e., the total number of electrons is $N=L+1$,
and we investigate a paramagnetic situation, $N_{\uparrow}=N_{\downarrow}=
(L+1)/2$. Consequently, the number of bath sites~$L$ must be odd. From now on we
further assume that $(L+3)/2$ is even. 

There can be a local, possibly spin-dependent potential,
\begin{equation}
\hat{P} = -\sum_{\sigma} E_{d,\sigma}(\hat{n}_{d,\sigma}-1/2) \; .
\end{equation}
In the presence of an external 
magnetic field~${\cal H}_{\text{imp}}$ at the impurity we have 
$E_{d,\uparrow}=(g\mu_{\rm B}/2){\cal H}_{\text{imp}}=-E_{d,\downarrow}$.
In the magnetic Hartree-Fock approach, we have 
$E_{d,\uparrow}=Um=-E_{d,\downarrow}$
where the value of the $\hat{S}_z$ at the
impurity, $m=\langle \hat{n}_{d,\uparrow}-\hat{n}_{d,\downarrow}\rangle/2$, 
has to be determined self-consistently.

\subsection{Particle-hole symmetry}

To analyze particle-hole symmetry in the SIAM,
we set $E_{d,\sigma}=0$ and ${\cal H}_{\text{bath}}=0$ 
in the rest of this section,
i.e., we have no magnetic symmetry breaking.
Particle-hole symmetry for the SIAM was studied previously,
e.g., in Ref.~[\onlinecite{Annalenpaper}].

The ring geometry renders 
the analysis of particle-hole symmetry more cumbersome than the choice of
open boundary conditions. Since boundary conditions play no role
in the thermodynamic limit as investigated in Sects.~III--V,
the material presented in
this section is included for completeness rather than necessity.

\subsubsection{Wave numbers and particle-hole boundary conditions}

For a ring, the kinetic energy is diagonal in momentum space,
\begin{equation}
\hat{T}= \sum_{k,\sigma} \epsilon(k) 
\hat{c}_{k,\sigma}^+ \hat{c}_{k,\sigma}^{\vphantom{+}}
\; ,
\end{equation}
where
\begin{equation}
\hat{c}_{n,\sigma}^{\vphantom{+}} =\sqrt{\frac{1}{L}} \sum_{k} 
e^{\rmi k n}\hat{c}_{k,\sigma}^{\vphantom{+}} \quad ,\quad
\hat{c}_{k,\sigma}^{\vphantom{+}} =\sqrt{\frac{1}{L}} \sum_{n=0}^{L-1}
e^{-\rmi k n} \hat{c}_{n,\sigma}^{\vphantom{+}}\;, 
\end{equation}
and the dispersion relation is given by
\begin{equation}
\epsilon(k)=-\cos(k)/2 \; .
\label{eq:dispersion}
\end{equation}
In one dimension $Q=\pi$ is half a reciprocal lattice vector,
$\epsilon(k+2Q)=\epsilon(k)$.
For particle-hole symmetry we must demand that 
\begin{equation}
\epsilon(\pi-k)= - \epsilon(k) 
\end{equation}
for all accessible~$|k|\leq \pi$.
In particular, this equation implies that with $k$, also $\pi-k$
is an accessible $k$-value. This is not difficult to fulfill for even~$L$ 
but poses a problem for odd~$L$.

Let 
\begin{equation}
k_m= \frac{2\pi}{L}m +\varphi 
\; , \; m=-(L-1)/2, \ldots, (L-1)/2 \; ,
\label{eq:kvaluesallowed}
\end{equation}
where $0\leq \varphi < 2\pi/L$ and $k_m$ are defined modulo~$2\pi$.
Then, the set of $k$-values must also be given by
\begin{eqnarray}
k_m'&=& \pi-\frac{2\pi}{L}m -\varphi 
=\frac{2\pi}{L}\left(\frac{L-1}{2}+\frac{1}{2} -m\right) + \varphi 
- 2\varphi \nonumber \\
&=&  
\frac{2\pi}{L}\left(\frac{L-1}{2}-m\right) + \varphi
+\frac{\pi-2\varphi L}{L} \; .
\end{eqnarray}
Using the definition of the accessible $k$-values, we see that
we must set 
\begin{equation}
\varphi=\pm \pi/(2L)
\label{eq:varphipihalf}
\end{equation}
to make the sets $\{k\}$ and $\{k'\}$ identical.
Particle-hole symmetry for odd~$L$ destroys inversion symmetry because
the energy levels $\epsilon(k)$ are not degenerate, i.e., if $k$ is an allowed value,
$k'=-k$ is not accessible.

The accessible $k$-values belong to the boundary conditions
\begin{equation}
e^{\rmi k_m L}= e^{\rmi\pi/2}= \rmi \; ,
\end{equation}
i.e., they are neither periodic nor anti-periodic. 
We call these boundary conditions {\sl particle-hole periodic}.
They imply $\hat{c}_L^{\vphantom{+}}
=\rmi \hat{c}_0^{\vphantom{+}}$ ($\hat{c}_L^+=-\rmi \hat{c}_0^+$) in position space
so that we may write for the kinetic energy in eq.~(\ref{eq:defHzero})
\begin{eqnarray}
\hat{T}&=& -\frac{1}{4} \sum_{n=0,\sigma}^{L-2}\left(  
\hat{c}_{n,\sigma}^+ \hat{c}_{n+1,\sigma}^{\vphantom{+}}
+ \hat{c}_{n+1,\sigma}^+ \hat{c}_{n,\sigma}^{\vphantom{+}}\right) \nonumber \\
&& -\frac{1}{4} \left(\rmi \hat{c}_{L-1,\sigma}^+ \hat{c}_{0,\sigma}^{\vphantom{+}} 
- \rmi\hat{c}_{0,\sigma}^+ \hat{c}_{L-1,\sigma}^{\vphantom{+}}\right) 
\;. \label{eq:goodform}
\end{eqnarray}
Eq.~(\ref{eq:goodform}) shows that the kinetic energy is indeed
invariant under the particle-hole transformation
\begin{equation}
{\rm ph: } \quad \hat{c}_{n,\sigma}^{\vphantom{+}} 
\mapsto (-1)^n \hat{c}_{n,\sigma}^+ \quad \hbox{for}
\quad n=0,1,\ldots, L-1
\end{equation}
because either $n$ or $n+1$ is even when the other is odd,
and the origin and $L-1$ are both even numbers for odd~$L$.

\subsubsection{Model properties at particle-hole symmetry}

For the one-dimensional model~(\ref{eq:defHzero}) with 
particle-hole boundary conditions and $E_{d,\sigma}=0$, 
we define the particle-hole transformation 
\begin{eqnarray}
{\rm ph:}\;  \hat{c}_{n,\sigma}^{\vphantom{+}} 
&\mapsto& (-1)^n \hat{c}_{n,\sigma}^+ 
\quad \hbox{for}
\quad n=0,1,\ldots, L-1 \nonumber \; , \\
\hat{d}_{\sigma}^{\vphantom{+}} 
&\mapsto& (-1)\hat{d}_{\sigma}^+ \; .
\end{eqnarray}
It is readily seen that the transformation leaves the Hamiltonian $\hat{H}$ invariant.
The particle-number operators transform according to
\begin{equation}
\hat{n}_{d,\sigma} \mapsto  1- \hat{n}_{d,\sigma} \quad , \quad
\hat{c}_{n,\sigma}^+ \hat{c}_{n,\sigma}^{\vphantom{+}}
\mapsto 1- \hat{c}_{n,\sigma}^+ \hat{c}_{n,\sigma}^{\vphantom{+}}\; ,
\end{equation}
so that the $N$-particle sector maps onto the sector with $2(L+1)-N$ particles.
At half band-filling, $N=L+1$, the normalized ground state maps onto itself,
$|\Psi_0\rangle \mapsto |\Psi_0\rangle$, up to 
a global phase. Therefore,
particle-hole symmetry guarantees
\begin{equation}
\langle \Psi_0 |\hat{n}_{d,\sigma} |\Psi_0 \rangle =1/2
\label{eq:averagedisonehalf}
\end{equation}
for all interaction strengths~$U$ and hybridizations~$V$.
Moreover, we obtain 
\begin{eqnarray}
\langle \Psi_0 |\hat{c}_{n,\sigma}^+\hat{d}_{\sigma}^{\vphantom{+}} |\Psi_0 \rangle &=& 
(-1)^n \langle \Psi_0 |\hat{d}_{\sigma}^+ \hat{c}_{n,\sigma}^{\vphantom{+}}|\Psi_0 \rangle
\nonumber \\ 
&=&(-1)^n
\langle \Psi_0 |\hat{c}_{n,\sigma}^+\hat{d}_{\sigma}^{\vphantom{+}} |\Psi_0 \rangle^*
\label{eq:dcmatrixelement}
\end{eqnarray}
for the hybridization matrix element between impurity and bath electrons
at site~$n$. Therefore, the matrix elements are alternately real or purely
imaginary. In momentum space, eq.~(\ref{eq:dcmatrixelement})
reads
\begin{equation}
M(k) \equiv 
\langle \Psi_0 |\hat{c}_{k,\sigma}^+\hat{d}_{\sigma}^{\vphantom{+}} |\Psi_0 \rangle = 
\langle \Psi_0 |\hat{c}_{\pi-k,\sigma}^+\hat{d}_{\sigma}^{\vphantom{+}} |\Psi_0 \rangle^*
\; .
\label{eq:dcmatrixelementkspace}
\end{equation}
Since the wave numbers~$k$ enter the single-impurity Anderson model
only via the dispersion relation, $M(k)\equiv M(\epsilon(k))$,
eq.~(\ref{eq:dcmatrixelementkspace}) implies 
\begin{equation}
{\rm Re}M(-\epsilon)= {\rm Re}M(\epsilon)\quad , \quad
{\rm Im}M(-\epsilon)= -{\rm Im}M(\epsilon)
 \end{equation}
because $\epsilon(\pi-k)=-\epsilon(k)$.
We shall use this relation in Sect.~\ref{sec:nonint}.

\subsubsection{Phase shifts and periodic boundary conditions}

Instead of using particle-hole periodic boundary conditions,
we may distribute the phase shift $\Phi=\pm \pi/2$ evenly 
and use periodic boundary conditions.
We rewrite
\begin{equation}
\hat{c}_{n,\sigma}^{\vphantom{+}} 
= \exp\left(\rmi \varphi n\right)\hat{b}_{n,\sigma}^{\vphantom{+}} 
\end{equation}
for $n=0,1,\ldots, L-1$.
Then,
\begin{equation}
\hat{T}=
-\frac{1}{4}  \sum_{n=0,\sigma}^{L-1}\left(  e^{\rmi \varphi} 
\hat{b}_{n,\sigma}^+ \hat{b}_{n+1,\sigma}^{\vphantom{+}}
+ e^{-\rmi \varphi} \hat{b}_{n+1,\sigma}^+ \hat{b}_{n,\sigma}^{\vphantom{+}}\right) \;,
\label{eq:goodformagain}
\end{equation}
where $\hat{b}_{L,\sigma}^{\vphantom{+}}=\hat{b}_{0,\sigma}^{\vphantom{+}}$, 
i.e., the $b$-electrons obey periodic boundary conditions.

When we Fourier transform into momentum space,
we use the wave numbers
\begin{equation}
\tilde{k}_m=\frac{2\pi}{L}m \quad , \quad m=0,1,\ldots,L-1 \; .
\end{equation}
The kinetic energy becomes
\begin{equation}
\hat{T} =\sum_{\tilde{k},\sigma} (-2t\cos(\tilde{k}+\varphi)) 
\hat{b}_{\tilde{k},\sigma}^+\hat{b}_{\tilde{k},\sigma}^{\vphantom{+}} \; .
\end{equation}
Therefore, the dispersion relation 
and the set of accessible $k$-values are still given by eqs.~(\ref{eq:dispersion}) 
and~(\ref{eq:kvaluesallowed}).

The kinetic energy operator~(\ref{eq:goodformagain}) is particle-hole symmetric
under the transformation
\begin{equation}
{\rm ph: } \quad \hat{b}_{n,\sigma}^{\vphantom{+}} \mapsto (-1)^n e^{-2\rmi \varphi n}
\hat{b}_{n,\sigma}^+ \quad \hbox{for}
\quad n=0,1,\ldots, L-1 \; .
\end{equation}
This is readily seen for all electron transfers between sites $n$ and $(n+1)$
for $n=0,1,\ldots,(L-2)$, where the value of $\varphi$ is actually irrelevant.
For the electron transfer between the last and first site, however,
we find
\begin{eqnarray}
&&e^{\rmi\varphi} \hat{b}_{L-1,\sigma}^+\hat{b}_{0,\sigma}^{\vphantom{+}}
+e^{-\rmi\varphi} \hat{b}_{0,\sigma}^+\hat{b}_{L-1,,\sigma}^{\vphantom{+}}
\mapsto \nonumber \\
&&e^{\rmi\varphi} e^{2\rmi\varphi (L-1)}
\hat{b}_{L-1,\sigma}^{\vphantom{+}}\hat{b}_{0,\sigma}^+ 
+e^{-\rmi\varphi} \hat{b}_{0,\sigma}^{\vphantom{+}}
e^{-2\rmi \varphi(L-1)}\hat{b}_{L-1,\sigma}^+
\nonumber \\
\end{eqnarray}
because both the origin and the last site are even. For the transformed term
to become equivalent to the original term, we must impose
\begin{equation}
e^{2\rmi\varphi L}=-1
\end{equation}
which again gives $\varphi=\pm \pi/(2L)$ as in eq.~(\ref{eq:varphipihalf}).

\subsection{Mapping onto a chain problem}
\label{sec:mappingringtochain}

\subsubsection{Canonical transformation}

\begin{figure}[t]
\includegraphics[width=8.5cm]{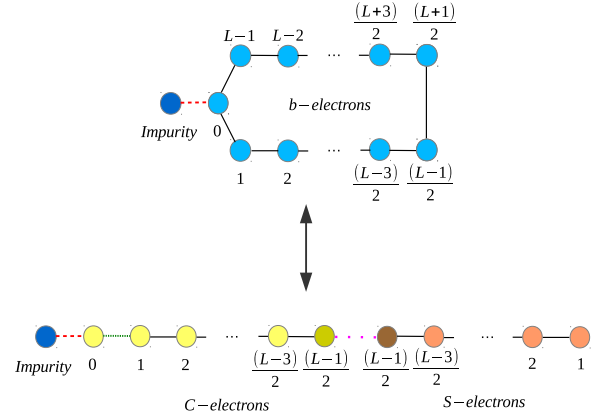}
\caption{(Color online) SIAM in ring geometry, eq.~(\ref{eq:SIAMH}), with 
the kinetic energy from
eq.~(\ref{eq:goodformagain}), and in two-chain geometry with
the kinetic energy from eq.~(\ref{eq:TfortwochainSIAM}).
Bonds
with the same color have the same electron transfer amplitudes.
The dotted bond between the $C$-electron and $S$-electron chains
has a complex hopping amplitude.\label{fig:ringandchains}}
\end{figure}

For $n=1,2,\ldots,(L-1)/2$ we perform the canonical 
transformation~\cite{PhysRevB.74.195112,PhysRevB.84.115403}
\begin{eqnarray}
\hat{C}_{n,\sigma}^{\vphantom{+}} &=& \sqrt{\frac{1}{2}} 
\left(e^{\rmi n \varphi}\hat{b}_{n,\sigma}^{\vphantom{+}}
+e^{-\rmi n \varphi}\hat{b}_{L-n ,\sigma}^{\vphantom{+}}\right) 
\; , \nonumber \\
\hat{S}_{n,\sigma}^{\vphantom{+}} &=& \sqrt{\frac{1}{2}} 
\left(e^{\rmi n\varphi}\hat{b}_{n,\sigma}^{\vphantom{+}}
-e^{-\rmi n\varphi}\hat{b}_{L-n ,\sigma}^{\vphantom{+}}\right) 
\end{eqnarray}
with the inverse transformation
\begin{eqnarray}
\hat{b}_{n,\sigma}^{\vphantom{+}} &= &\sqrt{\frac{1}{2}} e^{-\rmi n \varphi}
\left(\hat{C}_{n,\sigma}^{\vphantom{+}}
+\hat{S}_{n,\sigma}^{\vphantom{+}}\right) \; , \nonumber\\
\hat{b}_{L-n,\sigma}^{\vphantom{+}} &=& \sqrt{\frac{1}{2}} 
e^{\rmi n \varphi}\left(\hat{C}_{n,\sigma}^{\vphantom{+}}
-\hat{S}_{n,\sigma}^{\vphantom{+}}\right) \; .
\end{eqnarray}
The kinetic energy becomes
\begin{eqnarray}
-4\hat{T}&=& \sum_{\sigma} \left(\hat{B}_{1,\sigma} +\hat{B}_{2,\sigma}\right)
\nonumber \\
&& +\sum_{n=1,\sigma}^{(L-3)/2} \left( 
 \hat{C}_{n,\sigma}^+ \hat{C}_{n+1,\sigma}^{\vphantom{+}}
+ \hat{C}_{n+1,\sigma}^+ \hat{C}_{n,\sigma}^{\vphantom{+}}\right)\nonumber\\
&& +\sum_{n=1,\sigma}^{(L-3)/2} \left( 
\hat{S}_{n,\sigma}^+ \hat{S}_{n+1,\sigma}^{\vphantom{+}}
+ \hat{S}_{n+1,\sigma}^+ \hat{S}_{n,\sigma}^{\vphantom{+}}\right)  \; ,
\label{eq:TfortwochainSIAM}
\end{eqnarray}
where the boundary term at the left chain end reads
\begin{equation}
\hat{B}_{1,\sigma}=  \sqrt{2} \left(\hat{c}_{0,\sigma}^+ \hat{C}_{1,\sigma}^{\vphantom{+}}
+\hat{C}_{1,\sigma}^+\hat{c}_{0,\sigma}^{\vphantom{+}}\right) \; .
\end{equation}
In contrast to open boundary conditions or boundary conditions that violate
  particle-hole symmetry, the connection term 
between the $C$-electrons  and $S$-electrons is finite. The term at $n=(L-1)/2$
is given by
\begin{equation}
\hat{B}_{2,\sigma} = \rmi \left( 
\hat{S}_{(L-1)/2,\sigma}^+ \hat{C}_{(L-1)/2,\sigma}^{\vphantom{+}} -
\hat{C}_{(L-1)/2,\sigma}^+ \hat{S}_{(L-1)/2,\sigma}^{\vphantom{+}}\right)  .
\label{eq:rightboundary} 
\end{equation}
The ring and two-chain geometries are shown in Fig.~\ref{fig:ringandchains}.

Note that the particle-hole transformation for the kinetic energy in the
two-chain formulation is non-trivial,
\begin{eqnarray}
{\rm ph:} \;
(-1)^n e^{-\rmi n \varphi }
\hat{C}_{n,\sigma}^{\vphantom{+}} &\mapsto &
\cos(\varphi n)\hat{C}_{n,\sigma}^+ -\rmi \sin(\varphi n)\hat{S}_{n,\sigma}^+ \; , 
\nonumber \\
(-1)^n e^{-\rmi n \varphi }\hat{S}_{n,\sigma}^{\vphantom{+}} &\mapsto & 
\cos(\varphi n)\hat{S}_{n,\sigma}^+ -\rmi \sin(\varphi n)\hat{C}_{n,\sigma}^+ \nonumber \\
\end{eqnarray}
for $n=1,\ldots, L-1$, and $\hat{c}_{0 ,\sigma}^{\vphantom{+}}
\mapsto \hat{c}_{0 ,\sigma}^+$ as before.

For comparison, we give in appendix~\ref{app:A}
the standard derivation of the chain geometry from the ring geometry
via the Lanczos procedure.~\cite{PhysRevB.88.245113}
The chains of $C$-electrons and $S$-electrons do not decouple because
particle-hole symmetry for odd chain lengths~$L$ is not compatible 
with inversion symmetry.
Apparently, it is not advantageous numerically to investigate
a ring geometry at particle-hole symmetry. It is more favorable
to start from an inversion-symmetric chain where
the $C$-electron and $S$-electron chains decouple. In the following
we shall investigate the consequences
of an ad-hoc decoupling of the two chains.
Note that this does not influence the results in the thermodynamic limit
where boundary conditions become irrelevant.

\subsubsection{Chain separation}

For large rings, the inter-chain coupling is 
small for 
two reasons.
First, 
as seen from eq.~(\ref{eq:rightboundary}), 
the chains for the $C$-electrons and $S$-electrons are coupled
at a single site only, namely, at the 
chain 
center $n=(L-1)/2$.
Second, in the SIAM
the interesting physics happens at and around the origin, i.e.,
at the left boundary of the $C$-electron chain. 
Because of their large separation, 
we can expect that the right 
half of the chain has little effect
on the physics at the left boundary.

The chain-separated SIAM reads
\begin{equation}
\hat{H}= \hat{H}^C +\hat{T}^S \; .
\end{equation}
The undisturbed chain of anti-symmetric standing waves of 
length~$(L-1)/2$ is described by 
\begin{equation}
\hat{T}^S= 
-\frac{1}{4}\sum_{n=1,\sigma}^{(L-3)/2} \left( 
 \hat{S}_{n,\sigma}^+ \hat{S}_{n+1,\sigma}^{\vphantom{+}}
+ \hat{S}_{n+1,\sigma}^+ \hat{S}_{n,\sigma}^{\vphantom{+}}\right)
\end{equation}
The electrons on the chain of symmetric standing waves 
of length $(L+1)/2$
couple to the impurity at the origin,
\begin{eqnarray}
\hat{H}^C&=& \hat{H}_0^C +U\left(\hat{n}_{d,\uparrow}-\frac{1}{2}\right)
\left(\hat{n}_{d,\downarrow}-\frac{1}{2}\right)  \nonumber \; ,\\
\hat{H}_0^C &=& \hat{T}^C +\hat{V} \nonumber \; ,\\
\hat{T}^C&=& -\frac{\sqrt{2}}{4} \sum_{\sigma}
\left(\hat{C}_{0,\sigma}^+ \hat{C}_{1,\sigma}^{\vphantom{+}}
+\hat{C}_{1,\sigma}^+\hat{C}_{0,\sigma}^{\vphantom{+}}\right) \nonumber \\
&& -\frac{1}{4}\sum_{n=1,\sigma}^{(L-3)/2} \left( 
 \hat{C}_{n,\sigma}^+ \hat{C}_{n+1,\sigma}^{\vphantom{+}}
+ \hat{C}_{n+1,\sigma}^+ \hat{C}_{n,\sigma}^{\vphantom{+}}\right)\; ,
\nonumber \\
\hat{V}&=& V \sum_{\sigma} \left( \hat{d}_{\sigma}^{+}
\hat{C}_{0,\sigma}^{\vphantom{+}} 
+ \hat{C}_{0,\sigma}^{+} \hat{d}_{\sigma}^{\vphantom{+}} \right) \; ,
\label{eq:defHzerosinglechain}
\end{eqnarray}
where we identified $\hat{c}_{0,\sigma}^{\vphantom{+}}
\equiv \hat{C}_{0,\sigma}^{\vphantom{+}}$ to keep the notation consistent.

When we ignore the chain coupling 
term $\hat{B}_{2,\sigma}$, 
we can factorize the ground state
into the contributions from the chains $C$ and $S$,
\begin{equation}
|\Psi_0\rangle = |\Psi_0^C\rangle |\Psi_0^S\rangle  \; ,
\label{eq:factorize}
\end{equation}
where the upper index refers to the two commuting parts of the Hamiltonians
for the $C$-electrons and $S$-electrons and 
$|\Psi_0^{C,S}\rangle$ are normalized to unity.

The mapping is advantageous for the DMRG treatment because 
we do not have to treat a ring geometry of $L$ sites with periodic boundary conditions
but a chain with $(L+1)/2+1$ sites 
where open boundary conditions apply. The 
$C$-electron chain has only about half as many sites 
as the ring which essentially
doubles the system sizes that can be treated numerically for the ring geometry.

Note, however, that $\hat{H}^C$ does not obey particle-hole symmetry
for finite~$L$ but only in the thermodynamic limit.
Deviations from particle-hole symmetry can be monitored by
investigating the site occupancy of the impurity.
Deviations from the exact value of one half, see
eq.~(\ref{eq:averagedisonehalf}), can be used to quantify
the violation of particle-hole symmetry, see Sect.~\ref{sec:DMRG}.

In the Numerical Renormalization Group approach, the SIAM is 
directly considered in energy space. After an appropriate discretization,
the resulting Wilson chain is treated numerically.~\cite{RevModPhys.80.395}
In our approach, we map the Hamiltonian on {\em finite} rings 
to a 
chain
while keeping particle-hole symmetry and the band-width finite. 
The use of a Hamiltonian on a ring geometry permits the direct application,
comparison, and assessment of lattice-based variational 
methods such as Hartree-Fock, Gutzwiller, and DMRG, as done in this work.
Our results also permit to assess the quality of 
other present, and conceivable future, 
many-body methods for lattice Hamiltonians.

\subsection{Spin correlation function}

In this work we visualize the Kondo screening cloud
for the single-impurity Anderson model. To this end, we calculate
the spin correlation function between the impurity and bath sites.

\subsubsection{Definition and general properties}

Due to the spin-rotational invariance of the model
it is sufficient to study the spin correlation function
along the spin quantization axis. 
The local correlation function is defined by
\begin{eqnarray}
C_{dd}^S &=& \langle \Psi_0 | \hat{S}^z_d \hat{S}^z_d | \Psi_0 \rangle
= \frac{1}{4} 
\langle \Psi_0 | \left(\hat{n}_{\uparrow}^d -\hat{n}_{\downarrow}^d\right)^2 
| \Psi_0 \rangle \nonumber \\
&=&  \frac{1}{4}- \frac{1}{2}  
\langle \Psi_0 | \hat{n}_{\uparrow}^d \hat{n}_{\downarrow}^d
| \Psi_0 \rangle \; , 
\end{eqnarray}
where we used particle-hole symmetry~(\ref{eq:averagedisonehalf})
in the last step.
The value for the on-site spin correlation interpolates between
the itinerant limit, $C_{dd}(U=0)=1/8$, and the atomic limit,
$C_{dd}(W=0)=1/4$.

The correlation function between the impurity site and the 
bath site~$r$ is defined by
\begin{eqnarray}
C_{dc}^S (r) &=& \langle \Psi_0 | \hat{S}^z_d \hat{S}^z_{r,c} | \Psi_0 \rangle
\label{eq:defCDC}
\\
&=& \frac{1}{4} 
\langle \Psi_0 | \left(\hat{n}_{\uparrow}^d -\hat{n}_{\downarrow}^d\right)
\left(\hat{c}_{r,\uparrow}^+ \hat{c}_{r,\uparrow}^{\vphantom{+}}-
\hat{c}_{r,\downarrow}^+ \hat{c}_{r,\downarrow}^{\vphantom{+}}\right) 
 | \Psi_0 \rangle \; . \nonumber 
\end{eqnarray}
Due to inversion symmetry we have 
\begin{equation}
C_{dc}^S (L-r)= C_{dc}^S (r)
\label{eq:inversion}
\end{equation}
for $1\leq r \leq (L-1)/2$. 

To visualize the screening of the impurity spin,
we define ${\cal S}(0)=C_{dd}^S + C_{dc}^S (0)$ and, for $R\geq 1$,
\begin{equation}
{\cal S}(R)
= C_{dd}^S + C_{dc}^S (0) + \sum_{r=1}^R 
\left( C_{dc}^S (r) + C_{dc}^S (L-r)\right) \; .
\label{eq:calS}
\end{equation}
It describes the amount of the unscreened spin 
at distance $R$ from the impurity site.~\cite{PhysRevB.80.205114}
The impurity is completely screened by all bath electrons.
To see this we consider ${\cal S}\left((L-1)/2\right)$ on finite systems,
\begin{eqnarray}
{\cal S}((L-1)/2)&=&
\langle \Psi_0 | \hat{S}^z_d \biggl(\hat{S}^z_d +\sum_{r=0}^{L-1}\hat{S}^z_{r,c}\biggr)| 
\Psi_0 \rangle\nonumber \\
&=&\langle \Psi_0 | \hat{S}^z_d \hat{S}^z|\Psi_0 \rangle =0
\label{eq:totalscreening}
\end{eqnarray}
because $|\Psi_0 \rangle$ is an eigenstate of the operator $\hat{S}^z$
for the total spin in $z$-direction with eigenvalue zero.

\subsubsection{Spin correlations in two-chain geometry}

For the first site of the chain we have
\begin{equation}
C_{dc}^S (0)=\frac{1}{4} \langle \Psi_0^C | 
\left(\hat{n}_{\uparrow}^d -\hat{n}_{\downarrow}^d\right) 
\left(\hat{C}_{0,\uparrow}^+ \hat{C}_{0,\uparrow}^{\vphantom{+}}-
\hat{C}_{0,\downarrow}^+ \hat{C}_{0,\downarrow}^{\vphantom{+}}\right) 
| \Psi_0^C\rangle\;,
\label{eq:requalsone} 
\end{equation}
where we used eq.~(\ref{eq:factorize}) 
and the normalization of $|\Psi_0^S\rangle$.

For the spin correlation function between the impurity site
and a bath site at distance~$1\leq r\leq (L-1)/2$
we use inversion symmetry~(\ref{eq:inversion}) to write
\begin{eqnarray}
C_{dc}^S (r) &=& \langle \Psi_0 | \hat{S}^z_d \hat{S}^z_r | \Psi_0 \rangle
\nonumber \\
&=& \frac{1}{2} \langle \Psi_0 | \hat{S}^z_d \left(
\hat{S}^z_r  + \hat{S}^z_{L-r}\right)| \Psi_0 \rangle \label{eq:generalr}\\
&=& 
\frac{1}{8} 
\langle \Psi_0^C | 
\left(\hat{n}_{\uparrow}^d -\hat{n}_{\downarrow}^d\right) 
\left(\hat{C}_{r,\uparrow}^+ \hat{C}_{r,\uparrow}^{\vphantom{+}}-
\hat{C}_{r,\downarrow}^+ \hat{C}_{r,\downarrow}^{\vphantom{+}}\right) 
| \Psi_0^C\rangle\;, \nonumber 
\end{eqnarray}
where we used the mapping onto the chain operators in the second step,
\begin{equation}
\hat{c}_{r,\sigma}^+ \hat{c}_{r,\sigma}^{\vphantom{+}}
+ 
\hat{c}_{L-r,\sigma}^+ \hat{c}_{L-r,\sigma}^{\vphantom{+}}
= 
\hat{C}_{r,\sigma}^+ \hat{C}_{r,\sigma}^{\vphantom{+}}
+ 
\hat{S}_{r,\sigma}^+ \hat{S}_{r,\sigma}^{\vphantom{+}} \; ,
\end{equation}
and the factorization~(\ref{eq:factorize}) in the last step; 
recall that the $S$-electron system is a paramagnetic Fermi sea,
$\langle \Psi_0^S | 
\hat{S}_{r,\uparrow}^+ \hat{S}_{r,\uparrow}^{\vphantom{+}}
- 
\hat{S}_{r,\downarrow}^+ \hat{S}_{r,\downarrow}^{\vphantom{+}}
| \Psi_0^S\rangle=0$.

Equations~(\ref{eq:requalsone}) and~(\ref{eq:generalr})
must be evaluated using DMRG, in general.
For $U=0$, the ground-state energy and the spin correlation function 
can be evaluated analytically to a 
large extent, as we show next.

\section{Non-interacting SIAM}
\label{sec:nonint}

It is instructive to discuss the non-interacting SIAM.
Moreover, it provides the basis for the Gutzwiller approach
in Sect.~\ref{sec:Gutzwiller}.
We defer the details of the derivation to the appendix,
and merely summarize the relevant results.

\subsection{Ground-state energy}

The ground-state energy sums the band contribution
and the energy of the doubly occupied bound state.
The total energy reads
\begin{eqnarray}
e_0(V)
&=&e_0^{\rm band} (V)+e_0^{\rm b}(V) \nonumber \\
&=&
\frac{1}{2\pi}\biggl[
-\pi +2v_+\arctan\left(\frac{1}{v_-}\right) \nonumber \\
&& \hphantom{\frac{1}{2\pi}\biggl[}
+ v_-\ln\left(\frac{v_+-1}{v_++1}
\right)
\biggr] +(1-v_+) \; ,
\label{eq:finalenergyTDL}
\end{eqnarray}
where
\begin{equation}
v_{\pm}(V)\equiv v_{\pm}=\frac{\sqrt{\sqrt{1+64V^4}\pm 1}}{\sqrt{2}} \; .
\label{eq:defvplusminus}
\end{equation}
The small-$V$ expansion becomes
\begin{equation}
e_0^{\rm small}(V)=
\frac{4V^2}{\pi} \left(\ln(V^2)+\ln(2)-1\right) -4V^4 \; .
\label{eq:Vsmallgsenergynonint}
\end{equation}
Corrections are of the order $V^6\ln(V^2)$. 
For $V=0.1$, the approximate formula works very well.
We have $e_0(0.1)=-0.06291$
whereas the approximation gives
$e_0^{\rm small}(0.1)=-0.06294$, with a relative error of
less than one per mill.

To determine the Gutzwiller variational energy we also need the 
derivative of the ground-state energy.
We have
\begin{eqnarray}
e_0'(x)&=&\frac{4x}{\pi (v_+(x)^2+v_-(x)^2)}
\nonumber \\
&& \biggl[ 
2\pi v_-(x) \left(\frac{\arccot(v_-(x))}{\pi}-1\right) \nonumber \\
&& \hphantom{\biggl[ }
+ v_+(x) \ln\left(\frac{v_+(x)-1}{v_+(x)+1}\right) \biggr] \; .
\end{eqnarray}
For small~$x$ this reduces to
\begin{equation}
e_0'(x\ll 1)\approx (8x/\pi)\ln\left( 2x^2\right) \; .
\label{eq:smallVderivative}
\end{equation}

\subsection{Magnetization and zero-field magnetic susceptibility}
\label{sec:magsusUzero}

We introduce the magnetic energy scale
$B_{\text{imp}}\equiv B=(g\mu_{\rm B}/2){\cal H}$ where ${\cal H}$ 
is the external magnetic field at the impurity, 
and express the impurity magnetization $M(V,{\cal H})=g\mu_{\rm B} m(V,B)$
in terms of the impurity spin in $z$-direction,
\begin{equation}
m(V,B)=\langle \hat{S}^z_d \rangle =
(\langle \hat{n}_{d,\uparrow}-\hat{n}_{d,\downarrow}\rangle)/2 \; .
\end{equation}
The magnetic susceptibility follows from
\begin{equation}
\chi(V,B)= \frac{\partial M({\cal H})}{\partial {\cal H}}
= \left(\frac{g\mu_{\rm B}}{2}\right)^2 
\frac{\partial [2 m(V,B)]}{\partial B} \; .
\end{equation}
We give closed expressions for $m(V,B)$ and $\chi(V,B)$ 
for the non-interacting SIAM
in one dimension.

\subsubsection{Magnetization}

For the one-dimensional non-interacting SIAM we find
for a magnetic field that acts solely at the impurity
\begin{eqnarray}
2m(V,B)&=& Z[v_{\rm b}(V,B)]-Z[v_{\rm b}(V,-B] \nonumber \\
&& +\!\sum_{\sigma_n=\pm 1} \int_{-1/2}^0
\frac{\rmd \omega}{\pi} 
\frac{\sigma_n \Gamma\sqrt{1-4\omega^2}}{
(\omega+\sigma_n B)^2(1-4\omega^2)+ \Gamma^2}
\nonumber \\
\label{eq:magB}
\end{eqnarray}
with $\Gamma=2V^2$. Here, $v_{\rm b}(V,B)<-1/2$ is the energy of the
bound state outside the band. It is the root of $P_+(\omega,B)$, i.e.,
$P_+(v_{\rm b}(V,B))=0$, with
\begin{equation}
P_+(\omega,B)=\omega+B +\frac{2V^2}{\sqrt{4\omega^2-1}} \; .
\end{equation}
Moreover, the weight of the bound state in the $d$-electron spectral function
is given by
\begin{equation}
Z[v_{\rm b}(V,B)]= \left[
1-\frac{8V^2 v_{\rm b}(V,B)}{(4[v_{\rm b}(V,B)]^2-1)^{3/2}}
\right]^{-1} \; .
\label{eq:Zmag}
\end{equation}
In general, the magnetization must be determined numerically from
eqs.~(\ref{eq:magB}) and~(\ref{eq:Zmag}).

\subsubsection{Small hybridizations}

In the limit $V\ll 1$, we ignore the bound-state contribution of order~$V^4$,
and simplify the magnetization to
\begin{eqnarray}
m(V,B)&=& \int_{-\infty}^0 \frac{\rmd \omega}{2\pi}
\left[ \frac{\Gamma}{(\omega+B)^2+\Gamma^2}-
\frac{\Gamma}{(\omega-B)^2+\Gamma^2} \right]
\nonumber \\
&=& \int_0^B \frac{\rmd \omega}{\pi} \frac{\Gamma}{\omega^2+\Gamma^2}
= \frac{1}{\pi} \tan^{-1}(B/\Gamma) \; .
\label{eq:magsmallVnonint}
\end{eqnarray}
The width~$\Gamma$ of the $d$-electron spectral function
is the relevant energy scale for magnetic excitations.

For small hybridizations, the susceptibility becomes
\begin{equation}
\chi(V,B)=\left( \frac{g\mu_{\rm B}}{2}\right)^2 
\frac{2}{\pi}\frac{\Gamma}{B^2+\Gamma^2}
\label{eq:chiBnonint}
\end{equation}
with the zero-field limit
\begin{equation}
\chi_0(V) = \left(\frac{g\mu_{\rm B}}{2}\right)^2 \frac{2}{\pi \Gamma} \; .
\end{equation}
As seen from Eq.~(\ref{eq:chiBnonint}), in the limit $V\to 0$
the magnetic susceptibility is proportional to the zero-field $d$-electron 
spectral function, $\mu(V,B) \propto D_{d,d,\sigma}(B)$.

\subsubsection{External magnetic field for impurity and bath electrons}

For the case $B_{\text{imp}}=B_{\text{bath}}\equiv B$, 
the bound states are shifted in energy,
\begin{equation}
v_{\rm b}(V,B)=-B -v_+(V)/2\; ,
\end{equation}
but their weights $Z[v_{\rm b}(V,B)]$ do not change because
$v_{\rm b}(V,\pm B)\pm B = -v_+(V)/2$ in both cases.
The rigid shift in single-particle energies by the magnetic field also
guarantees that the impurity remains half filled on average for all external fields.
The impurity magnetization becomes  ($B\ll W$) 
\begin{eqnarray}
2\tilde{m}(V,B)&=& \sum_{\tau=\pm1}
\int_{-\infty}^0 \frac{V^2\rho_0(\omega+\tau B)}{(\omega+\tau B)^2
+(\pi V^2 \rho_0(\omega+\tau B))^2} \nonumber\\
&=&2\int_0^{B}
\frac{\rmd \omega}{\pi} 
\frac{\Gamma\sqrt{1-4\omega^2}}{\omega^2(1-4\omega^2)+ \Gamma^2}\; .
\label{eq:magBfull}
\end{eqnarray}
For small hybridizations,
$\tilde{m}(V,B)$ reduces to the result for $m(V,B)$ 
in eq.~(\ref{eq:magsmallVnonint}).

\begin{figure}[t]
\includegraphics[width=8.5cm]{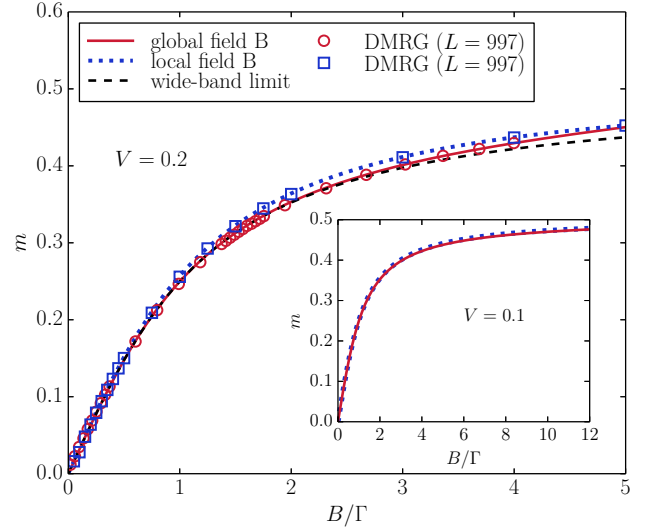}
\caption{(Color online) 
Impurity magnetization for the non-interacting symmetric SIAM for $V=0.2$ 
as a function of $B=g\mu_{\rm B}{\cal H}/2$. 
We show $m(0.2,B)$, eq.~(\ref{eq:magB}) (local field, blue dotted line),
$\tilde{m}(0.2,B)$, eq.~(\ref{eq:magBfull}) (global field, red straight line),
and the wide-band limit~(\ref{eq:magsmallVnonint})
(local field, black dashed line), together with the corresponding 
DMRG data (symbols, $L=997$ sites). 
Inset: impurity magnetization for $V=0.1$.\label{fig:magnonint}}
\end{figure}

We show the impurity magnetization as a function of $B/\Gamma$ 
in Fig.~\ref{fig:magnonint}.
Only for $V=0.2$ and $B\gtrsim 2\Gamma$, 
there is a discernible difference between the curves for $m(0.2,B)$, 
eq.~(\ref{eq:magB}), where the external field is confined to the impurity,
and $\tilde{m}(0.2,B)$, eq.~(\ref{eq:magBfull}), where the external field
polarizes all electrons. In both cases, the DMRG data, see Sect.~\ref{sec:DMRG}, 
faithfully reproduce the analytic results,
within small errors resulting from finite-size effects.

The wide-band limit closely follows the result for the global magnetic field. 
This indicates that the difference between applying the external field locally or globally
is mostly due to the polarization of the bound-states for a local field. 
The bound states have a noticeable weight for $V=0.2$.

Since the weight of the bound states is of the order $V^4$, their contribution 
is much smaller for $V=0.1$. Correspondingly, as seen from the inset
of Fig.~\ref{fig:magnonint}, the discrepancies between
the magnetization curves for local and global external fields
become very small. 
Since we shall work with $V\leq 0.1$ for the rest of the paper,
we will restrict ourselves to purely local external magnetic fields,
and shall safely ignore the influence of the magnetic field on the bath electrons.

\subsection{Spin correlation function}
\label{subsec:spinCFnonint}

\subsubsection{General properties}

Starting from eq.~(\ref{eq:defCDC}) we can use Wick's theorem 
and spin symmetry to show that
\begin{equation}
C_{dc}^S(r) = -\frac{1}{2}
\left| \langle \Phi_0 | \hat{c}_{r,\uparrow}^+\hat{d}_{\uparrow}^{\vphantom{+}}
| \Phi_0 \rangle \right|^2 \equiv -\frac{1}{2} M_r^2
\label{eq:wickfactorization}
\end{equation}
for the ground state $|\Phi_0\rangle$ of the non-interacting SIAM.
The matrix element is calculated in the appendix,
\begin{eqnarray}
M_r
&=& \sqrt{\frac{1}{L}}
\sum_k e^{-\rmi k r} 
\langle \Phi_0 | 
\hat{c}_{k,\uparrow}^+\hat{d}_{\uparrow}^{\vphantom{+}}
| \Phi_0 \rangle  \nonumber \\
&=& V \int_0^{\pi} \frac{\rmd k}{\pi} \cos(kr) M[-\cos(k)/2] \; ,
\label{eq:matrixelement}
\end{eqnarray}
where we took the thermodynamic limit and used $\epsilon(k)=-\cos(k)/2$
in one dimension. Since $M(\epsilon)$ is real,
particle-hole symmetry leads to $M(-\epsilon)=M(\epsilon)$ so that
the matrix element vanishes for odd sites, $M_{2m-1}=0$, $m\geq 1$.
For even sites we find the bound-state and band contributions
\begin{eqnarray}
M_{2m}^{\rm b}&=& -\frac{2VZ(V)}{\sqrt{v_+^2-1}}
\left(\sqrt{v_+^2-1}+v_+\right)^{-2m} \; , \label{eq:M2mintegration}\\
M_{2m}^{\rm band}&=& -\frac{2V(-1)^m}{\pi}
\int_0^{\pi} \rmd y \frac{\cos(y/2)}{\sin^2(y)+64 V^4}\nonumber \\
&&\hphantom{-\frac{2V(-1)^m}{\pi}}
\times [ \sin(y) \cos(my)+8V^2\sin(my)] \nonumber 
\end{eqnarray}
with the pole frequency $\omega_{\rm b}=-v_+/2$ and the pole weight
\begin{equation}
Z(V) =\frac{1}{1+4V^2v_+/v_-^3} 
\end{equation}
and $v_{\pm}$ from eq.~(\ref{eq:defvplusminus}).

\subsubsection{Small hybridizations}

The bound-state contribution $M_r^{\rm b}$ 
is of the order $V^3$ for small~$V$, 
and becomes exponentially small for $r\gg 1/(4V^2)$.
For small~$V$,~\cite{1742-5468-2014-4-P04011} 
the band contribution 
is dominated by the region $y\to 0$ in the integrand in eq.~(\ref{eq:M2mintegration}).
We thus approximate for small~$V$
\begin{eqnarray}
M_{2m}
&\approx& -\frac{2V}{\pi}  \int_0^{\infty} \rmd x
\frac{x\cos(8V^2m x)+\sin(8V^2m x)}{x^2+1}\nonumber \\
&=& 
 (-1)^m \frac{2V}{\pi} 
e^{\alpha}\Ei(-\alpha) 
\; , \quad \alpha = 8V^2m \; , 
\label{eq:M2mapprox}
\end{eqnarray}
where 
\begin{equation}
\Ei(x)= - \int_{-x}^{\infty}\rmd t \frac{e^{-t}}{t}
\end{equation}
is the exponential integral.
Thus, the spin correlation function approximately becomes ($m\neq 0$)
\begin{equation}
C_{dc}^S(2m) \approx -\frac{1}{2} 
\left(\frac{2V}{\pi} e^{\alpha}\Ei(-\alpha) \right)^2
\; , \quad \alpha = 8V^2m \; .
\label{eq:spinapprox}
\end{equation}

\begin{figure}[t]
\includegraphics[width=8.5cm]{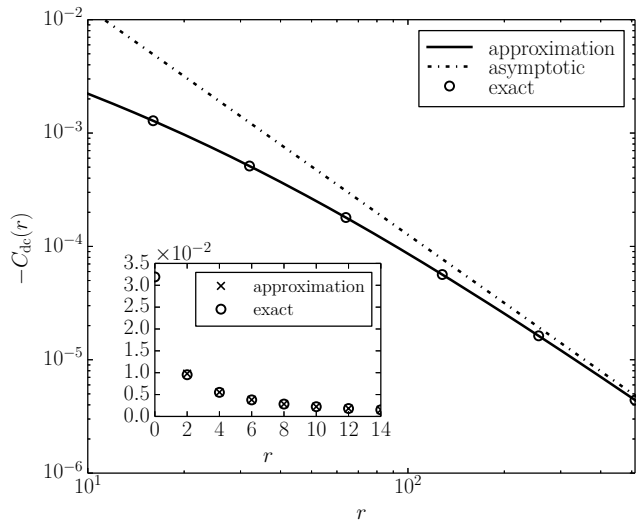}
\caption{Spin correlation function for the one-dimensional
non-interacting symmetric SIAM for $V=0.1$ (circles)
on a log-log scale. The analytic result~(\ref{eq:spinapprox}) 
is shown as a straight line. The asymptotics~(\ref{eq:analytdecay})
is shown as dash-dotted line, and the exact values~(\ref{eq:M2mintegration})
are shown as open symbols.
Inset: Spin correlation function 
for small distances on a linear scale.\label{fig:spinCFnonint}}
\end{figure}

In Figs.~\ref{fig:spinCFnonint} and~\ref{fig:screenedspin} 
we show the spin correlation function 
and the unscreened spin for the non-interacting symmetric SIAM
in one dimension for $V=0.1$. 
As seen from Fig.~\ref{fig:spinCFnonint},
the spin correlation function decays to
zero proportional to $1/m^2$.
The exact result~(\ref{eq:M2mintegration}) and the
approximate formula~(\ref{eq:M2mapprox}) yield almost identical results,
already for $m\geq 2$. For $m\ge 10$, the relative error is of the order 
$10^{-4}$ for $V=0.1$.

Correspondingly, the unscreened spin shown in Fig.~\ref{fig:screenedspin} 
decays to zero proportional to $1/m$.
For small~$V$, the screening is fairly inefficient and, correspondingly, 
the screening cloud extends very far into the host metal, even in the case
of the non-interacting SIAM.

\subsubsection{Small hybridizations and large distances}

Here, we work out the long-range behavior of the spin correlation function.
The asymptotic regime is reached for $\alpha\gg 1$, i.e., 
for $m\gg 1/(8V^2)$, where $\exp(\alpha)\Ei(\alpha)\approx -1/\alpha$
in eq.~(\ref{eq:spinapprox}).
For the correlation function we find in this region 
\begin{equation}
C_{dc}^S(2m\gg 1/(4V^2)) \approx 
-\frac{1}{2}\left(\frac{2V}{\pi\alpha}\right)^2 =
-\frac{1}{32\pi^2 V^2} \frac{1}{m^2}\; .
\label{eq:analytdecay}
\end{equation}
For the non-interacting symmetric SIAM in one dimension, 
the spin correlations between the impurity and a
bath electron at site $2m$ asymptotically decays
proportional to $1/(2m)^2$, see Fig.~\ref{fig:spinCFnonint}.

\begin{figure}[t]
\includegraphics[width=8.5cm]{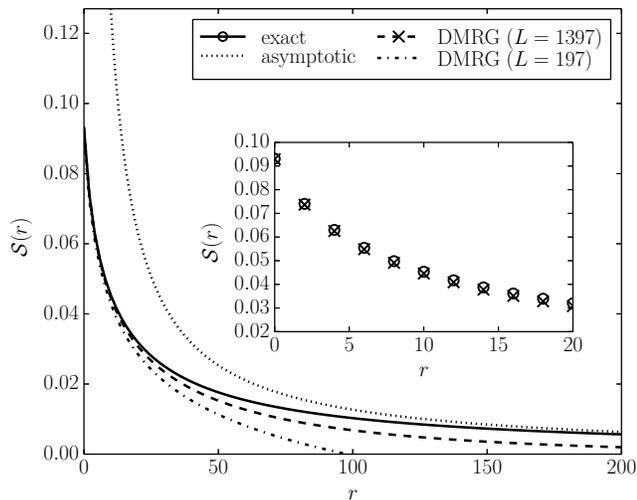}
\caption{Unscreened spin ${\cal S}(r)$ at distance~$r$ from the impurity site,
see eq.~(\ref{eq:calS}),
for the one-dimensional non-interacting 
symmetric single-impurity Anderson model for $V=0.1$.
The analytic result based on eq.~(\ref{eq:spinapprox})
is shown as a 
solid line.
The asymptotic result~(\ref{eq:unscreenddecay})
is presented by a 
dotted line. The DMRG data for $L=197, 1397$ sites
are given by dashed lines, see Sect.~\ref{sec:DMRG}.
Inset: Unscreened spin for small distances, DMRG data for $L=1397$ 
sites.\label{fig:screenedspin}}
\end{figure}

The matrix element $M_{2m}$ at $\alpha=1$ ($2m=1/(4V^2)$)
is already very small, of the order $V^2$ in the asymptotic region.
Nevertheless, the contribution to the screening is finite even for $V\to 0$.
The spins for $|m|>1/(8V^2)$ ($\alpha>1$) contribute approximately
\begin{eqnarray}
\frac{\Delta S}{C_{dd}} &\approx& 8\left(\frac{2V}{\pi}\right)^2
\int_{1/(8V^2)}^{\infty}\rmd m \left(e^{\alpha}\Ei(-\alpha) \right)^2\nonumber \\
&=& \frac{4}{\pi^2} \int_1^{\infty}\rmd \alpha \left(e^{\alpha}\Ei(-\alpha) \right)^2
\approx 0.23 \; .
\end{eqnarray}
The sites for $|m|>1/(8V^2)$ contribute about 25\% 
to the total screening of the spin at the impurity site where $C_{dd}=1/8$.

Indeed, for large distances~$r=2m$ from the impurity, the unscreened spin
decays only proportional to $1/r$,
\begin{equation}
{\cal S}(r\gg 1) \sim \frac{1}{16\pi^2V^2} \frac{1}{r} \;,
\label{eq:unscreenddecay}
\end{equation}
as follows from eq.~(\ref{eq:totalscreening}) when we employ
the Euler-Maclaurin formula for the asymptotic 
expression~(\ref{eq:analytdecay}). This is shown in 
Fig.~\ref{fig:screenedspin}.

\section{Gutzwiller variational approach}
\label{sec:Gutzwiller}

In this section we define the Gutzwiller variational state and 
determine its variational parameters from minimizing the 
variational ground-state energy.~\cite{Schoenhammer1990,PhysRevB.41.9452}
Moreover, we determine the 
variational magnetization, zero-field susceptibility, and
spin-spin correlation function between the impurity site and
the host electrons.

\subsection{Ground-state energy}

\subsubsection{Definition}

The Gutzwiller wave function for the symmetric SIAM reads
\begin{eqnarray}
|\Psi_{\rm G} \rangle &=&
\biggl[ \lambda_d \bigl(\hat{n}_{\uparrow}^d\hat{n}_{\downarrow}^d
+(1- \hat{n}_{\uparrow}^d)(1-\hat{n}_{\downarrow}^d)\bigr) \nonumber \\
&& +\lambda_{\sigma}\bigl( 
\hat{n}_{\uparrow}^d(1-\hat{n}_{\downarrow}^d) +
(1-\hat{n}_{\uparrow}^d)\hat{n}_{\downarrow}^d\bigr) \biggr]
| \Phi_0 \rangle \; ,
\end{eqnarray}
where $|\Phi_0\rangle$ is a normalized single-particle product state.

The Gutzwiller wave function is normalized,
\begin{equation}
\langle \Psi_{\rm G} | \Psi_{\rm G}\rangle =1 \; ,
\end{equation}
and symmetric,
\begin{equation}
\langle \Psi_{\rm G} | \hat{n}_{d,\sigma} |\Psi_{\rm G}\rangle =\frac{1}{2}
\; ,
\end{equation}
if we use a symmetric single-particle product state,
\begin{equation}
\langle \Phi_0 | \hat{n}_{d,\sigma} |\Phi_0\rangle =\frac{1}{2}
\; ,
\end{equation}
and if we set
\begin{equation}
\lambda_d^2=1-\sqrt{1-q^2} \quad , \quad 
\lambda_{\sigma}^2= 2-\lambda_d^2 =1+\sqrt{1-q^2} \; .
\label{eq:deflambdadlambdasigma}
\end{equation}
Here, we introduced the remaining variational parameter~$0\leq q \leq 1$
that characterizes the Gutzwiller wave function.

\subsubsection{Optimizing the variational parameters}

The Gutzwiller variational ground-state energy with respect to the 
energy of the bare band is the minimum of 
\begin{equation}
E_{\rm var}(q) = e_0(qV)+ \frac{U}{4} \left(1-\sqrt{1-q^2}\right)
\label{eq:Evarres}
\end{equation}
over the variational parameter $0\leq q \leq 1$. Here,
$e_0(V)$ is the ground-energy of the non-interacting symmetric SIAM,
$\hat{H}_0$ in eq.~(\ref{eq:defHzero}), see eq.~(\ref{eq:finalenergyTDL}).
The minimum cannot be obtained analytically in general
but we can derive an implicit equation.

The minimization condition $(\rmd E_{\rm var}(q))/(\rmd q)=0$
leads to the equation
\begin{equation}
U(q,V)=-2\Gamma \sqrt{1-q^2}e_0'(qV)/(qV)
\label{eq:Utildeofq}
\end{equation}
with $\Gamma=\pi \rho_0(0)V^2=\pi d_0 V^2=2 V^2$ 
on a chain with nearest-neighbor hopping.
Therefore, we know $U(q,V)$ for every $0\leq q \leq 1$.
The variational ground-state energy is thus given implicitly by
eq.~(\ref{eq:Evarres}).

\begin{figure}[t]
\includegraphics[width=8.5cm]{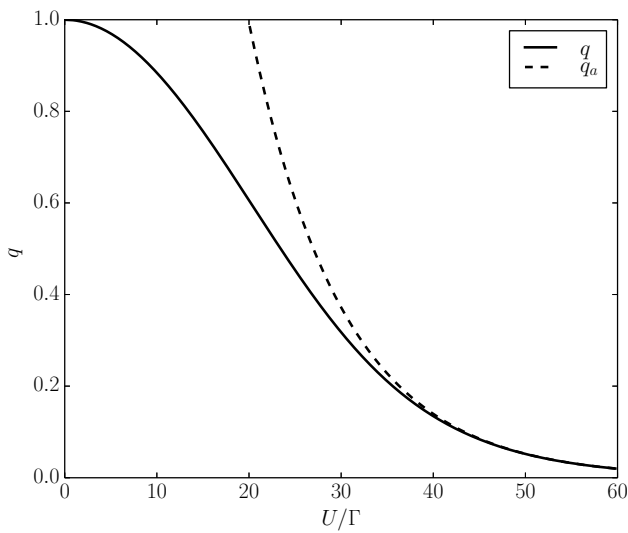}
\caption{Optimal Gutzwiller variational parameter as a function of $U/\Gamma$ 
for $V=0.1$ and the one-dimensional symmetric SIAM
($\Gamma=\pi d_0 V^2=2V^2$).
The asymptotic result~(\ref{eq:qasymptotic}) is shown with a dashed 
line.\label{fig:strongcouplingGutz}}
\end{figure}

In Fig.~\ref{fig:strongcouplingGutz} we show 
the Gutzwiller parameter as a function of~$U/\Gamma$
for $V=0.1$, and compare to the analytic expression in the strong-coupling limit.

\subsubsection{Strong coupling limit}

For strong couplings, we find $q\to 0 $ so that we may use the small-$V$
expression to derive the variational
ground-state energy analytically. Using eq.~(\ref{eq:smallVderivative}) in
eq.~(\ref{eq:Utildeofq}) gives $q(U)\approx q_a(U)$ with
\begin{equation}
[q_a(U)]^2=\frac{1}{\Gamma}\exp\left(
-\frac{\pi U}{16\Gamma}\right) \;,
\label{eq:qasymptotic}
\end{equation}
and the variational ground-state energy becomes
\begin{equation}
E_{\rm opt}(q\ll 1,V)\approx -\frac{2}{\pi}\exp\left(
-\frac{\pi U}{16\Gamma}\right)\propto  
\exp\left( -\frac{1}{4d_0J_{\rm K}}\right)
\end{equation}
with the Kondo energy $J_{\rm K}=4V^2/U$.

The ground-state energy becomes exponentially small, corresponding to
the exponentially small Abrikosov-Suhl resonance 
in the spectral function.~\cite{Hewson}
However, the Gutzwiller exponent is too small by a factor of two,
$T_{\rm K}\propto \exp[ -1/(2 d_0J_{\rm K})]$,~\cite{Hewson} i.e.,
the Gutzwiller approach overestimates the width of the resonance.
As seen from Fig.~\ref{fig:strongcouplingGutz}, for $V=0.1$ 
the asymptotic behavior sets in around $U/\Gamma\approx 35$,
for $q\lesssim 0.2$.

\subsection{Magnetization and magnetic susceptibility}

In the Gutzwiller variational approach, the impurity spin in $z$-direction
is given by
\begin{equation}
m^{\text{G}}(V,B)=\frac{\lambda_{\sigma}^2}{2}\langle \Phi_0 | 
\hat{n}_{d,\uparrow}-\hat{n}_{d,\downarrow}| \Phi_0\rangle 
\end{equation}
with $\lambda_{\sigma}$ from eq.~(\ref{eq:deflambdadlambdasigma}).
Here, we keep a spin-dependent the $q$-factor and only consider the
magnetic-field induced changes in the single-particle product state $|\Phi_0\rangle$.
Therefore, the Gutzwiller variational result for
the magnetization can be obtained from the non-interacting expression
by replacing $q$ by $qV$, see Sect.~\ref{sec:magsusUzero},
\begin{equation}
m^\text{G}(V,B)=(1+\sqrt{1-q^2}) m(qV,B) \; .
\end{equation}
The zero-field susceptibility in Gutzwiller theory reads
\begin{equation}
\chi_0^{\text{G}}(V,U)
=(1+\sqrt{1-q^2}) 
\left(\frac{g\mu_{\rm B}}{2}\right)^2 
\frac{2}{\pi \Gamma} \frac{1}{q^2}
\label{eq:Gutzwillersusceptibility}
\end{equation}
so that the variational Wilson ratio becomes
\begin{equation}
R^{\text{G}}(V,U)= 
\frac{\chi_0^{\text{G}}(V,U)}{\left(g\mu_{\rm B}/2\right)^2 
D_{d,d}^{\text{G}}(\omega=0)} 
= 1+\sqrt{1-q^2} \; .
\label{eq:WilsonGutzwiller}
\end{equation}
Here, we used the fact that the Gutzwiller approach describes a Fermi liquid where
the density of states at the Fermi level is enhanced by a factor $1/q^2$.
Eq.~(\ref{eq:WilsonGutzwiller}) shows that the Gutzwiller approach correctly
reproduces the weak-coupling and strong-coupling limit,
$R(U=0)=1$ and $R(U\gg \Gamma)=2$.
In the strong-coupling limit,
the Wilson ratio deviates from two algebraically in $1/U$
due to the presence of charge fluctuations.~\cite{Hewson}
In contrast, the Gutzwiller Wilson ratio
is exponentially close to two because the Gutzwiller approach does not
describe charge excitations properly.

For strong couplings, the zero-field susceptibility becomes
\begin{equation}
\Gamma \chi_0^{\text{G}}(V,u\gg 1)
\approx 
\left(\frac{g\mu_{\rm B}}{2}\right)^2 
\frac{4\Gamma }{\pi} \exp\left(\frac{\pi u}{16}
\right)
\end{equation}
with $u=U/\Gamma$. The Gutzwiller approach correctly reproduces
the exponentially large zero-field susceptibility for strong interactions, 
see Sect.~\ref{sec:exactspins}.

\subsection{Spin correlation function}
\label{eq:spinGUTZ}

\subsubsection{Local correlation function}

The spin correlation function on the impurity site reads
\begin{eqnarray}
C_{dd}^S(q) &=& \frac{1}{4} -\frac{1}{2} 
\langle \Psi_{\rm G} | 
\hat{n}_{d,\uparrow} +\hat{n}_{d,\downarrow}| \Psi_{\rm G} \rangle \nonumber \\
&=& \frac{1}{4} -\frac{\lambda_d^2 }{8}
= \frac{1+\sqrt{1-q^2}}{8} \; .
\end{eqnarray}
The value for the on-site spin correlation correctly interpolates between
the itinerant limit, $C_{dd}^S(q=1)=1/8$, and the atomic limit,
$C_{dd}^S(q=0)=1/4$.

\subsubsection{Correlation function between impurity and bath sites}

We continue with the spin correlation function between the impurity site
and a bath site at distance~$r$,
\begin{eqnarray}
C_{dc}^S(q,r) 
&=& \frac{1+\sqrt{1-q^2}}{4} 
\langle \Phi_0 | 
\left(\hat{d}_{\uparrow}^+\hat{d}_{\uparrow}^{\vphantom{+}}
 -\hat{d}_{\downarrow}^+\hat{d}_{\downarrow}^{\vphantom{+}}\right)
\nonumber \\
&& \hphantom{ \frac{1+\sqrt{1-q^2}}{4} \langle \Phi_0 | }
\left(\hat{c}_{r,\uparrow}^+\hat{c}_{r,\uparrow}^{\vphantom{+}}
 -\hat{c}_{r,\downarrow}^+\hat{c}_{r,\downarrow}^{\vphantom{+}}\right)
| \Phi_0 \rangle 
\nonumber \\
&=& -
\frac{1+\sqrt{1-q^2}}{2} \left| 
\langle \Phi_0 | 
\hat{c}_{r,\uparrow}^+\hat{d}_{\uparrow}^{\vphantom{+}}
| \Phi_0 \rangle 
\right|^2
\; ,
\end{eqnarray}
where we applied spin symmetry  and Wick's theorem
in the last step.
The matrix element is evaluated in Sect.~\ref{subsec:spinCFnonint},
and we merely have to replace $V\to (qV)$ in all expressions
there.

\section{Interacting SIAM}
\label{sec:DMRG}

In this section we compare our Gutzwiller variational results to those from
the DMRG method that provides essentially exact numerical data
for the SIAM on large rings.
For comparison we also include results from magnetic Hartree-Fock theory,
as derived in the appendix, and compare to 
the ground-state energy from 
the Bethe Ansatz solution.~\cite{KAWAKAMI1981483}

\subsection{DMRG Method}

We study the symmetric SIAM in the effective single-chain 
representation~(\ref{eq:defHzerosinglechain}) using the DMRG method.
The mapping leads to an effective system size that is 
about half of the ring size, and
it provides open boundary conditions that are more favorable
for the DMRG method than periodic boundary 
conditions.~\cite{PhysRevLett.69.2863,PhysRevB.48.10345}
However, particle-hole symmetry is recovered only in the thermodynamic limit,
$L\to\infty$.

\subsubsection{Technicalities}

We study the effective Hamiltonian of $C$-electrons on a chain
up to system 700 sites that corresponds to $L=1397$ in the ring
geometry. This allows us to study systems with periodic boundary
conditions that are three times longer than used in previous
studies with open boundary conditions.~\cite{PhysRevB.80.205114}
The accuracy of the calculations is controlled using the 
dynamic block-state selection (DBSS) 
scheme.~\cite{PhysRevB.67.125114,PhysRevB.70.205118} 
Setting the control parameter to $\chi=10^{-5}$, the truncation error yields 
around $10^{-7}$ while the number of maximally kept DMRG block-states 
can grow up to $M=5000$ for large system sizes.
For strong interactions, we target multiple states to stabilize
convergence.

On finite lattices, the calculation of the magnetization 
as a function of a globally applied field ${\cal H}$ is more subtle
because $S^z$ is a good quantum number. Therefore,
the spin quantum number $S^z$ changes from $S^z=0$ 
for ${\cal H}=0$ to $S^z=1,2,3,\ldots$ for increasing external fields
in steps of $g\mu_{\rm B}{\cal H}_n$ when 
\begin{equation}
g\mu_{\rm B}{\cal H}_n=E_0(S^z=n)-E_0(S^z=n-1)
\end{equation}
for $n=1,2,3,\ldots\ $. 
Thus, the impurity magnetization $\tilde{m}(V,B)$
is recorded only at discrete values of
the external field whereby expectation values are calculated with
the ground state for $S^z=n$.
Since the energy differences are of the order $1/L$, the smallest accessible
magnetic energy scale is of the order $W/L$.
In this work, we include only the results for $U=0,V=0.2$, see Fig.~\ref{fig:magnonint},
to demonstrate the applicability of the approach.
 
For most of the results below, we apply the magnetic field only at the impurity.
Since $\hat{P}$ in the Hamiltonian~(\ref{eq:defHzero}) is not
conserved, standard DMRG ground-state calculations 
provide the results for the impurity magnetization.

\subsubsection{Tests}

\begin{figure}[t]
\includegraphics[width=8.5cm]{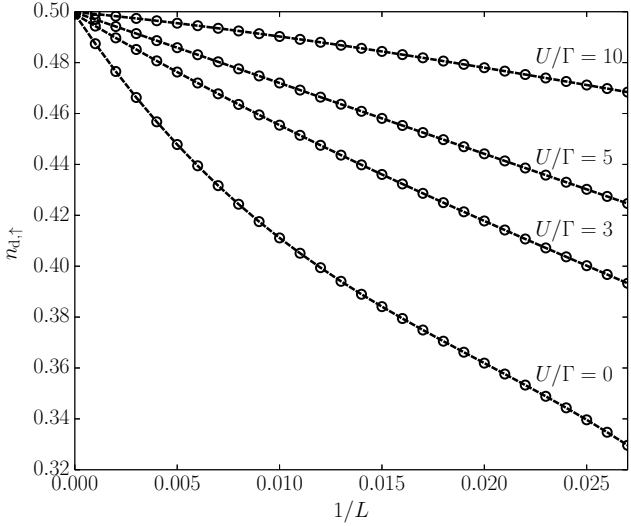}
\caption{DMRG results for the impurity occupation $n_{d,\uparrow}$
of the open-chain SIAM
as a function of inverse system size $1/L$ for various values of
$U/\Gamma$ and $V=0.1$ 
($\Gamma=\pi d_0 V^2=2V^2$). Lines are only 
guides to the eyes.\label{fig:nimp-onehalf}}
\end{figure}

To test the accuracy of our open-chain approach,
in Fig.~\ref{fig:nimp-onehalf} we show 
the finite-size scaling
of the impurity occupancy $n_{d,\sigma}(L)=\langle \hat{n}_{d,\sigma} \rangle$
for the open-chain SIAM~(\ref{eq:defHzerosinglechain})
for various values of $U/\Gamma$ at $V=0.1$. 
It is seen that the occupation extrapolates
to its value in the presence of 
particle-hole symmetry, $n_{d,\sigma}(\infty)=1/2$.
For $U>0$, the electrons repel each other on the impurity.
Thus, the Hubbard interaction 
suppresses charge fluctuations and shifts 
$n_{d,\sigma}(L)$ towards one half already at small system sizes.
Apparently, for numerical treatments
the choice of open boundary conditions 
is favorable over the ring geometry because particle-hole symmetry
holds also for finite system sizes. Of course, the boundary conditions
play no role in the thermodynamic limit, as seen for the extrapolated
impurity occupancy in Fig.~~\ref{fig:nimp-onehalf}.

\begin{figure}[t]
\includegraphics[width=8.5cm]{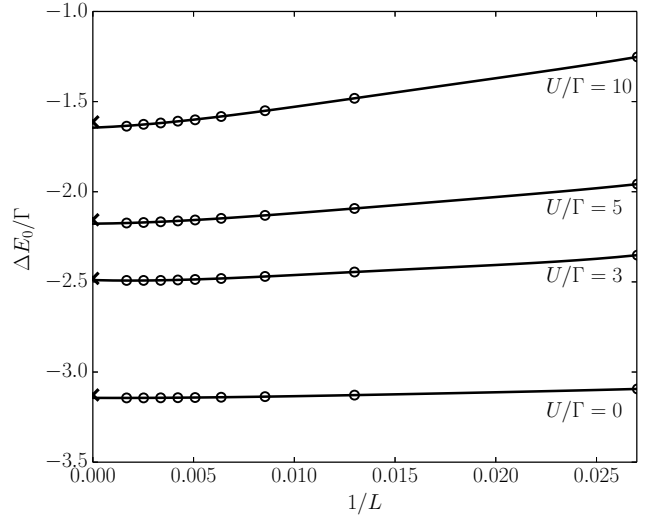}
\caption{DMRG result for the ground-state energy $\Delta E_0(U,V)$ 
of the one-dimensional 
symmetric SIAM
as a function of inverse system size $1/L$ for various values of
$U/\Gamma$ and $V=0.1$ ($\Gamma=\pi d_0 V^2=2V^2$).
The crosses
denote the values from the Bethe Ansatz, becoming exact 
in the wide-band limit.
The solid lines represent the second-order polynomial fit.
\label{fig:finitesize}}
\end{figure}

As another test, we present the ground-state energy $\Delta E_0(U,V)$
as a function of inverse system size for $V=0.1$ and various values of
the interaction strength~$U/\Gamma$ in Fig.~\ref{fig:finitesize}.
Here, we measure the ground-state energy 
with respect to the case $V=0$, 
\begin{equation}
\Delta E_0(U,V)
=E_0(U,V) -E_0(U,0) \; ,
\end{equation}
i.e., we subtract the band contribution of the
free host electrons and the term $-U/4$ for the singly occupied impurity site.
Therefore, $\Delta E_0(U,V)$ is of the order unity
and tends to zero for large interaction strengths.
Using a second-order polynomial fit in the inverse system size,
the DMRG energies extrapolated to the thermodynamic limit
coincide with the values from Bethe Ansatz. Note that
the Bethe Ansatz approach covers the wide-band limit, $U\ll W$, 
and also ignores corrections of order $V^4$. Therefore, 
the extrapolated DMRG energies are slightly below the Bethe-Ansatz
energies.

\subsection{Ground-state energy}

\subsubsection{Small interaction strengths}

For the symmetric SIAM, the ground-state energy
is known for weak coupling, $U\ll \pi \Gamma$, ~\cite{Hewson,Yamada1975}
\begin{equation}
\Delta E_0(U,V) =
e_0(V) + \frac{U}{4} +\pi\Gamma \sum_{n=1}^{\infty} (-1)^n 
e^{(2n)}(V) 
\left(\frac{U}{\pi \Gamma}\right)^{2n}
\label{eq:gsenergyasymptotics}
\end{equation}
where $\Gamma=\pi d_0V^2=2V^2$. 
Due to particle-hole symmetry, there are no odd-order corrections 
in the weak-coupling series beyond the Hartree term. 

For $V\ll 1$, $e^{(2n)}(V)$ weakly
depends on~$V$. We find 
\begin{eqnarray}
e^{(2)}(0.1)&=&0.0374447 \; , \nonumber \\
e^{(2)}(0.05)&=&0.0369271 \; , 
\label{eq:exactsecondorderV01}
\end{eqnarray}
in very good agreement with the analytical result
obtained by Yamada,~\cite{Yamada1975}
\begin{equation}
e^{(2)}(V=0)=\frac{1}{4}-\frac{7}{4\pi^2}\zeta(3) \approx 0.0368608 \; .
\label{eq:exactsecondorder}
\end{equation}
Moreover, the fourth-order coefficient is known to be very small,
$e^{(4)}(V=0)\approx 0.0008$.~\cite{Yamada1975}

The Gutzwiller approach leads to
\begin{equation}
e_{\rm G}^{(2)}(V)
= \frac{\pi\Gamma}{[-32V e_0'(V)]}
\approx -\frac{\pi^2}{128\ln(2V^2)} \;. 
\label{eq:Gutzsecondorder}
\end{equation}
In contrast to the exact expression,
the prefactor of the second-order term vanishes logarithmically for $V\to 0$. 
For $V=0.1$, 
we find $e_{\rm G}^{(2)}(0.1)=0.01943$, about half of the exact value
in eq.~(\ref{eq:exactsecondorderV01}).
The paramagnetic Fermi sea remains the Hartree-Fock ground state until
magnetic  order sets in at about $U_{{\rm c,HF}}\approx \pi \Gamma$.
Therefore, there is no second-order term in the ground-state energy
in Hartree-Fock theory.

\subsubsection{Wide-band limit}

For small hybridizations and $V\ll U\ll W$, the SIAM can be solved analytically
using the Bethe Ansatz because the dispersion relation 
of the host electrons can be linearized around the
Fermi wave vector.~\cite{PhysRevLett.45.379,0022-3719-14-10-014,%
0022-3719-16-12-017,0022-3719-16-12-018}
For the symmetric SIAM in the absence of a magnetic field,
the ground-state energy can be calculated analytically,~\cite{KAWAKAMI1981483}
\begin{equation}
\frac{\Delta E_0^{\rm BA}(U_{\rm BA},V_{\rm BA})}{t_{\rm BA}}=\frac{U_{\rm BA}}{2}
+  \int_{-\infty}^{A^2}\rmd \Lambda 2x(\Lambda)\sigma_S(\Lambda) 
\end{equation}
with
\begin{eqnarray}
x(\Lambda) &=& -\frac{\sqrt{2}}{2} \sqrt{\Lambda
+\sqrt{\Lambda^2 +U_{\rm BA}^2 V_{\rm BA}^4/4}}
\; , \nonumber \\
\sigma_S(\Lambda) &=& \int_{-\infty}^{\infty}\frac{\rmd k}{4\pi}
\frac{V_{\rm BA}^2}{(k+U_{\rm BA}/2)^2+V_{\rm BA}^4/4} \nonumber \\
&& \hphantom{\int_{-\infty}^{\infty}\frac{\rmd k}{4\pi}}
\times 
\frac{1}{U_{\rm BA}V_{\rm BA}^2}\sech\left[
\frac{\pi(k^2-\Lambda)}{U_{\rm BA}V_{\rm BA}^2}
\right] \, ,
\label{eq:KOBetheAnsatzenergy}
\end{eqnarray}
where $\sech(x)=1/\cosh(x)$ is the hyperbolic secant function.
The energy shift $U_{\rm BA}/2$ takes our definition into account that
$\Delta E_0(U,V)$ is measured with respect to the limit of vanishing 
hybridization, $\Delta E_0^{\rm BA}(U,0)=0$.

Note that in eq.~(\ref{eq:KOBetheAnsatzenergy}) 
all energies are expressed in units of $t_{\rm BA}$
so that the Fermi velocity is $v_{\rm F}^{\rm BA}=t_{\rm BA}$.
In our energy units 
we have $v_{\rm F}=W/2$ so that we must set $t_{\rm BA}=W/2\equiv 1/2$,
i.e., we must scale all energies by a factor of two.
Moreover, in the Hamiltonian used in the Bethe Ansatz, 
only the symmetric linear combination of right-movers and left-movers
couples to the impurity whereas the hybridization in
the lattice Hamiltonian~(\ref{eq:defV})
is expressed in terms of left-movers and right-movers.
This implies $V_{\rm BA}/t_{\rm BA}=2\sqrt{2}V/W$
and $U_{\rm BA}=2 U/W$ in our energy units.

We adjust the bandwidth cutoff-parameter~$A$
to reproduce the ground-state energy~(\ref{eq:Vsmallgsenergynonint}) 
of the non-interacting SIAM to orders $V^2\ln(V^2)$ and $V^2$.
For $A=2e$ we indeed find
$\Delta E_0^{\rm BA}(0,V)= (4V^2/\pi)[\ln(V^2)+\ln(2)-1] + {\cal O}(V^4)$,
see eq.~(\ref{eq:Vsmallgsenergynonint}).

Ignoring terms of order $V^4$ and higher that are beyond the wide-band
limit, the ground-state energy reads
\begin{eqnarray}
\frac{\Delta E_0^{\rm BA}(U,V)}{\Gamma}&=&  \frac{u}{2} 
+ \sum_{\sigma} \int_{-e/\Gamma}^0
\frac{\rmd p}{\pi}
\frac{1}{(p+\sigma_n u/2)^2+1} \nonumber \\
&& \hphantom{\frac{u}{2}+}
\times \!
\int_{-\infty}^{\infty} \frac{\rmd y}{\pi} \sech(y) \tilde{x}(p^2-2uy/\pi,u) \, ,
\nonumber \\
\tilde{x}(\lambda,u) &=& \frac{\sqrt{2}}{2}\sqrt{\lambda+\sqrt{\lambda^2 + u^2}}
\label{eq:BetheAnsatzgsenergy}
\end{eqnarray}
with $\Gamma=2V^2$ and $u=U/\Gamma$. These expressions
are amenable to a numerical evaluation of the integrals.

To extract the limiting behavior and to show the equivalence 
with the Hartree-Fock energy for $U/\Gamma\gg 1$, we write
\begin{equation}
\Delta E_0^{\rm BA}(U,V)=
\Delta E_0^{\rm BA,1}(U,V)
+\Delta E_0^{\rm BA,2}(U,V) \; ,
\end{equation}
where
\begin{equation}
\frac{\Delta E_0^{\rm BA,1}(U,V)}{\Gamma} 
= \frac{u}{2} +  \sum_{\sigma} \int_{-e/\Gamma}^0
\frac{\rmd p}{\pi}
\frac{p}{(p+\sigma_n u/2)^2+1}
\end{equation}
and
\begin{eqnarray}
\frac{\Delta E_0^{\rm BA,2}(U,V)}{\Gamma}  &=&
\sum_{\sigma} \int_{-\infty}^0
\frac{\rmd k}{\pi}
\frac{u}{(k\sqrt{u}+\sigma_n u/2)^2+1} \nonumber \\
&& 
\times
\int_{-\infty}^{\infty} \frac{\rmd y}{\pi} \sech(y) X(k^2-2y/\pi)\nonumber \\
\label{eq:BetheAnsatzgsenergytwoterm}
\end{eqnarray}
with 
\begin{equation}
X(\lambda) = -k+\frac{\sqrt{2}}{2} \sqrt{\lambda+\sqrt{\lambda^2 + 1}}
\; .
\end{equation}
The second term gives  for $u\gg1$
\begin{equation}
\frac{\Delta E_0^{\rm BA,2}(U\gg \Gamma,V)}{\Gamma}
= \frac{\pi}{u} +{\cal O}(1/u^2) \; .
\end{equation}

The first term is equivalent 
to the Hartree-Fock expression in the limit $m\to 1/2$
that is reached for $U/\Gamma\gg 1$.
Moreover, the integral is readily evaluated and gives 
in the intermediate coupling regime ($\Gamma=2V^2/W \ll U\ll W$)
\begin{eqnarray}
\frac{\Delta E_0^{\rm BA,1}(U,V)}{\Gamma}&=& \frac{2}{\pi}\ln(\Gamma/e) 
+\frac{u}{2}\left(1-\frac{2}{\pi}\tan^{-1}(u/2)\right) \nonumber \\
&&+\frac{1}{\pi} \ln(1+u^2/4) \; .
\label{eq:intermediateU}
\end{eqnarray}
It is seen that the ground-state energy 
increases logarithmically, i.e., as a function
of $\ln(u)$, in the intermediate coupling regime.~\cite{KAWAKAMI1981483}

One may wonder whether or not the Kondo energy scale can be extracted
from the Bethe-Ansatz energy expression~(\ref{eq:BetheAnsatzgsenergy}).
Indeed, the region~$[|p+ u/2|\leq \nu_1, |y-p^2\pi/(2u)|\leq \nu_2]$ with $\nu_{1,2}$
of order unity gives rise to a contribution of the order of
\begin{equation}
T_{\text{L}}(U)
=\Gamma \sqrt{\frac{u}{2}} \exp\left[-\frac{\pi u}{8}
+\frac{\pi}{2u}\right]
\label{eq:Tkdef}
\end{equation}
with $u=U/\Gamma$. $T_{\text{L}}(U)$
is proportional to the Kondo temperature for the symmetric SIAM
in the strong-coupling limit.~\cite{Hewson}
Note, however, that the integration over all $(p,y)$ region wipes out
this term in the ground-state energy. It is only in magnetic properties that
the energy scale~$T_{\text{L}}$ becomes visible,~\cite{Hewson}
see Sect.~\ref{sec:susceptibilityBA-DMRG}.

\subsubsection{Comparison}

\begin{figure}[t]
\includegraphics[width=8.5cm]{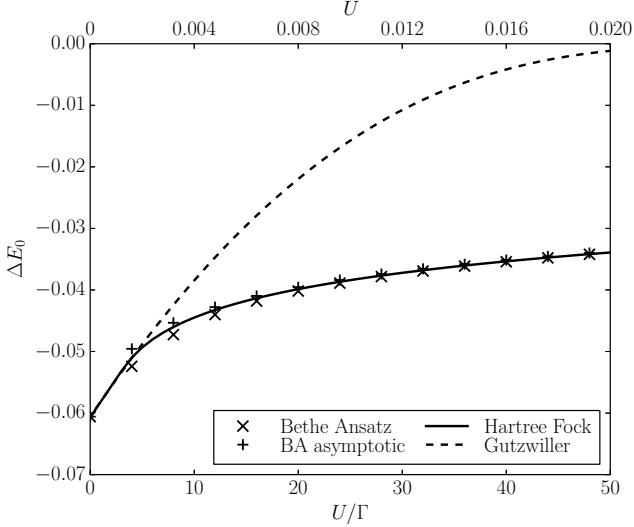}
\caption{Ground-state energy $\Delta E_0(U,V)$ 
for the symmetric SIAM in one dimension
as a function $U/\Gamma$ for $V=0.01$ 
($\Gamma=\pi d_0 V^2=2V^2=0.0002$).
The full (dashed) lines display 
the Hartree-Fock (Gutzwiller) 
variational upper bound, the open symbols give the Bethe Ansatz 
results from eq.~(\ref{eq:BetheAnsatzgsenergy}),
and the crosses denote the asymptotic 
result~(\ref{eq:intermediateU}).\label{fig:gsenergiesBA}}
\end{figure}

\begin{figure}[t]
\includegraphics[width=8.5cm]{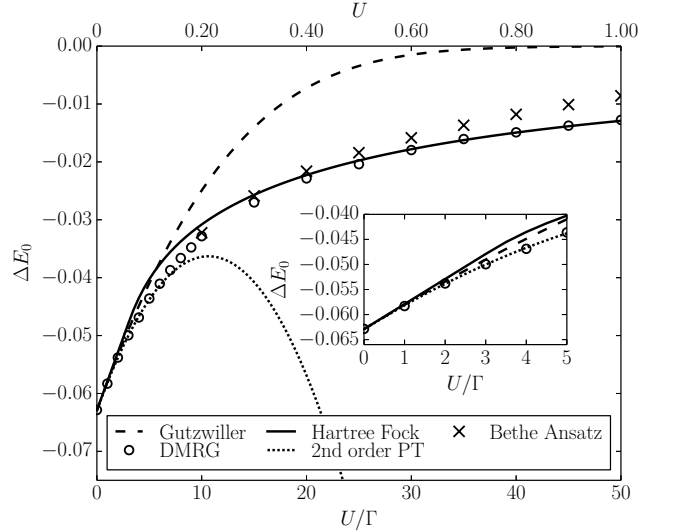}
\caption{Ground-state energy $\Delta E_0(U,V)$ 
for the symmetric SIAM on a ring
as a function $U/\Gamma$ for $V=0.1$ 
($\Gamma=\pi d_0 V^2=2V^2=0.02$).
The second-order weak-coupling result~(\ref{eq:exactsecondorderV01}) 
is shown as a short-dashed line, the full (dashed) lines display 
the Hartree-Fock (Gutzwiller) 
variational upper bound, the open symbols are the DMRG
data, extrapolated to the thermodynamic limit, and the crosses
give the Bethe-Ansatz values.
Inset: ground-state energy for small interaction 
strengths.\label{fig:gsenergies}}
\end{figure}

In Fig.~\ref{fig:gsenergiesBA} we compare the Gutzwiller, Hartree-Fock, and
Bethe Ansatz energies for $V=0.01$ ($\Gamma=0.0002$).
For such small hybridizations, DMRG calculations would require system sizes
that are an order of magnitude larger because even at $U=0$
the relevant energy scale $\Gamma=2V^2/W$ becomes very small.
In Fig.~\ref{fig:gsenergies} we show the ground-state energies
for $V=0.1$ ($\Gamma=0.02$) 
from weak-coupling perturbation theory, 
Gutzwiller, Hartree-Fock, finite-size extrapolated data from DMRG,
and Bethe Ansatz. Since the Bethe Ansatz approach covers 
the wide-band limit, $U\ll W$, the extrapolated DMRG energies 
and the Hartree-Fock energies lie below the Bethe-Ansatz energies,
as becomes discernible at $V=0.1$ for $U/\Gamma\gtrsim 10$ 
in Fig.~\ref{fig:gsenergies}. 

As seen from the two figures, the Gutzwiller 
energy curve deviates noticeably 
from the exact results for $U/\Gamma>10$.
Like second-order perturbation theory, it provides a good estimate only
for $U\lesssim 5\Gamma$. At large interactions, the Gutzwiller
variational energy becomes exponentially small. Since 
the wave function does not properly describe charge fluctuations, i.e.,
the Hubbard bands, the Gutzwiller
variational energy bound is poor. 

In Hartree-Fock theory, a magnetic moment is
formed only for $U_{\rm c}^{\rm HF}\approx \pi \Gamma$, 
and the ground-state energy contains a cusp at $U_{\rm c}$. 
More importantly, above $U\approx 5 \Gamma$ the Hartree-Fock theory
provides an excellent bound on the exact ground-state energy.
For small hybridizations,
the exact Bethe-Ansatz and DMRG energies are in almost perfect agreement
with the Hartree-Fock upper bounds.
Since the quasi-particle 
peak provides an exponentially small energy contribution
for $U\gg \Gamma$, the energy is solely determined by the lower Hubbard band
which gives rise to a $\ln(u)$ increase of the ground-state energy,
see eq.~(\ref{eq:intermediateU}).
The result for $\Delta E_0^{\rm BA,1}(U,V)/\Gamma$
is also shown in Fig.~\ref{fig:gsenergiesBA}.

Obviously, the lower Hubbard band for $u\gg 1$ 
is qualitatively well captured
by Har\-tree-Fock theory. Therefore,
Hartree-Fock theory provides an excellent starting point for 
analytical theories like the Local Moment Approach that covers both the
high-energy and low-energy parts of the single-particle 
spectrum.~\cite{Logan1998,Logan2000,Logan2009}

\subsection{Magnetization and magnetic susceptibility}
\label{sec:susceptibilityBA-DMRG}

The Bethe Ansatz permits the exact calculation of the impurity magnetization 
in the presence of a magnetic field on the impurity. 
Hereby, it is implicitly understood that the effect of the magnetic field 
on the conduction electrons is negligibly small so that it does not make
a difference whether or not the magnetic field is also applied to the 
bath electrons.

\subsubsection{Magnetization}

The analysis of the Bethe Ansatz equations depends on the value of 
the external field~$b=B/\Gamma$. First,
region~I, $b\leq b_0(u)$, covers the weak-field regime $b\to 0$
and the Kondo regime for $u\gg 1$. Region~II,
$b\geq b_0(u)$, covers the large-field regime $b\to \infty$.
The magnetization and the magnetic susceptibility are continuous 
at $b=b_0(u)$.
The boundary value is determined by ($\ln(e)=1$)
\begin{eqnarray}
b_0(u)&=& \sqrt{2u} \frac{1}{\sqrt{\pi}}
\sum_{n=0}^{\infty}\frac{1}{n!}\left(\frac{2n+1}{2e}\right)^{n+1/2}
\frac{1}{(2n+1)^{3/2}} \nonumber \\
&\approx& 0.398942 \sqrt{2u} \; .
\end{eqnarray}
For $u\to 0$, only region~II exists, whereas in the Kondo limit,
for $u\to \infty$, only region~I remains.

\paragraph{Magnetization in region I:}

The magnetization and the magnetic field parametrically depend on each other.
For $p\geq 0$, Tsvelik and Wiegmann give~\cite{0022-3719-16-12-017}
\begin{equation}
b_I(p,u)=\sqrt{\frac{2u}{\pi}}
\sum_{n=0}^{\infty}\frac{1}{n!}\left(\frac{2n+1}{2e}\right)^{n+1/2}
\frac{e^{-\pi(2n+1)p}}{(2n+1)^{3/2}}
\label{eq:bregionI}
\end{equation}
for the applied magnetic field so that $b_I(0,u)=b_0(u)$.

The impurity magnetization contains two terms,
\begin{equation}
m_I(p,u)= m_{\rm K}(p,u)+ m_{\rm reg}(p,u) \; ,
\end{equation}
namely, the `Kondo term' $m_{\rm K}(p,u)$ and the `regular term' $m_{\rm reg}(p,u)$.
We discuss them separately.

The Kondo term is given by ($\eta=0^+$),~\cite{0022-3719-16-12-017,correcttypo}
\begin{eqnarray}
m_{\rm K}(p,u)
&=& \frac{(-\rmi)}{4\pi^{3/2}}\int_{-\infty}^{\infty}
\frac{\rmd \omega}{\omega-\rmi\eta}
\left(\frac{\rmi \omega +\eta}{e}\right)^{\rmi \omega}
\Gamma\left(\frac{1}{2}-\rmi \omega\right)\nonumber \\
&& \hphantom{\frac{-\rmi}{4\pi^{3/2}}\int} 
\times \exp\left[-2\pi\rmi \omega (p-J^{-1}(u))\right] \; ,
\label{eq:mKondoregionI}
\end{eqnarray}
where
\begin{equation}
J^{-1}(u)=\frac{u^2-4}{8u}
\end{equation}
is the inverse Kondo coupling, and $\Gamma(x)$ denotes the Gamma function.
The Kondo contribution cannot be obtained in weak-coupling perturbation 
theory because of the $1/u$-singularity in the exponent. Moreover, it
gives rise to a diverging zero-field susceptibility for $u\gg 1$,
see Sect.~\ref{subsubsec:zfsusBA} below.

The regular contribution reads
\begin{eqnarray}
m_{\rm reg}(p,u)&=& \frac{1}{\sqrt{\pi}}\sum_{n=0}^{\infty}
\frac{1}{n!}\left(\frac{2n+1}{2e}\right)^{n+1/2}
\frac{e^{-\pi p(2n+1)}}{2n+1} \nonumber \\
&& \hphantom{\frac{1}{\sqrt{\pi}}\sum_{n=0}^{\infty}}
\times F\left(\frac{\pi(2n+1)}{2u},u\right)
\label{eq:mregregionI}
\end{eqnarray} 
with
\begin{eqnarray}
F(a,u)&=& \int_{-\infty}^{\infty}\frac{\rmd y}{\pi} e^{-ay^2} 
\frac{1}{1+(\rmi y+u/2)^2} \nonumber \\
&=& \int_0^{\infty}\frac{2\rmd y}{\pi} e^{-ay^2}
\frac{1-y^2+u^2/4}{u^2y^2+(1-y^2+u^2/4)^2} \nonumber \\ 
&=& e^{-a(1-u^2/4)}\int_0^{a}\frac{\rmd x}{\sqrt{\pi x}} 
\exp\left[x-\frac{a^2u^2}{4x}\right]
\end{eqnarray}
with the analytic expression
\begin{eqnarray}
F(a,u)&=& -e^{-a(1-u^2/4)}
\sin(au) \nonumber \\
&& + e^{-a(1-u^2/4)}\text{Re}\left[
e^{\rmi a u} \Erfi\left(\sqrt{a}(2-\rmi u)/2\right)
\right] \; ,  \nonumber \\
\end{eqnarray}
where $\Erfi(x)$ is the complex error function.
The analytic formula is helpful for the derivation
of series expansions of $F(a,u)$ for small and large arguments~$a$.

\paragraph{Magnetization in region II:}

For the applied magnetic field,
Tsvelik and Wiegmann give~\cite{0022-3719-16-12-017}
\begin{equation}
b_{II}(p,u)=b_0(u) + \sqrt{\frac{u}{4\pi}}
\int_0^{\infty}\frac{\rmd x}{x^{3/2}}
\frac{\left(1-e^{-2\pi p x}\right)}{\Gamma\left(\frac{1}{2}+x\right)}
\left(\frac{x}{e}\right)^x
\label{eq:bregionII}
\end{equation}
for $p\geq 0$. The magnetization in region~II is given by
\begin{equation}
m_{II}(p,u)= \frac{1}{2} -\frac{1}{2}
\int_0^{\infty}
\frac{\rmd x}{\sqrt{\pi}x} 
\frac{e^{-2\pi p x}}
{\Gamma\left(\frac{1}{2}+x\right)}
\left(\frac{x}{e}\right)^x F\left(\frac{\pi x}{u},u\right) \; .
\label{eq:mregionII}
\end{equation}
As shown in the appendix,
the result for the non-interacting SIAM is readily recovered from 
eqs.~(\ref{eq:bregionII}) and~(\ref{eq:mregionII}).
There, we also derive an explicit formula for the large-field limit,
\begin{equation}
m_{II}(b\gg \sqrt{u/\pi})\approx \frac{1}{2}-\frac{1}{\pi b} 
+\frac{u}{2\pi b^2} +\frac{4\pi-12u-3\pi u^2}{12 \pi^2 b^3}
\label{eq:strongfields} 
\end{equation}
up to and including all terms of the order $1/b^3$,
and the low-energy
Kondo scale $T_{\rm L}(U)$ is absent for large fields.

In fact, there are no logarithmic terms to all orders of the $1/b$ expansion
because, for $b\gg 1$, both $b(p,u)$ and $m_{II}(p,u)$ can be expressed 
in terms of a series with odd powers in the parameter~$1/\sqrt{z}$ 
where $z$ obeys $p=z-\ln(2\pi e z)/(2\pi)$.~\cite{0022-3719-16-12-017} 
Therefore, at large values of the external field, the impurity
magnetization does not show any signs of the logarithmic
Doniach-\v{S}unji\'c-Hamann tails in the impurity spectral 
function.~\cite{Logan1998,Logan2000,Logan2009,0022-3719-3-2-010}

\begin{figure}[h]
\includegraphics[width=8.5cm]{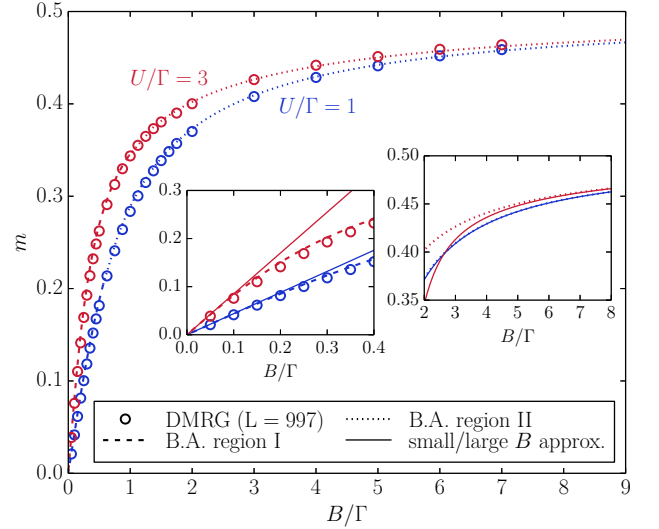}
\caption{(Color online) Impurity magnetization $m(V=0.1,B)=\langle \hat{S}_d^z\rangle$
as a function of the external magnetic
field~$b=B/\Gamma=g\mu_{\rm B}{\cal H}/(2\Gamma)$
for weak interactions, $u=U/\Gamma=1,3$, 
from the Bethe Ansatz for the symmetric SIAM. The insets show
the behavior for small and large fields 
with the small-field asymptotics~(\ref{eq:smallBBA})
and the large-field asymptotics~(\ref{eq:strongfields}).
The circles are DMRG results for rings with $L=997$ 
sites.\label{fig:magnetizationweak}}
\end{figure}

\begin{figure}[t]
\includegraphics[width=8.5cm]{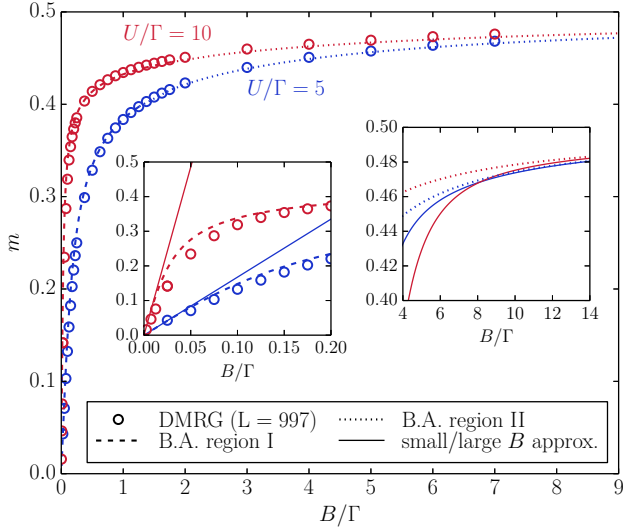}
\caption{(Color online) Same as Fig.~\ref{fig:magnetizationweak} for
moderate couplings $u=5,10$.\label{fig:magnetizationlargeU}}
\end{figure}

We show the impurity magnetization as a function of the local external field
for small and moderate interactions strengths in 
Fig.~\ref{fig:magnetizationweak} and Fig.~\ref{fig:magnetizationlargeU},
respectively. The DMRG reproduces the magnetization curves very well,
particularly at strong magnetic fields.
DMRG requires very large system sizes to resolve the 
steep initial slope of the magnetization curves, especially 
for moderate to large interaction strengths.
This behavior is reflected in the zero-field susceptibility
that becomes exponentially large for large interactions
in an exponentially narrow region of external fields, as we discuss next.

\subsubsection{Zero-field magnetic susceptibility}
\label{subsubsec:zfsusBA}

For $b_I(p,u)\equiv b\to 0$, we have $p\to\infty$ 
in eq.~(\ref{eq:bregionI}) so that
we only retain the first term in the series. Thus,
\begin{equation}
p=-\frac{1}{\pi}\ln\left(b\sqrt{\frac{\pi e}{u}}\right)\gg 1 \; .
\end{equation}
For the case $[p-J^{-1}(u)]\geq 0$, we may represent $m_{\rm K}(p,u)$
in terms of a sum by performing a contour integral 
in the lower complex $\omega$-plane,
\begin{eqnarray}
m_{\rm K}(p,u) &=& \frac{1}{\sqrt{\pi}} \sum_{n=0}^{\infty}
\frac{(-1)^n}{2n+1} \frac{1}{n!}\left(\frac{2n+1}{2e}\right)^{n+1/2}\nonumber\\
&& \hphantom{\frac{1}{\sqrt{\pi}} \sum_{n=0}^{\infty}}
\times \exp\left[-\pi(p-J^{-1}(u))(2n+1)\right] \; .\nonumber \\
\end{eqnarray}
In this sum for the Kondo contribution and in the sum for the regular
contribution, eq.~(\ref{eq:mregregionI}),
we keep only the first term in the series and find ($m_I(p,u)\equiv m(b,u)$)
\begin{equation}
m(b\to 0,u) \approx \frac{b}{\sqrt{2u}}e^{\pi/J(u)}
\left(1+e^{-\pi/J(u)}F(\pi/(2u),u)\right)
\label{eq:smallBBA}
\end{equation}
for small~$b$,
with corrections of the order $b^3$.~\cite{0022-3719-16-12-017}

Thus, the Bethe Ansatz provides an explicit expression 
for the impurity susceptibility
in the wide-band 
limit,~\cite{0022-3719-16-12-017,Kawakamisusceptibility,Hewson}
\begin{eqnarray}
\chi_0^{\text{BA}}(U,V)&= &
\left(\frac{g\mu_{\rm B}}{2}\right)^2 
 \frac{1}{T_{\text{L}}(U) } \label{eq:BAsusceptibility}
\\
&& \left[1+ \int_0^{\pi/(2u)}\!\! \frac{{\rm d}x}{\sqrt{\pi x}}\exp\left(
x-\frac{\pi^2}{16x}\right)\right]
\nonumber
\end{eqnarray}
with $u=U/\Gamma$ and $T_{\text{L}}(U)$ from eq.~(\ref{eq:Tkdef}).
Since the integral vanishes for $u\to\infty$, the exponential term gives
the result in the Kondo limit.

\begin{figure}[t]
\includegraphics[width=8.5cm]{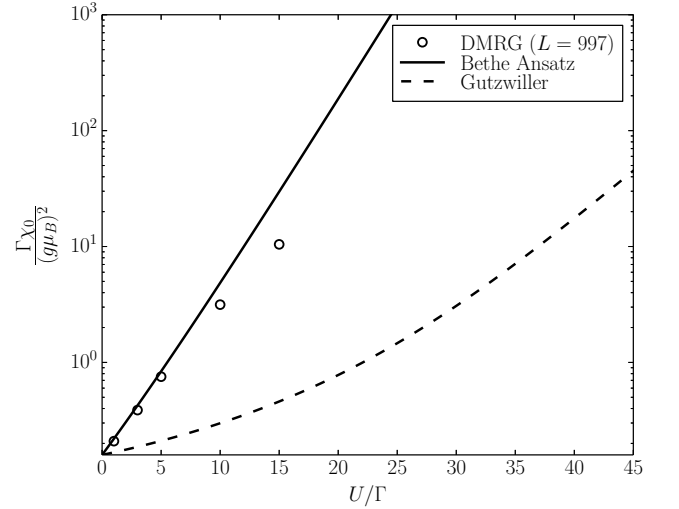}
\caption{(Color online) 
Zero-field magnetic susceptibility, $\Gamma \chi_0/(g\mu_{\rm B})^2$,
as a function of the interaction strength~$u=U/\Gamma$
from Bethe Ansatz, eq.~(\ref{eq:BAsusceptibility}), and
the Gutzwiller variational approach, eq.~(\ref{eq:Gutzwillersusceptibility}),
for the SIAM, in comparison with DMRG results for $L=997$ sites.
Note the logarithmic scale for the ordinate.\label{fig:suscept}}
\end{figure}

We show the zero-field susceptibility in Fig.~\ref{fig:suscept}.
As seen from eqs.~(\ref{eq:Tkdef}) and~(\ref{eq:BAsusceptibility}),
the zero-field susceptibility increases exponentially as a function of $u$.
This behavior is difficult to reproduce in DMRG because,
as the magnetization is bounded from above, the magnetic-field
region where the susceptibility is exponentially large 
is exponentially small.
Therefore, it is hard
to calculate the zero-field susceptibility for $u\gtrsim 10$ from DMRG,
and other numerical method such as the NRG must be employed 
for large interaction strengths.
For $u=10,15$ we choose $B=0.0025\Gamma$ and calculate 
$\chi_0^\text{DMRG}(U,V=0.1)=(g\mu_{\rm B}/2)^2[2m(B,u)/B]$.
As seen from Fig.~\ref{fig:suscept}, the agreement between the Bethe Ansatz
results and DMRG is very good for $u\lesssim 10$, and quite acceptable 
for $u\lesssim 15$ where the zero-field susceptibility is enhanced by more than
a factor of 100 over its non-interacting value.

The Gutzwiller variational theory reproduces the exponential 
behavior of the zero-field susceptibility but with an exponent 
that is too small by
a factor of two. Therefore, the Gutzwiller approach also underestimates the
value of the zero-field spin susceptibility,
see Fig.~\ref{fig:suscept}.

\subsection{Spin correlation function}
\label{sec:exactspins}

\subsubsection{Local moment}

\begin{figure}[t]
\includegraphics[width=8.5cm]{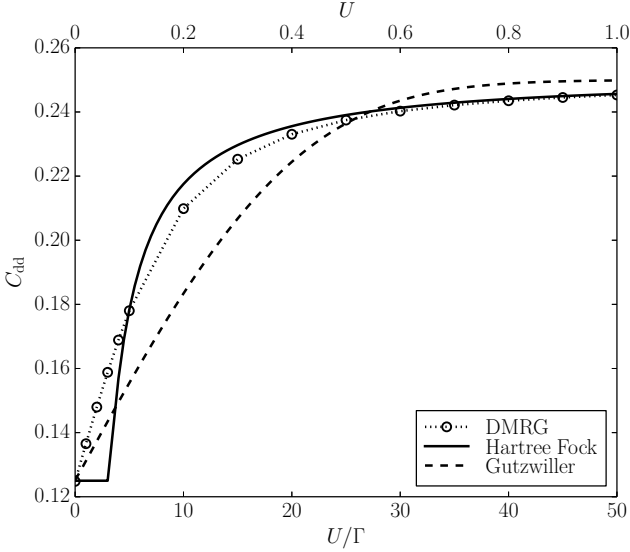}
\caption{Local spin correlation 
$C_{dd}=\langle (n_{d,\uparrow}-n_{d,\downarrow})^2\rangle/4$
for $V=0.1$ ($\Gamma=2V^2=0.02)$
as a function of $U/\Gamma=5$ in Gutzwiller and Hartree-Fock theory
for the SIAM compared with DMRG results with $L=1397$ 
sites.\label{fig:localmoment}}
\end{figure}

In Fig.~\ref{fig:localmoment} we show the local moment on the impurity site,
$C_{dd}=\langle (n_{d,\uparrow}-n_{d,\downarrow})^2\rangle/4$ from
Gutzwiller, Hartree-Fock, and DMRG. 
The Gutzwiller approach provides a reasonable
estimate for the local moment for all interaction strengths. However,
it underestimates 
its value for weak interactions and slightly overestimates it in the strong-coupling limit. 

Hartree-Fock theory uses the non-interacting Fermi-sea ground state for weak
interactions, and starts with the interaction-driven build-up
of the local moment at $U_{\rm c}^{\rm HF}\approx \pi \Gamma$.
Similarly to the Hartree-Fock energy curve shown in Fig.~\ref{fig:gsenergiesBA}, 
a kink in $C_{dd}^{\text{HF}}(U)$ is observed at the critical Hartree-Fock interaction. 
For moderate to
strong interactions, $U\gtrsim 5 \Gamma$, it
provides an excellent estimate for the local spin correlation.
For $U/\Gamma=5$ and for $U/\Gamma=50$, 
the magnitude of the local moment is of the same magnitude both in
the Gutzwiller approach and in Hartree-Fock theory.

\subsubsection{Unscreened spin}

\begin{figure}[t]
\includegraphics[width=8.5cm]{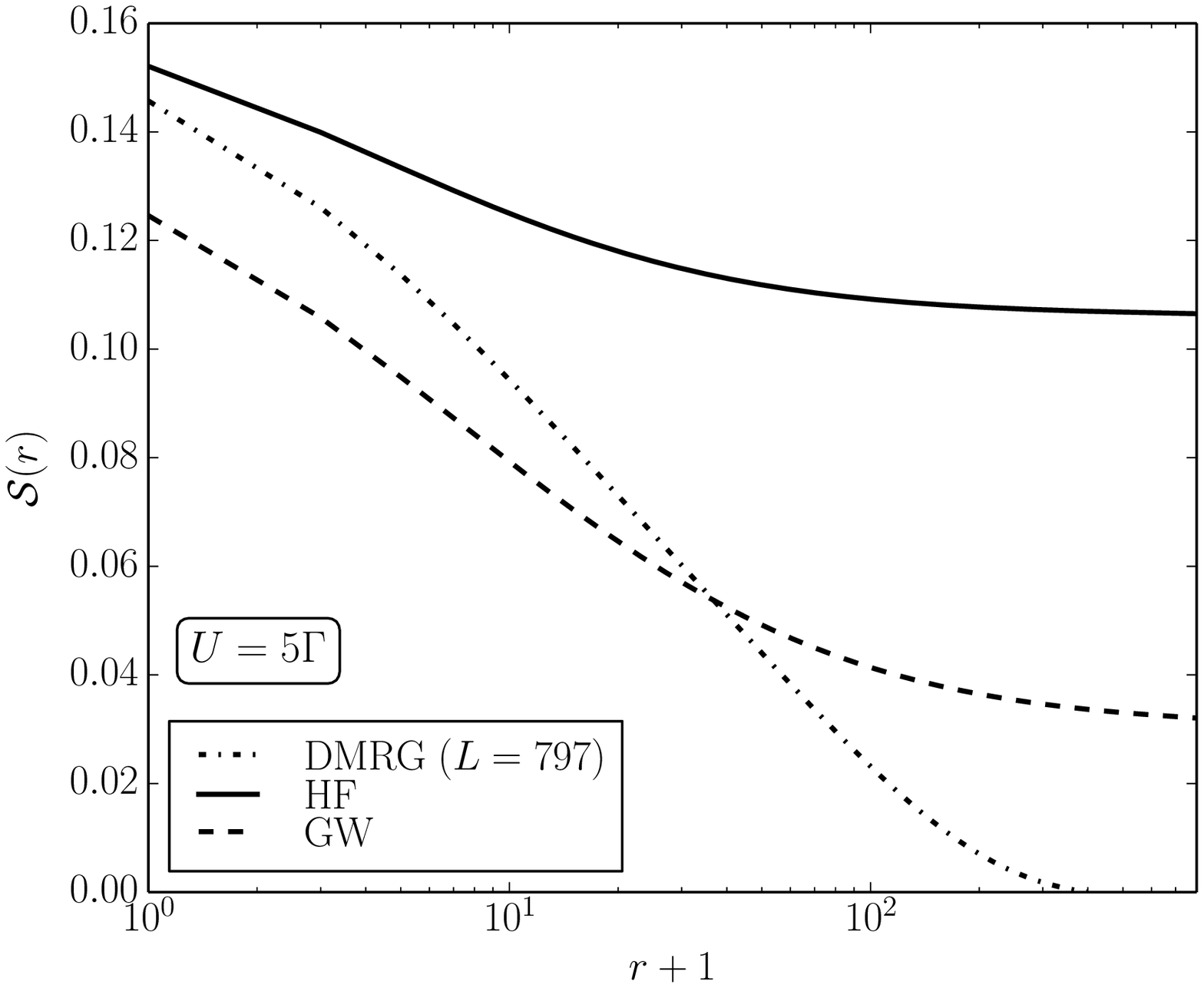}
\includegraphics[width=8.5cm]{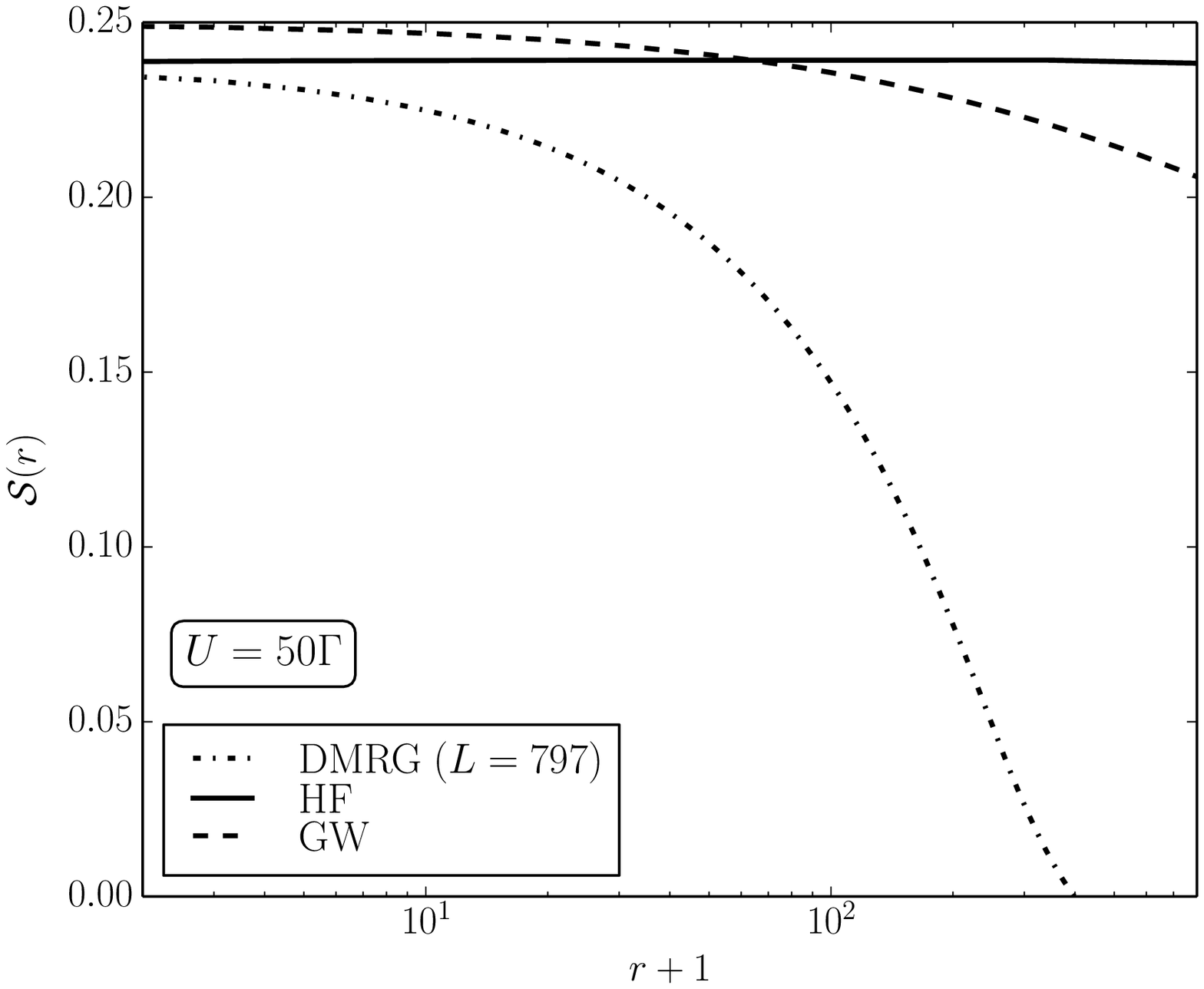}
\caption{Unscreened spin ${\cal S}(r)$ at distance~$r$ from the impurity site,
see eq.~(\ref{eq:calS}),
for $V=0.1$ ($\Gamma=2V^2=0.02)$, 
$U/\Gamma=5$ (upper panel) and $U/\Gamma=50$ (lower panel).
The Gutzwiller and magnetic Hartree-Fock results for the SIAM are compared with
DMRG results with $L=797$ sites.\label{fig:screenedspin-DMRG}}
\end{figure}

In Fig.~\ref{fig:screenedspin-DMRG} we show the unscreened spin ${\cal S}(r)$
for $U/\Gamma=5$ (upper panel) and $U/\Gamma=50$ (lower panel)
at $V=0.1$ ($\Gamma=0.02$).
Even for $U/\Gamma=5$, 
the asymptotic region is not yet reached for $L=797$ in DMRG
where by construction the spin is screened at $r=(L-1)/2=398$.
For $U/\Gamma=5$, finite-size effects are unimportant up to $r\approx 30$.
When the interaction is very large, $U=50\Gamma$,
finite-size effects dominate the DMRG data for all~$r>1$.

Despite its failure to describe the ground-state energy properly,
the Gutzwiller approach reproduces the mesoscopically large Kondo screening
cloud. For $r\to \infty$, the impurity spin is perfectly screened,
${\cal S}^{\rm G}(r\to\infty)=0$, but the Kondo cloud extends over 
many thousands of sites even for moderately strong interactions, $U/\Gamma=5$. 
In contrast, in magnetic Hartree-Fock theory the screening is never complete,
${\cal S}^{\rm HF}(r\to\infty)>0$.
Therefore, among the three approaches discussed here,
the Gutzwiller wave functions provides the best {\em qualitative\/}
description of the Kondo screening cloud for strong couplings.

\subsubsection{Fermi liquid regime}

The calculation of the static spin correlation function $C_{dc}(r)$ 
in eq.~(\ref{eq:defCDC})
poses a difficult many-body problem. We separate
the spin correlation function into its Fermi-liquid contribution (dressed bubble)
and a part that contains vertex parts in a diagrammatic approach,
\begin{eqnarray}
C_{dc}^S(r) &=& C_{dc}^{S,{\rm FL}} + C_{dc}^{S, x} \; , \nonumber \\
C_{dc}^{S, {\rm FL}}&=& -\frac{1}{2}
\left| \langle \Psi_0 | \hat{c}_{r,\uparrow}^+\hat{d}_{\uparrow}^{\vphantom{+}}
| \Psi_0 \rangle \right|^2 =- \frac{1}{2} M_r^2\; .
\label{eq:wickfactorizationagain}
\end{eqnarray}
The vertex part vanishes for the non-interacting case, 
see eq.~(\ref{eq:wickfactorization}).
For the Fermi-liquid part, we used that the $d$-electron density is one half
and that the system is unpolarized.

Due to particle-hole symmetry for the translational invariant system,
the Fermi-liquid contribution vanishes for odd sites,
$C_{dc}^{S, {\rm FL}}(2m+1)=0$.
The spin correlation function at odd sites remain small even for 
substantial interactions.
However, from the DMRG data we infer that, for large interactions and
intermediate length scales, the vertex term for even sites
is (much) larger than the Fermi-liquid contribution.
In this regime, the ground state is very far from a single Slater determinant.

In the limit of very large distances and thus small excitation energies, 
we expect that the Fermi liquid picture description is applicable.
For the Fermi liquid contribution we can write quite generally
\begin{eqnarray}
M_{2m}&=& -\frac{V}{\pi} \int_{-\pi/2}^0 \rmd p\, {\rm Im}\Bigl[ 
\bigl[\sin(2mp)-\rmi \cos(2mp)\bigr] \nonumber \\
&& \hphantom{-\frac{V}{\pi} \int_{-\pi/2}^0 \rmd p\, {\rm Im}\Bigl[ }
\tilde{G}_{d,d}^{\rm ret}(\sin(p)/2)\Bigr] \; ,
\end{eqnarray}
where we neglected the contributions from the bound states
because their contribution vanishes exponentially for large distances.
Here, $\tilde{G}_{d,d}^{\rm ret}(\omega)$ is the exact retarded 
impurity Green function. 

At very large distances, only the region of small~$p$ contributes to
the integration because of the vastly oscillating sine and cosine functions.
Thus, we may approximate the impurity spectral function
by its Fermi-liquid form,
\begin{equation}
D_{d,d}(\omega) \approx \frac{1}{\pi} \frac{\Gamma_*}{\omega^2+\Gamma_*^2}
\label{eq:FLGammastar}
\end{equation}
with $\Gamma_*\propto T_{\text{L}}$ with the Kondo scale from 
eq.~(\ref{eq:Tkdef}).
Thus, we recover the result~(\ref{eq:analytdecay}) for the decay of the
spin correlation function at large distances, 
\begin{equation}
C_{dc}^S(2m\gg 1/(2\Gamma_*^2)) \approx 
-\frac{V^2}{8\pi^2 \Gamma_*^2} \frac{1}{m^2}\; .
\label{eq:analytdecayexact}
\end{equation}
The long-range decay of the correlation function is algebraic
but, since $\Gamma_*$ is exponentially small, this decay only sets in 
at exponentially large length scales.
The Gutzwiller approach reproduces this result qualitatively.

Note that the subtleties of the Kondo screening,
e.g., the Doniach-\v{S}unji\'c-Hamann tails in the impurity spectral 
function,~\cite{Logan1998,Logan2000,Logan2009,0022-3719-3-2-010}
contribute to the Kondo screening cloud ${\cal S}(r)$ for
intermediate to large distances that are well below $1/\Gamma_*^2$.
The visualization of the screening cloud requires the calculation of a two-particle
correlation function which is very demanding; for a variational
approach to the Kondo model, see Ref.~[\onlinecite{PhysRevB.92.195106}].

\section{Conclusions}
\label{sec:conclusions}

In this work, we studied the ground-state energy, the impurity magnetization
and susceptibility, and the Kondo screening cloud 
for the symmetric single-impurity Anderson model (SIAM) using
the results from the Gutzwiller, magnetic Hartree-Fock, 
and DMRG variational approaches. We restricted our study to
the case of a regular metal with a constant density of states around the
Fermi energy; Kondo screening for other host density of states, e.g., in graphene,
is studied in Ref.~\onlinecite{PhysRevB.88.245113,0034-4885-76-3-032501}.
For the ground-state energy and
magnetic properties, we compared our results to those
from the Bethe Ansatz that become exact in the wide-band limit.
For further reference, we defer
many technical details to the appendix.

Each of the three variational methods has its merits and limitations. 
\begin{itemize}
\item[--] 
The Hartree-Fock approach provides an excellent description of the 
ground-state energy for intermediate to strong couplings. 
However, since it displays a gap for (magnetic) excitations, 
Hartree-Fock theory
fails to reproduce the large magnetic susceptibility for strong couplings.
Concomitantly, it is unable to screen the impurity spin.

The Hartree-Fock theory correctly describes the charge excitations
of the symmetric SIAM. This makes it the perfect starting point for
more elaborate analytical approximations such as 
the local-moment approach that introduces the missing low-energy
spin-flip processes into the Hartree-Fock 
description.~\cite{Logan1998,Logan2000,Logan2009}
\item[--]  
Gutzwiller theory provides 
a rather poor upper bound for the ground-state energy.
However, it qualitatively describes the exponentially large magnetic
zero-field susceptibility for strong couplings because 
it retains
an exponentially small resonance in the impurity density of states
in the Kondo limit. Consequently, the impurity spin is completely
screened by the bath electrons at infinite distance from the impurity.

As an inherent Fermi-liquid description, Gutzwiller theory 
correctly reproduces the long-range behavior of the Kondo cloud.
However, for short and intermediate distances,
its description of the Kondo cloud is too simple-minded.
\item[--]  
  The DMRG method is numerically 
highly accurate for finite systems.
In this work, we map the SIAM on a ring onto a two-chain 
geometry with open boundary
conditions where we 
disregard the inter-chain coupling. 
Therefore, we can treat rings with up to $L=1400$ sites, and extrapolations 
of the ground-state energy to the thermodynamic limit are unproblematic.
We see that the Bethe Ansatz description is applicable
for interactions up to about half the bandwidth even at $V=0.1W$.

The intrinsic energy resolution is limited to $\Delta \omega= W/L$.
Therefore, the DMRG encounters problems to resolve the Abrikosov-Suhl resonance
in the impurity density of states in the strong-coupling limit,
and a reliable description of the impurity magnetization
and of the magnetic susceptibility
is limited to moderate interaction strengths. Correspondingly,
DMRG properly describes the short-range region of the Kondo cloud 
but does not cover the long-distance asymptotics because the
Kondo cloud in DMRG cannot exceed half the system size.
The NRG is best suited to resolve small energy scales, and
thus overcomes the DMRG limitations.
\end{itemize}
At the end of our presentation, we emphasize that
the Kondo screening cloud is amazingly large, even in the non-interacting 
limit, for reasonably small hybridization strengths, e.g., $V=0.1$,
and in one dimension where a larger fraction of the bath electrons can couple 
to the impurity than in higher dimensions.
This implies that magnetic impurities in metals can be correlated over
mesoscopic distances. 

Note, however, that this behavior depends on a number of assumptions,
namely,
(i), a perfect metallic host without impurities, 
(ii), zero temperature,
and, (iii), the Kondo regime which is guaranteed for the SIAM
by particle-hole and spin-flip symmetry for $U/\Gamma \gg 1$.
Deviations from these exceptional conditions, 
especially a finite temperature, will
drastically limit the range over which the impurity spin is screened.

Nevertheless, we can expect that two magnetic impurities in a metal
can sense each others' presence over quite some distance so that
they will bind into magnetic singlet (or triplet) pairs.
An investigation of this pairing requires the analysis of the
two-impurity Anderson model (TIAM), see, e.g., 
Ref.~[\onlinecite{Linneweberetal2017}]
for a recent Gutzwiller variational study, and references therein. 
A DMRG study of the TIAM is currently under way.

\begin{acknowledgments}
This research has been supported in part by
the Hungarian National Research, 
Development and Innovation Office (NKFIH) through Grant No.\ 
K120569 and PD-17-125261 
and the Hungarian Quantum Technology National Excellence Program 
(Project No. 2017-1.2.1-NKP-2017-00001).
\"O.L.\ also acknowledges support from the Alexander von Humboldt Foundation.
\end{acknowledgments}

\appendix

\section{Chain mapping using the Lanczos construction}
\label{app:A}

The Lanczos algorithm provides another way to derive the chain geometry from
the ring geometry in Sect.~\ref{sec:mappingringtochain}.
Dropping spin indices, we start from the seed state
\begin{equation}
|\Psi_0\rangle=\hat{d}^+|\text{vac}\rangle
\end{equation}
and find the Lanczos basis from
\begin{eqnarray}
|\Psi_1\rangle&=&\hat{H}_0|\Psi_0\rangle-a_0 |\Psi_0\rangle 
= V\hat{b}_0^+|\text{vac}\rangle
\nonumber \; ,\\
|\Psi_{n+1}\rangle &=& \hat{H}_0|\Psi_n\rangle -a_n |\Psi_n\rangle -b_n^2|\Psi_{n-1}\rangle
\; ,
\end{eqnarray}
for $1\leq n\leq L-1$, where
\begin{eqnarray}
a_n&=& \frac{\langle \Psi_n| \hat{H}_0|\Psi_n\rangle}{\langle \Psi_n |\Psi_n\rangle }
\nonumber \; ,\\
b_n^2 &= & \frac{\langle \Psi_n| \Psi_n\rangle}{\langle \Psi_{n-1} |\Psi_{n-1}\rangle} \; .
\end{eqnarray}
It is readily shown that $a_n=0$ ($0\leq n\leq L$), $b_1^2=2t^2$,
$b_n^2=t^2$ ($2\leq n\leq L$) and 
\begin{equation}
|\Psi_n\rangle = V (-t)^{n-1} \bigl[ e^{\rmi (n-1)\varphi}\hat{b}_{L-n+1}^+
+ e^{-\rmi (n-1)\varphi}\hat{b}_{n-1}^+\bigr] |\text{vac}\rangle \; .
\end{equation}
The algorithm automatically terminates after $n=L$, i.e., $|\Psi_{L+1}\rangle\equiv 0$.

After a proper normalization we find the $L+1$ basis states
\begin{eqnarray}
|d\rangle &=& |\Psi_0\rangle = \hat{d}^+ |\text{vac}\rangle
\; , \nonumber \\
|0\rangle &=& \hat{C}_0^+ |\text{vac}\rangle
\; , \nonumber \\
|n\rangle &=& \sqrt{\frac{1}{2}} \left[  e^{-\rmi n\varphi}\hat{b}_{n}^+
+ e^{\rmi n\varphi}\hat{b}_{L-n}^+\right]|\text{vac}\rangle \nonumber \\
 &=& 
\left\{ \begin{array}{@{}lcr@{}}
\hat{C}_n^+ |\text{vac}\rangle & \text{for} & 1\leq n\leq (L-1)/2\; , \\[3pt]
\rmi \hat{S}_{L-n}^+ |\text{vac}\rangle & \text{for} &(L+1)/2\leq n \leq L\; .
\end{array}\right.
\end{eqnarray}
Up to a phase factor for the $S$-electrons, 
this is the same basis as used in the canonical transformation.
The extra phase factor accounts for the electron transfer $\rmi t$ between
the $C$-electron and $S$-electron chains in Sect.~\ref{sec:mappingringtochain}.

\section{Equation-of-motion approach}

\subsection{Causal and retarded Green functions}

The causal Green function for the fermionic
Heisenberg operators $\hat{A}^{\vphantom{+}}(t)=\exp(\rmi \hat{H_0}t) 
\hat{A}^{\vphantom{+}}\exp(-\rmi \hat{H_0}t)$
and $\hat{B}^+$ is defined by 
\begin{eqnarray}
G_{A,B}^{\rm c}(t)
&=& (-\rmi) \hat{T} \langle \hat{A}^{\vphantom{+}}(t) \hat{B}^+\rangle 
\nonumber \\
&=&
(-\rmi) \Theta(t) \langle \hat{A}^{\vphantom{+}}(t) \hat{B}^+\rangle 
+\rmi \Theta(-t) \langle \hat{B}^+\hat{A}^{\vphantom{+}}(t)  \rangle 
\; .
\end{eqnarray}
Equal-time expectation values 
$\langle \hat{B}^+ \hat{A}^{\vphantom{+}}\rangle$
in the ground state $|0\rangle\equiv|\Phi_0\rangle$
can be directly calculated
from the causal Green functions by taking the limit $t\to 0^-$.

For the equation-of-motion approach, it is more convenient
to study the retarded Green function,
\begin{equation}
G_{A,B}^{\rm ret}(t)
= (-\rmi) \Theta(t) \langle 
\left[\hat{A}^{\vphantom{+}}(t), \hat{B}^+\right]_+\rangle \; . 
\end{equation}
Its Fourier transformation is defined by
\begin{equation}
\tilde{G}_{A,B}^{\rm ret}(\omega)= 
\int_{0}^{\infty} \!\!\rmd t e^{(\rmi \omega-\eta) t}G_{A,B}^{\rm ret}(t)
= \int_{-\infty}^{\infty} \!\!\rmd \omega' \frac{D_{A,B}(\omega')}{
\omega-\omega'+\rmi \eta} \, .
\end{equation}
Here, the spectral function is defined by 
\begin{eqnarray}
D_{A,B}(\omega)
&=& \sum_m \Bigl[
\langle 0 | \hat{B}^+ | m\rangle 
\langle m | \hat{A}^{\vphantom{+}} | 0\rangle
\delta(\omega-E_0+E_m) \nonumber \\
&&  
+ 
\langle 0 | \hat{A}^{\vphantom{+}} | m\rangle
\langle m | \hat{B}^+ | 0\rangle 
\delta(\omega+E_0-E_m) \Bigr] ,\;\;
\label{eq:defDOSAB}
\end{eqnarray}
where $|m\rangle$ denotes the eigenstates of $\hat{H}_0$ with
energy $E_m$ (Lehmann representation).
Using the Lehmann representation it is readily shown that
\begin{equation}
\tilde{G}_{A,B}^{\rm c}(\omega)
= \int_{-\infty}^{\infty} \rmd \omega' \frac{D_{A,B}(\omega')}{
\omega-\omega'+\rmi \eta \sgn(\omega)}
\label{eq:lehmann}
\end{equation}
with the sign function $\sgn(x)=|x|/x$.
Therefore, the causal Green function is obtained from the
retarded Green function by replacing $\omega+\rmi \eta $
by $\omega+\rmi \eta \sgn(\omega)$.

When $\hat{A}\neq \hat{B}$, the spectral function $D_{A,B}(\omega)$
is not necessarily real. We separate the real and imaginary part,
\begin{equation}
D_{A,B}(\omega)
= \frac{D_{A,B}(\omega) +D_{A,B}^*(\omega) }{2} 
+ \rmi 
\frac{D_{A,B}(\omega) -D_{A,B}^*(\omega) }{2\rmi} 
\end{equation}
and use $D_{A,B}^*(\omega) =D_{B,A}(\omega)$ to find
\begin{eqnarray}
D_{A,B}(\omega) &=& -\frac{1}{\pi}
{\rm Im}\left[ \frac{G_{A,B}^{\rm ret}(\omega)+G_{B,A}^{\rm ret}(\omega)}{2}\right] 
\nonumber \\
&& -\frac{\rmi }{\pi}
{\rm Im}\left[ \frac{G_{A,B}^{\rm ret}(\omega)-G_{B,A}^{\rm ret}(\omega)}{2\rmi }\right] 
\; .
\label{eq:DOSAB}
\end{eqnarray}
For $\hat{A}=\hat{B}$ we recover the standard expression
\begin{equation}
D_{A,A}(\omega) = -\frac{1}{\pi}
{\rm Im}\left[ G_{A,A}^{\rm ret}(\omega)\right] \; .
\label{eq:DOSAA}
\end{equation}

\subsection{Green functions for the non-interacting SIAM}

\subsubsection{Time domain}
For the non-interacting symmetric SIAM, we study the 
four retarded Green functions
\begin{eqnarray}
G_{k,p;\sigma}^{\rm ret}(t) &=& (-{\rm i}) \Theta(t) \langle 
\left[ \hat{c}_{k,\sigma}^{\vphantom{+}}(t),\hat{c}_{p,\sigma}^+\right]_+
\rangle \; , \label{eq:defGFretarded1}\\
G_{d,p;\sigma}^{\rm ret}(t) &=& (-{\rm i}) \Theta(t) \langle 
\left[\hat{d}_{\sigma}^{\vphantom{+}}(t),\hat{c}_{p,\sigma}^+\right]_+
\rangle \; , \label{eq:defGFretarded2}\\
G_{k,d;\sigma}^{\rm ret}(t) &=& (-{\rm i}) \Theta(t) \langle 
\left[ \hat{c}_{k,\sigma}^{\vphantom{+}}(t),\hat{d}_{\sigma}^+\right]_+
\rangle \; , \label{eq:defGFretarded3}\\
G_{d,d;\sigma}^{\rm ret}(t) &=& (-{\rm i}) \Theta(t) \langle 
\left[ \hat{d}_{\sigma}^{\vphantom{+}}(t),\hat{d}_{\sigma}^+\right]_+
\rangle \; . 
\label{eq:defGFretarded4}
\end{eqnarray}
Taking the time derivative leads to
\begin{eqnarray}
{\rm i} \dot{G}^{\rm ret}_{k,p;\sigma}(t) &=& \delta(t) \delta_{k,p}
+(-{\rm i}) \Theta(t) 
\langle \Bigl[ \left[ \hat{c}_{k,\sigma}^{\vphantom{+}}(t),\hat{H}_0\right]_-,
\hat{c}_{p,\sigma}^+\Bigr]_+\rangle \nonumber \\
&=&
\delta(t) \delta_{k,p}+ \epsilon(k) G_{k,p;\sigma}^{\rm ret}(t) 
+\frac{V_k^*}{\sqrt{L}} G_{d,p;\sigma}^{\rm ret}(t)  \; ,
\nonumber \\
\label{eq:GFdot1}\\
{\rm i} \dot{G}_{d,p;\sigma}^{\rm ret}(t) &=& 
(-{\rm i}) \Theta(t) 
 \langle  
\Bigl[ \left[\hat{d}_{\sigma}^{\vphantom{+}}(t),\hat{H}_0\right]_-,
\hat{c}_{p,\sigma}^+\Bigr]_+\rangle  \nonumber \\
&=& 
\sum_k\frac{V_k}{\sqrt{L}} G_{k,p;\sigma}^{\rm ret}(t) 
-E_{d,\sigma} G_{d,p;\sigma}^{\rm ret}(t)\; ,
\label{eq:GFdot2}\\
{\rm i} \dot{G}_{k,d;\sigma}^{\rm ret}(t) &=& 
(-{\rm i})\Theta(t)  \langle  
\Bigl[\left[ \hat{c}_{k,\sigma}^{\vphantom{+}}(t),\hat{H}_0\right]_-,
\hat{d}_{\sigma}^+\Bigr]_+\rangle \nonumber \\
&=& 
\epsilon(k) G_{k,d;\sigma}^{\rm ret}(t) +\frac{V_k^*}{\sqrt{L}} G_{d,d;\sigma}^{\rm ret}(t)  \; ,
\label{eq:GFdot3}
\end{eqnarray}
and
\begin{eqnarray}
{\rm i} \dot{G}_{d,d;\sigma}^{\rm ret}(t) &=&
\delta(t) 
+ (-{\rm i}) \Theta(t) \langle 
\Bigl[\left[ \hat{d}_{\sigma}^{\vphantom{+}}(t),\hat{H}_0\right]_-,
\hat{d}_{\sigma}^+ \Bigr]_+\rangle  \nonumber \\
&=& \delta(t) + \sum_k\frac{V_k}{\sqrt{L}} G_{k,d;\sigma}^{\rm ret}(t) 
-E_{d,\sigma} G_{d,d;\sigma}^{\rm ret}(t)\, .\nonumber \\
\label{eq:GFdot4}
\end{eqnarray}
Here, we used the anticommutation relations of the Fermi operators 
and the commutation relations
\begin{eqnarray}
\left[\hat{c}_{k,\sigma}^{\vphantom{+}},\hat{T}\right]_- 
= \epsilon(k) \hat{c}_{k,\sigma}^{\vphantom{+}} &\; , \;  & 
\left[\hat{d}_{\sigma}^{\vphantom{+}},\hat{T}\right]_- 
= 0 \nonumber \; , \\
\left[\hat{c}_{k,\sigma}^{\vphantom{+}},\hat{V}\right]_- 
= \frac{V_k^*}{\sqrt{L}} \hat{d}_{\sigma}^{\vphantom{+}} &, & 
\left[\hat{d}_{\sigma}^{\vphantom{+}},\hat{V}\right]_- 
= \sum_k \frac{V_k}{\sqrt{L}} \hat{c}_{k,\sigma}^{\vphantom{+}}
\; ,\nonumber \\
\left[\hat{d}_{\sigma}^{\vphantom{+}},\hat{P}\right]_- =
-E_{d,\sigma} \hat{d}_{\sigma}^{\vphantom{+}} &.&
\end{eqnarray}
In the presence of an external magnetic field for the bath electrons, $\epsilon(k)$
must be replaced by $\epsilon(k)-\sigma_nB_{\text{bath}}$ 
with $\sigma_n=1$ ($\sigma_n=-1$) for $\sigma=\uparrow$ ($\sigma=\downarrow$).
For non-interacting electrons, the equations of motion
lead to a closed 
set of differential equations~(\ref{eq:GFdot1})--(\ref{eq:GFdot4}).

\subsubsection{Explicit solution in the frequency domain}

The equation-of-motion method works in the frequency domain.
The Fourier transformation
of the time derivative of retarded Green functions are given by
\begin{eqnarray}
{\rm FT}\left\{{\rm i}\dot{G}_{A,B}^{\rm ret}(t)\right\}(\omega)&=&
\int_{-\infty}^{\infty}{\rm d} t e^{-\eta |t|} e^{{\rm i}\omega t} 
\left({\rm i} \dot{G}_{A,B}^{\rm ret}(t)\right) \nonumber \\
&=& {\rm i}\biggl[ 
\left. G_{A,B}^{\rm ret}(t)e^{-\eta |t|}e^{{\rm i}\omega t}\right|_{-\infty}^{\infty}
\nonumber \\
&& - \int_0^{\infty} {\rm d}t G_{A,B}^{\rm ret}(t)
\frac{{\rm d}}{{\rm d}t} 
\left( e^{-\eta t}e^{{\rm i}\omega t}\right)
\biggr] \nonumber \\
&=& (\omega+{\rm i}\eta) \tilde{G}_{A,B}^{\rm ret}(\omega) \; ,
\label{eq:FTGdot}
\end{eqnarray}
where we used partial integration in the first step and
the fact that $G_{A,B}^{\rm ret}(t<0)=0$.

To solve the equations~(\ref{eq:GFdot1})--(\ref{eq:GFdot4})
we transformation them into frequency space.
We find
\begin{eqnarray}
\left(\omega+E_{d,\sigma}+{\rm i}\eta\right)\tilde{G}_{d,d;\sigma}^{\rm ret}(\omega) &=&
1\! + \!\sum_k\frac{V_k}{\sqrt{L}} \tilde{G}_{k,d;\sigma}^{\rm ret}(\omega),
\label{eq:GFdot4om}\\
\left(\omega+E_{d,\sigma}+{\rm i}\eta\right)\tilde{G}_{d,p;\sigma}^{\rm ret}(\omega)&=& 
\sum_k\frac{V_k}{\sqrt{L}} \tilde{G}_{k,p;\sigma}^{\rm ret}(\omega),
\label{eq:GFdot2om}\\
\left(\omega-\epsilon(k)+{\rm i}\eta\right)\tilde{G}_{k,d;\sigma}^{\rm ret}(\omega)&=& 
\frac{V_k^*}{\sqrt{L}} \tilde{G}_{d,d;\sigma}^{\rm ret}(\omega),
\label{eq:GFdot3om}\\
\left(\omega-\epsilon(k) +{\rm i}\eta\right)
 \tilde{G}_{k,p;\sigma}^{\rm ret}(\omega)
&=& \delta_{k,p}
+\frac{V_k^*}{\sqrt{L}} \tilde{G}_{d,p;\sigma}^{\rm ret}(\omega) .
\label{eq:GFdot1om}
\end{eqnarray}
The resulting set of equations is readily solved.
We define the retarded hybridization function
\begin{equation}
\Delta^{\rm ret} (\omega)= 
\frac{1}{L} 
\sum_{k}\frac{|V_k|^2}{\omega-\epsilon(k)+{\rm i}\eta}\; ,
\label{eq:Deltacandret}
\end{equation}
and find
\begin{equation}
\tilde{G}_{d,d;\sigma}^{\rm ret}(\omega) =
\frac{1}{\omega+E_{d,\sigma}-\Delta^{\rm ret}(\omega)}
\; ,
\label{eq:EOMimpurityGF4}
\end{equation}
\begin{equation}
\tilde{G}_{k,d;\sigma}^{\rm ret}(\omega) =
\sqrt{\frac{1}{L}}
\frac{V_k^*}{(\omega-\epsilon(k)+{\rm i}\eta)
(\omega+E_{d,\sigma}-\Delta^{\rm ret}(\omega))}
\; , \label{eq:EOMimpurityGF3}
\end{equation}
\begin{equation}
\tilde{G}_{d,p;\sigma}^{\rm ret}(\omega) = 
\sqrt{\frac{1}{L}}
\frac{V_p}{(\omega-\epsilon(p)+{\rm i}\eta)
(\omega+E_{d,\sigma}-\Delta^{\rm ret}(\omega))}
\; , \label{eq:EOMimpurityGF2}
\end{equation}
and 
\begin{eqnarray}
 \tilde{G}_{k,p;\sigma}^{\rm ret}(\omega) &=& 
\frac{1}{\omega-\epsilon(k)+{\rm i}\eta}\biggl( \delta_{k,p}
\nonumber\\
&& +\frac{1}{L}
\frac{V_pV_k^*}{(\omega-\epsilon(p)+{\rm i}\eta)
(\omega+E_{d,\sigma}-\Delta^{\rm ret}(\omega))}\biggr)\; . \nonumber \\
\label{eq:EOMimpurityGF1}
\end{eqnarray}
The equations for the causal Green functions are obtained by replacing
$\eta$ by $\eta{\rm sgn}(\omega)$.
In the presence of a magnetic field for the bath electrons, we must replace
$\Delta^{\rm ret}(\omega)$ by $\Delta_{\sigma}^{\rm ret}(\omega)=
\Delta(\omega+\sigma_nB_{\text{bath}})$.

\section{Spectral properties}

To simplify the analysis, we shall consider the case $V_k=V>0$.
Moreover, we study the case of a one-dimensional ring
with electron transfers between nearest-neighbors and bandwidth $W\equiv 1$,
\begin{equation}
\epsilon(k)=-\cos(k)/2 \quad \hbox{for} \; |k|\leq \pi \; .
\end{equation}
The non-interacting density of states becomes
\begin{equation}
\rho_0(\epsilon)=\frac{2}{\pi\sqrt{1-4\epsilon^2}} \quad \hbox{for}\;
|\epsilon|\leq 1/2
\label{eq:DOS1d}
\end{equation}
so that $d_0=\rho_0(0)=2/\pi$ for the density of states in one dimension.
\subsection{Impurity spectral function}

First, we work out the impurity spectral function
from the impurity Green function~(\ref{eq:EOMimpurityGF4}).
We have 
\begin{eqnarray}
\Delta^{\rm ret}(\omega) &=& V^2\int_{-1/2}^{1/2} {\rm d}\epsilon
\frac{\rho_0(\epsilon)}{\omega-\epsilon+{\rm i}\bar{\eta}}\nonumber \\
&=&V^2 \Lambda_0(\omega)-{\rm i}\pi 
V^2\rho_0(\omega)\; .
\label{eq:DeltaReandIm}
\end{eqnarray}
In one dimension, $\Lambda_0(\omega)=0$ for $|\omega|\leq 1/2$
and
\begin{equation}
\Lambda_0(\omega)=\frac{2\sgn(\omega)}{\sqrt{4\omega^2-1}}
\end{equation}
for $|\omega|> 1/2$.
Thus, we have
\begin{eqnarray}
\omega+E_{d,\sigma}-\Delta^{\rm ret}(\omega) &=& 
\omega+E_{d,\sigma}+\rmi \pi V^2\rho_0(\omega) \nonumber \\
&& \hphantom{\omega+E_{d,\sigma}}
(|\omega|\leq 1/2) \nonumber \; ,\\[6pt]
&=& \omega+E_{d,\sigma}- \frac{2V^2\sgn(\omega)}{\sqrt{4\omega^2-1}}+\rmi \eta
\nonumber \\
&&\hphantom{\omega+E_{d,\sigma}} (|\omega|>1/2) \; . 
\label{eq:omminusDeltaom}
\end{eqnarray}
We keep the infinitesimal imaginary part $\eta=0^+$ because of 
the (anti-)bound states outside the host-electron band.

The density of states for the $d$-electrons (`$d$-electron spectral function') 
follows from eq.~(\ref{eq:defDOSAB}) as
\begin{eqnarray}
D_{d,d;\sigma}(\omega) &=& -\frac{1}{\pi} 
{\rm Im}\left(\frac{1}{\omega+E_{d,\sigma}-\Delta_{\sigma}^{\rm ret}(\omega)}\right)
\nonumber \\
&=& \frac{V^2\rho_0(\omega_{\sigma})}{(\pi\rho_0(\omega_{\sigma})V^2)^2
+(\omega+E_{d,\sigma})^2} \quad \hbox{for} \; |\omega_{\sigma}|\leq \frac{1}{2}
\nonumber \\
\label{eq:Dddinside}
\end{eqnarray}
with $\omega_{\sigma}=\omega+\sigma_nB_{\text{bath}}$.
For $ |\omega_{\sigma}|>1/2$ we have
\begin{eqnarray}
D_{d,d;\sigma}(\omega)
&=& \delta\left(\omega+E_{d,\sigma}+\frac{2V^2}{\sqrt{4\omega_{\sigma}^2-1}}\right) 
\nonumber \\
&& 
+ \delta\left(\omega+E_{d,\sigma}-\frac{2V^2}{\sqrt{4\omega_{\sigma}^2-1}}\right) \; .
\label{eq:Dddoutside}
\end{eqnarray}
For $B_{\text{bath}}=E_{d,\sigma}=0$, we can further write
\begin{equation}
D_{d,d;\sigma}(\omega)=
Z(V)\delta\left(\omega+\frac{v_+}{2}\right) +
Z(V)\delta\left(\omega-\frac{v_+}{2}\right)
\end{equation}
with 
\begin{eqnarray}
Z(V) &=& \left[\frac{\partial}{\partial \omega}
\left( \omega-\Delta^{\rm ret}(\omega)
\right)_{\pm v_+/2}\right]^{-1}\!\!
=\frac{1}{1+4V^2v_+/v_-^3} \; , \nonumber \\
v_{\pm}(V)&=&\frac{\sqrt{\sqrt{1+64V^4}\pm 1}}{\sqrt{2}} \equiv v_{\pm}\; .
\end{eqnarray}
The bound and anti-bound states at $\omega=\pm v_+/2$ contribute
two poles of strength $Z(V)$ to the impurity spectral function.
For small $V$, we have $v_{+}/2\approx 1/2+4V^4$,
$v_{-}/2\approx 2V^2$, and
$Z(V)\approx 16 V^4$. The pole contributions are very small for small~$V$,
of order $V^4$.

For $E_{d,\sigma}\neq 0$, it is best to determine 
the energies of (anti-)bound state $v_{\rm b}(V,E_{d\sigma})<-1/2$
[$v_{\rm ab}(V,E_{d\sigma}>1/2$], and their strengths $Z_{\rm b}(V)$
[$Z_{\rm ab}(V)]$ numerically from the equations
\begin{equation}
P_{\pm}(v_{\rm b/ab}) =0 \quad , \quad
Z_{\rm b/ab}(V) =\left[P_{\pm}'(v_{\rm b/ab})\right]^{-1}\; , 
\label{eq:Zvbcalcl}
\end{equation}
where
\begin{equation}
P_{\pm}(\omega)=\omega+E_{d,\sigma}\pm \frac{2V^2}{\sqrt{4\omega_{\sigma}^2-1}} 
\; .\label{eq:Ppmdef}
\end{equation}

\subsection{Density of states}
\label{sec:DOSprops}

The single-particle density of states is defined by
\begin{equation}
D_{\sigma}(\omega) = \sum_m \delta(\omega-E_m) \; .
\end{equation}
To make contact with the retarded Green functions,
we write the single-particle density of states in the form
\begin{eqnarray}
D_{\sigma}(\omega)&=& - \frac{1}{\pi} {\rm Im}\biggl(\sum_m 
\langle \hat{a}_{m,\sigma}^+ \frac{1}{\omega-(\hat{H}_0-E_0)+{\rm i}\eta}
\hat{a}_{m,\sigma}^{\vphantom{+}}\rangle \nonumber \\
&& \hphantom{- \frac{1}{\pi}}
+\langle \hat{a}_{m,\sigma}^{\vphantom{+}}
\frac{1}{\omega-(\hat{H}_0-E_0)+{\rm i}\eta}
\hat{a}_{m,\sigma}^+ 
\rangle
\biggr) \, , 
\end{eqnarray}
where we used the fact that $\hat{a}_{m,\sigma}^+$
($\hat{a}_{m,\sigma}^{\vphantom{+}}$)
creates (annihilates) an electron with exact single-particle energy $E_m$ 
in the ground state.
The sum over $m$ runs over all single-particle
excitations of the ground state and thus represents
the trace over all single-particle eigenstates,
\begin{equation}
D_{\sigma}(\omega)= - \frac{1}{\pi} {\rm Im}{\rm Tr}_1
\Bigl(\frac{1}{\omega-(\hat{H}_0-E_0)+{\rm i}\eta}\Bigr) \; .
\end{equation}
We can equally use the excitations
$\hat{c}_{k,\sigma}^+|\Phi_0\rangle$, 
$\hat{c}_{k,\sigma}^{\vphantom{+}}|\Phi_0\rangle$,
and 
$\hat{d}_{\sigma}^+|\Phi_0\rangle$, 
$\hat{d}_{\sigma}^{\vphantom{+}}|\Phi_0\rangle$, respectively,
to perform the
trace over the single-particle excitations of the ground state.
Therefore, we may write
\begin{eqnarray}
D_{\sigma}(\omega)&=& - \frac{1}{\pi} {\rm Im}
\biggl[\sum_k \biggl(
\langle \hat{c}_{k,\sigma}^+ \frac{1}{\omega-(\hat{H}_0-E_0)+{\rm i}\eta}
\hat{c}_{k,\sigma}^{\vphantom{+}}\rangle 
\nonumber \\
&& \hphantom{ - \frac{1}{\pi} {\rm Im}\biggl( \sum_k}
+ \langle \hat{c}_{k,\sigma}^{\vphantom{+}}
\frac{1}{\omega-(\hat{H}_0-E_0)+{\rm i}\eta}
\hat{c}_{k,\sigma}^+ 
\rangle\biggr)
\nonumber \\
&& 
\hphantom{ - \frac{1}{\pi} {\rm Im}\biggl( }
+ 
\langle \hat{d}_{\sigma}^+ \frac{1}{\omega-(\hat{H}_0-E_0)+{\rm i}\eta}
\hat{d}_{\sigma}^{\vphantom{+}}\rangle
\nonumber \\
&& \hphantom{ - \frac{1}{\pi} {\rm Im}\biggl( }
+\langle \hat{d}_{\sigma}^{\vphantom{+}}
\frac{1}{\omega-(\hat{H}_0-E_0)+{\rm i}\eta}
\hat{d}_{\sigma}^+ 
\rangle \biggr] \nonumber \\
&=& - \frac{1}{\pi} {\rm Im}
\biggl[
\sum_k G_{k,k}^{\rm ret}(\omega)+ G_{d,d}^{\rm ret}(\omega) 
\biggr] \; .
\label{eq:totalDOSforHF}
\end{eqnarray}
Equation~(\ref{eq:EOMimpurityGF1}) shows that the band Green function
consists of the undisturbed host Green function for $V_k\equiv 0$
and a $1/L$ correction due to the hybridization.
Therefore, using eqs.~(\ref{eq:EOMimpurityGF4}) and~(\ref{eq:EOMimpurityGF1}),
the contribution due to a finite hybridization is given by
\begin{eqnarray}
D_{{\rm imp},\sigma}(\omega)&=& -\frac{1}{\pi}{\rm Im}
\biggl[
\frac{1}{\omega+E_{d,\sigma}-\Delta^{\rm ret}(\omega)}\nonumber \\
&& \hphantom{-\frac{1}{\pi}{\rm Im}}\times 
\left(
1 + \sum_k 
\frac{|V_k|^2/L}{(\omega-\epsilon(k)+{\rm i}\eta)^2}
\right)
\biggr] \nonumber \\
&=& 
- \frac{1}{\pi} {\rm Im}
\left[
\frac{1-(\partial \Delta^{{\rm ret}}(\omega))/(\partial\omega)}{
\omega+E_{d,\sigma}-\Delta^{\rm ret}(\omega)}\right]\nonumber \\ 
&=& 
- \frac{1}{\pi} \frac{\partial }{\partial \omega}
{\rm Im}
\left[\ln\left(\omega+E_{d,\sigma}-\Delta^{\rm ret}(\omega) \right)\right] .
\end{eqnarray}
We find the band contribution for $|\omega|< 1/2$
from the complex logarithm
\begin{equation}
D_{{\rm imp},\sigma}^{\rm band}(\omega)
= 
 - \frac{1}{\pi} 
\frac{\partial }{\partial \omega}
\left[
\Cot^{-1}\left( 
\frac{\omega+E_{d,\sigma}-V^2\Lambda_0(\omega)}{\pi\rho_0(\omega)V^2}
\right)
\right] \; ,
\label{eq:DOSfromGFagain}
\end{equation}
where $\Cot^{-1}(x)=\pi \Theta(-x)+\cot^{-1}(x)$ is continuous and differentiable
across $x=0$; $\Theta(x)$ is the Heaviside step function. We shall use this form 
for the calculation of the ground-state energy, see below.

Using eq.~(\ref{eq:omminusDeltaom}) and~(\ref{eq:Ppmdef}), 
the (anti-)bound states contribute 
\begin{eqnarray}
D_{{\rm imp},\sigma}^{\rm b/ab}(\omega)&=& - \frac{1}{\pi} 
\frac{\partial }{\partial \omega}
\left[
\Cot^{-1}\left( 
\frac{P_{\pm}(\omega)}{\eta}
\right)
\right]\nonumber \\
&=& 
\frac{1}{\pi}  \frac{\eta P_{\pm}'(\omega)}{P_{\pm}^2(\omega)+\eta^2} \; .
\label{eq:babDOScontributionA}
\end{eqnarray}
When we let $\eta\to 0$, we only retain a contribution where $P_{\pm}(\omega)=0$,
i.e., at $\omega_{\pm}=v_{\rm b/ab}(V,E_{d\sigma})$.
Thus,
\begin{eqnarray}
D_{{\rm imp},\sigma}^{\rm b/ab}(\omega)&=& \frac{1}{\pi} 
\frac{\eta Z_{\rm b/ab}(V)}{(\omega- v_{\rm b/ab})^2+(Z_{\rm b/ab}(V)\eta)^2} \nonumber \\
&=& 
\delta\left(\omega-v_{\rm b}(V,E_{d,\sigma})\right) +
\delta\left(\omega-v_{\rm ab}(V,E_{d,\sigma})\right) \; ,\nonumber \\
\label{eq:babDOScontribution}
\end{eqnarray}
where we let $\eta\to 0^+$ in the last step.
We see that the (anti-)bound states contribute poles of strength unity to
the density of states in the presence of the impurity. 
Recall that their contribution to the 
impurity spectral function was smaller by the weight 
factor~$Z_{\rm a/ab}(V)\ll 1$.

\section{Ground-state expectation values}

In this section we derive the ground-state expectation values
for the energy, the $d$-occupancy, and the hybridization matrix element
for the symmetric SIAM.

\subsection{Ground-state energy}
\label{sec:gsenergy}

\subsubsection{Symmetric SIAM}

We consider the case $E_{d,\sigma}=0$.
The ground-state energy can immediately be calculated using the density of states.
We subtract the energy for $V=0$ and write
\begin{equation}
e_0(V)=2\int_{-\infty}^0{\rm d}\omega \omega D_{{\rm imp},\sigma}(\omega)
=e_0^{\rm b}(V) + e_0^{\rm band}(V) \; .
\end{equation}
The upper limit of integration is zero because all single-particle states up to
the Fermi energy $E_{\rm F}=0$ are occupied.

Using eq.~(\ref{eq:babDOScontribution}),
 the contribution from the bound state is given by
\begin{equation}
e_0^{\rm b}(V)=2\left(\frac{1}{2}-\frac{v_+}{2}\right)=1-v_+ \; ,
\label{eq:boundenergycontribution}
\end{equation}
where we measure the energy contribution with respect to the lower band edge,
$\epsilon_{\rm edge}=-1/2$.
For small~$V$, the contribution is very small, $e_0^{\rm b}(V\ll 1)\approx -8V^4$.

The result for the band contribution 
to the impurity density of states~(\ref{eq:DOSfromGFagain})
and a partial integration
lead to the band contribution to the ground-state energy in the form
\begin{eqnarray}
e_0^{\rm band}(V)&=&R(V)+\frac{2}{\pi}\int_{-1/2}^0{\rm d}\omega 
\cot^{-1}\left( 
\frac{\omega-V^2\Lambda_0(\omega)}{\pi\rho_0(\omega)V^2}
\right)
\; , \nonumber \\
R(V)&=&-1 -\frac{2}{\pi} 
\left[ \omega\Cot^{-1}\left( 
\frac{\omega-V^2\Lambda_0(\omega)}{\pi\rho_0(\omega)V^2}\right)
\right]_{-1/2-\eta}^0
\nonumber \\
&& +\frac{2}{\pi}\int_{-1/2}^0\rmd \omega \, \pi \Theta\left( -
\frac{\omega-V^2\Lambda_0(\omega)}{\pi\rho_0(\omega)V^2}\right)\; .
\label{eq:ezeroSIAMtdl}
\end{eqnarray}
The first term in $R(V)$ accounts for the energy term introduced 
in the definition~(\ref{eq:boundenergycontribution})
of the bound-state energy. It is compensated by 
the last term in $R(V)$. The second term gives a zero contribution at $\omega=0$ 
because $\Cot^{-1}(0)$ is finite. It is more subtle to evaluate the second term
at $\omega=-1/2$ because, in one dimension, neither $\Lambda_0(\omega)$
nor $\rho_0(\omega)$ are continuous. This problem is circumvented
by using a finite $\bar{\eta}$ in eq.~(\ref{eq:DeltaReandIm}).
For $\bar{\eta}>0$ we thus 
see that $\Lambda_0(-1/2-\eta)\to-\infty$, $\rho_0(-1/2-\eta)\to 0^+$
so that we encounter $\Cot^{-1}(+\infty)=0$ so that the second term vanishes 
altogether. Therefore, we find $R(V)=0$.

In one dimension, using $\Lambda_0(\omega)=0$ and $\rho_0(\omega)$ for
$|\omega|<1/2$ from eq.~(\ref{eq:DOS1d}),
{\sc Mathematica}~\cite{Mathematica10} 
gives the band contribution
\begin{eqnarray}
e_0^{\rm band}(V)&=&\frac{2}{\pi} \int_{-1/2}^0 {\rm d} \epsilon\,
\cot^{-1} \left(
\frac{\epsilon\sqrt{1-4\epsilon^2}}{2V^2}
\right) \nonumber \\
&=& \frac{1}{2\pi}\biggl[
-\pi +2v_+\arctan\left(\frac{1}{v_-}\right) \nonumber \\
&& \hphantom{\frac{1}{2\pi}\biggl[}
+ v_-\ln\left(\frac{v_+-1}{v_++1}
\right)
\biggr] \; .
\end{eqnarray}
The total energy reads
\begin{eqnarray}
e_0(V)
&=&e_0^{\rm band}(V)+e_0^{\rm b}(V) \nonumber \\
&=&
\frac{1}{2\pi}\biggl[
-\pi +2v_+\arctan\left(\frac{1}{v_-}\right) \nonumber \\
&& \hphantom{\frac{1}{2\pi}\biggl[}
+ v_-\ln\left(\frac{v_+-1}{v_++1}
\right)
\biggr] +(1-v_+) \; .
\label{appeq:finalenergyTDL}
\end{eqnarray}

\subsubsection{Limit of small hybridization}

For $V\ll 1$ we Taylor expand $e_0(V)$ in eq.~(\ref{appeq:finalenergyTDL})
with the result
\begin{equation}
e_0^{\rm small}(V)=
\frac{4V^2}{\pi} \left(\ln(V^2)+\ln(2)-1\right) -4V^4 \; .
\end{equation}
Corrections are of the order $V^6\ln(V^2)$. 
For $V=0.1$, the approximate formula works very well.
We have $E_{\rm SIAM}^{(0)}(0.1)=-0.06291$
whereas the approximation gives
$E_{\rm SIAM}^{\rm small}(0.1)=-0.06294$, with a relative error of
less than one per mill.

The result agrees with the small-$V$ expression derived 
as eqs.~(E.1) and~(E.5) by Linneweber and collaborators.~\cite{Linneweberetal2017}
Using {\sc Mathematica}~\cite{Mathematica10} 
we find in one dimension using $d_0=\rho_0(0)=2/\pi$ and $\ln(e)=1$
\begin{equation}
C=\frac{e}{2} \exp\left( \int_{-1/2}^0{\rm d} \epsilon 
\frac{d_0-\rho_0(\epsilon)}{d_0\epsilon}
\right) = e
\end{equation}
so that 
\begin{eqnarray}
e_0^{\rm small}(V)&=&2V^2d_0 \ln\left(\frac{\pi V^2d_0}{C}\right)\nonumber \\
&=& \frac{4V^2}{\pi}\left( \ln(V^2)+\ln(2)-1\right) \; ,
\end{eqnarray}
with corrections of the order $V^4$.

\subsection{Expectation values from Green functions}
\label{sec:impocc}

The Green functions permit the calculation of ground-state expectation values.
By definition, we have ($\eta=0^+$)
\begin{eqnarray}
\langle \hat{B}^+ \hat{A}^{\vphantom{+}}\rangle 
&=& (-{\rm i}) G_{A,B}^{\rm c}(t=-\eta) \nonumber \\
&=& \int_{-\infty}^{\infty} \frac{{\rm d}\omega}{2\pi{\rm i}}
e^{{\rm i}\eta\omega}
\tilde{G}_{A,B}^{\rm c}(\omega) \nonumber \\
&=& \int_{-\infty}^{\infty} \rmd \omega' \! D_{A,B}(\omega') 
\int_{-\infty}^{\infty} \frac{\rmd \omega}{2\pi \rmi} 
\frac{e^{\rmi \eta \omega}}{\omega-\omega'+\rmi\eta'\sgn(\omega')}
\nonumber \\
&=& \int_{-\infty}^0 \!\rmd \omega' D_{A,B}(\omega') 
\; ,
\end{eqnarray}
where we used eq.~(\ref{eq:lehmann}) in the second step.
In the last step we extended the $\omega$-integral to a contour
integral in the upper complex plane that includes the real axis
and an arc around the origin with infinite radius. 
The residue theorem then results in
$\Theta(-\omega')$ because a pole of strength unity
appears in the upper complex plane only
for $\sgn(\omega')=-1$.

For the $d$-electron occupancy, we set
$\hat{A}=\hat{B}=\hat{d}_{\sigma}^{\vphantom{+}}$ and find
from eqs.~(\ref{eq:DOSAA}), (\ref{eq:Dddinside}), and~(\ref{eq:Dddoutside})
\begin{equation}
\langle \hat{d}_{\sigma}^+ \hat{d}_{\sigma}^{\vphantom{+}}\rangle = 
Z(V)+\int_{-1/2}^0{\rm d}\omega
\frac{\rho_0(\omega)V^2}{\omega^2 + (\pi\rho_0(\omega)V^2)^2} \; .
\end{equation}
Apparently, this expression is symmetric in $\omega$ so that
one readily recovers 
$\langle \hat{d}_{\sigma}^+ \hat{d}_{\sigma}^{\vphantom{+}}\rangle =1/2$.

\subsection{Hybridization}
\label{sec:hybridizationandMr}

\subsubsection{General expression}

The derivation of the hybridization matrix element
proceeds along the same lines. 
As in the previous subsection~\ref{sec:impocc} we find
\begin{equation}
M(\epsilon(k))\equiv \sqrt{L}
\langle \hat{c}_{k,\sigma}^+ \hat{d}_{\sigma}^{\vphantom{+}}\rangle
=\sqrt{L}  \int_{-\infty}^0 \rmd \omega D_{d,k}(\omega) \; .
\end{equation}
Using eqs.~(\ref{eq:defDOSAB}), (\ref{eq:EOMimpurityGF3}), 
and~(\ref{eq:EOMimpurityGF2}) we find
\begin{equation}
D_{d,k}(\omega) = -\frac{1}{\pi} \frac{V}{\sqrt{L}}{\rm Im}
\left[ \frac{1}{(\omega-\epsilon(k)+{\rm i}\eta)
(\omega-\Delta^{\rm ret}(\omega))} \right]
\end{equation}
because $\tilde{G}_{k,d;\sigma}^{\rm ret}(\omega)
=\tilde{G}_{d,k;\sigma}^{\rm ret}(\omega)$
for real hybridizations~$V_k$.
Particle-hole symmetry gives $M(\epsilon)=M^*(-\epsilon)=M(-\epsilon)$
because $M(\epsilon)$ is real.

For the hybridization matrix element in position space we thus find
\begin{eqnarray}
M_r&=&\int_0^{\pi}\frac{\rmd k}{\pi} \cos(kr) M(\epsilon(k)) \nonumber \\
&=& (1+(-1)^r) \int_{0}^{\pi/2} \frac{\rmd k}{\pi} \cos(kr) M(\epsilon(k)) 
\end{eqnarray}
so that $M_r$ is zero on for odd distances, $r=2m+1$.
For $\omega \leq 0$ we use
\begin{eqnarray}
Q_r(\omega) &=& \int_{0}^{\pi}\frac{\rmd k}{\pi} 
\frac{\cos(k r)}{\omega-\epsilon(k)+\rmi\eta} \nonumber \\
&=& (-\rmi)\rmi^r\int_0^{\infty} \rmd t e^{\rmi \omega t}J_r(t/2) \nonumber\\
&=& -\frac{2}{\sqrt{4\omega^2-1}}\left(\sqrt{4\omega^2-1}-2\omega\right)^{-r}
\nonumber \\
&& \hphantom{-\frac{(2\rmi) \rmi^r }{\cos(p)} }
\hbox{for $\omega <-1/2$}\; , \nonumber\\
&=&-\frac{(2\rmi) \rmi^r }{\cos(p)} 
\left[\cos(p r)+\rmi \sin(p r)\right]\nonumber \\
&&\hphantom{-\frac{(2\rmi) \rmi^r }{\cos(p)} }
\hbox{for $-1/2<\omega=\sin(p)/2 \leq 0$}
\label{eq:Qromega}
\end{eqnarray}
to find two contributions for $M_r$.
First, the pole at  energy $\omega=-v_+/2<-1/2$ in $D_{k,d}(\omega)$ gives
the bound-state contribution
\begin{eqnarray}
M_r^{\rm b}&=& V Z(V) Q_r(-v_+/2)\nonumber \\
&=&-\frac{2VZ(V)}{\sqrt{v_+^2-1}}
\left(\sqrt{v_+^2-1}+v_+\right)^{-r} \; .
\end{eqnarray}
The bound-state contribution to $M_r$ 
is of the order $V^3$ for small~$V$ 
and becomes exponentially small for $r\gg 1/(4V^2)$.
Second, the band contribution can be cast into the form
\begin{eqnarray}
M_{2m}^{\rm band}&=& -\frac{2V(-1)^m}{\pi}
\int_0^{\pi} \rmd u \frac{\cos(u/2)}{\sin^2u+64 V^4}\nonumber \\
&&\hphantom{-\frac{2V(-1)^m}{\pi}}
\times [ \sin(u) \cos(mu)+8V^2\sin(mu)]\; . \nonumber \\
\end{eqnarray}
Its limiting behavior is discussed in the main text.

\subsubsection{Matrix element in momentum space}

It is instructive to study the matrix element $M(\epsilon)$ in more detail.
Since we are interested in the small-$V$ limit, we ignore the bound-state contribution
and focus on the band contribution for $\epsilon>0$ so 
that $1/(\omega-\epsilon)$ remains finite,
\begin{equation}
H(\epsilon)=\int_{-1/2}^0\rmd \omega \frac{1}{\omega-\epsilon}
\frac{\rho_0(\omega)V^2}{\omega^2+(\pi V^2\rho_0(\omega))^2} \; .
\end{equation}
Note, however, that the integrand develops a singularity 
for $\epsilon\to 0^+$. We treat the singularity $1/(\omega-\epsilon)$ explicitly 
and write
\begin{eqnarray}
H(\epsilon) &=&H_1(\epsilon)  + \int_{-1/2}^0 
\frac{\rmd \omega}{\omega-\epsilon} \left(f_V(\omega)-f_V(\epsilon) \right)
\; , \nonumber \\
H_1(\epsilon) &=& 
f_V(\epsilon) \int_{-1/2}^0  \frac{\rmd \omega}{\omega-\epsilon} =
f_V(\epsilon) \ln\left|\frac{\epsilon}{\epsilon+1/2}\right| 
\label{eq:neededforB2}
\end{eqnarray}
with 
\begin{equation}
f_V(\epsilon)=\frac{2V^2}{\pi} 
\frac{\sqrt{1-4 \epsilon^2}}{4V^4+\epsilon^2(1-4\epsilon^2)} \; .
\end{equation}
We define
\begin{equation}
g_V(\epsilon)=\frac{2V^2}{\pi} 
\frac{1-4 \epsilon^2}{4V^4+\epsilon^2(1-4\epsilon^2)}
\end{equation}
and write
\begin{eqnarray}
H(\epsilon) &=&H_1(\epsilon)  + H_2(\epsilon)
+H_3(\epsilon) \nonumber \; , \\
H_2(\epsilon) &=&  \int_{-1/2}^0 \frac{\rmd \omega}{\omega-\epsilon} 
(g_V(\omega)-g_V(\epsilon)) \; , \label{eq:defH2}\\
H_3(\epsilon)&=& 
\int_{-1/2}^0 \frac{\rmd \omega}{\omega-\epsilon} 
\left(f_V(\omega)-g_V(\omega)-
f_V(\epsilon) +g_V(\epsilon)\right)
\; . \nonumber \\
\label{eq:defH3}
\end{eqnarray}
The term $H_2(\epsilon)$ can be cast into the form
\begin{eqnarray}
H_2(\epsilon)&=&- \frac{2V^2}{\pi} \frac{1}{4V^4+\epsilon^2(1-4\epsilon^2)}
\nonumber \\
&&\times
\int_{-1/2}^0 \rmd \omega (\epsilon+\omega) 
\frac{16 V^4+(1-4\omega^2)(1-4\epsilon^2)
}{4V^4+\omega^2(1-4\omega^2)} \; . \nonumber \\
\label{eq:H2aagain}
\end{eqnarray}
We thus have
\begin{eqnarray}
H_2(\epsilon)&=&
- \frac{2V^2}{\pi} 
\frac{1-4\epsilon^2}{4V^4+\epsilon^2(1-4\epsilon^2)}
\left[\epsilon\Lambda_1(V) +\Lambda_2(V)\right]
\nonumber \\
&& - \frac{2V^2}{\pi} \frac{1}{4V^4+\epsilon^2(1-4\epsilon^2)}
 \left[\epsilon\Lambda_3(V) +\Lambda_4(V)\right] \; .\nonumber \\
\label{eq:H2aagainbis}
\end{eqnarray}
{\sc Mathematica}~\cite{Mathematica10}  
provides closed formulae for $\Lambda_i(V)$,
\begin{eqnarray}
\Lambda_1(V)&=& \frac{v_+^3\arctan\left(1/v_-\right)}{2V^2\sqrt{1+64V^4}}
-\frac{v_-^3\arctanh\left(1/v_+\right)}{2V^2\sqrt{1+64V^4}}\nonumber\; , \\[6pt]
\Lambda_2(V)&=& \frac{\ln\left[(1 + 32 V^4 - \sqrt{1 + 64 V^4})/(32 V^4)\right]}{
2\sqrt{1 + 64 V^4}}\; ,\nonumber \\[6pt]
\Lambda_3(V)&=& \frac{8V^2\left(v_+\arctan\left(1/v_-\right)
+v_-\arctanh\left(1/v_+\right)\right)}{\sqrt{1+64V^4}}
\nonumber\; , \\[6pt]
\Lambda_4(V)&=& 
\frac{16V^4\ln\left[(1 + 32 V^4 - \sqrt{1 + 64 V^4})/(32 V^4)\right]}{
\sqrt{1 + 64 V^4}}\; .\nonumber \\
\end{eqnarray}
Lastly, we set $V=0$ in the integrand
of $H_3(\epsilon)$ after taking out the factor 
$\epsilon^2(1-4\epsilon^2)/(4V^4+\epsilon^2(1-4\epsilon^2))$. 
Note that the integrand in eq.~(\ref{eq:defH3})
is well behaved for $\omega \to 0$, $\epsilon\to 0$,
and $\omega\to \epsilon$
so that corrections are indeed small, of the order $V^4\ln(V^2),V^4$,
\begin{eqnarray}
H_3(\epsilon) &\approx& 
\frac{2V^2\epsilon^2(1-4\epsilon^2)}{\pi(4V^4+\epsilon^2(1-4\epsilon^2))}\times 
\nonumber \\
&& 
 \int\limits_{-1/2}^0 \frac{\rmd \omega}{\omega-\epsilon}
\Bigl[\frac{(1-\sqrt{1-4\omega^2})}{\omega^2\sqrt{1-4\omega^2}}-
\frac{(1-\sqrt{1-4\epsilon^2})}{\epsilon^2\sqrt{1-4\epsilon^2}}
\Bigr]\,.\nonumber \\
\end{eqnarray}
The integral can be done using {\sc Mathematica}~\cite{Mathematica10}. We find
\begin{eqnarray}
H_3(\epsilon) &\approx& 
\frac{2V^2\sqrt{1-4\epsilon^2}}{\pi(4V^4+\epsilon^2(1-4\epsilon^2))}
\times \nonumber \\
&&\left[(\ln(2)-2\epsilon)\sqrt{1-4\epsilon^2} +
\ln\left(\frac{1+2\epsilon}{1+\sqrt{1-4\epsilon^2}}\right)\right] \; .
\nonumber \\
\end{eqnarray}
The analytical approximation to $H(\epsilon)$ is thus given by
\begin{equation}
H(\epsilon)=H_1(\epsilon)+H_2(\epsilon)+H_3(\epsilon)
\end{equation}
with corrections of the order $V^4\ln(V^2)/(4V^4+\epsilon^2(1-4\epsilon^2))$.

We may perform the small-$V$ expansion for $H(\epsilon)$.
To leading order we have
\begin{eqnarray}
\Lambda_1(V\ll1) & =& \frac{\pi}{4V^2} -2 -2\pi V^2 +{\cal O}(V^4\ln(V^2)) \; ,
\; \nonumber \\
\Lambda_2(V\ll1) &= & \ln(4V^2)+{\cal O}(V^4\ln(V^2)) \; ,\nonumber \\
\Lambda_3(V\ll1) &= & 4\pi V^2+{\cal O}(V^4\ln(V^2)) \; ,\nonumber \\
\Lambda_4(V\ll1) &= & {\cal O}(V^4\ln(V^2)) \;.
\end{eqnarray}
Consequently, the leading contribution to $H_2(\epsilon)$ is 
given by
\begin{eqnarray}
H_2(\epsilon)&=& - \frac{1}{2} \frac{\epsilon(1-4\epsilon^2)}{
4V^4+\epsilon^2(1-4\epsilon^2)} 
+ \frac{4V^2\epsilon(1-4\epsilon^2)}{
\pi (4V^4+\epsilon^2(1-4\epsilon^2))} \nonumber \\
&&  -\frac{2V^2\ln(4V^2)(1-4\epsilon^2)}{
\pi (4V^4+\epsilon^2(1-4\epsilon^2))} 
+{\cal O}(V^4) \; .
\label{eq:H2aagainbisagain}
\end{eqnarray}
Therefore, the small-$V$ expansion for all $-1/2\leq \epsilon\leq 1/2$ becomes
\begin{eqnarray}
H(\epsilon) &\approx &
-\frac{1}{2} \,\frac{|\epsilon|(1-4\epsilon^2)}{4V^4+\epsilon^2(1-4\epsilon^2)} 
\nonumber \\
&& +\frac{2V^2}{\pi} 
\frac{\sqrt{1-4\epsilon^2}}{4V^4+\epsilon^2(1-4\epsilon^2)} 
\nonumber \\
&& \times
\left[ -\ln(2V^2)\sqrt{1-4\epsilon^2}
+\ln \left| \frac{2\epsilon}{1+\sqrt{1-4\epsilon^2}}
\right| \right] \; .\nonumber \\
\label{eq:onedHapprox}
\end{eqnarray}
Corrections are of the order $V^4\ln(V^2)/(4V^4+\epsilon^2(1-4\epsilon^2))$.
As required by particle-hole symmetry, $H(-\epsilon)=H(\epsilon)$.

For small-$V$ and low energies, $\epsilon\to 0$, the result can be cast into
\begin{eqnarray}
H(\epsilon\to 0^+,V\to 0) &\approx& H^{\rm low}(\epsilon)\; ,\nonumber \\
H^{\rm low}(\epsilon)&=& -\frac{1}{2} \frac{\epsilon}{4V^4+\epsilon^2}
+\frac{2V^2}{\pi} \frac{\ln[\epsilon/(2V^2)]}{4V^4+\epsilon^2} \; .\nonumber \\
\label{eq:Hlowonedim}
\end{eqnarray}
This form is sufficient to determine the long-range behavior
of the spin correlation function for small~$V$.

\section{Further results for the non-interacting SIAM}

In this section we collect some results for the semi-elliptic 
density of states, consider the 
limit of small hybridizations for a general density of states,
and provide results for non-interacting electrons in the limit
of high dimensions. 

\subsection{Semi-elliptic density of states}

For the semi-elliptic density of states,
\begin{equation}
\rho_0(\epsilon)=\frac{4}{\pi}\sqrt{1-4\epsilon^2} \quad \hbox{for}\;
|\epsilon|\leq 1/2 \; ,
\label{eq:DOSSE}
\end{equation}
we find
\begin{eqnarray}
\Lambda_0(\omega)&=&8\omega  \quad \hbox{for}\;
|\omega|\leq 1/2 \; , \nonumber \\
&=& 8\omega -4\sgn(\omega)\sqrt{4\omega^2-1} 
\quad \hbox{for}\;  |\omega|> 1/2 \; .
\end{eqnarray}
When we choose $0\leq V<1/4$, there is no (anti-)bound state outside the band.

\subsubsection{Ground-state energy}

The ground-state energy can be calculated analytically using 
{\sc Mathematica}~\cite{Mathematica10} ,
\begin{eqnarray}
e_0(V)&=&\frac{2}{\pi}\int_{-1/2}^0{\rm d}\omega 
\cot^{-1}\left( 
\frac{\omega-V^2\Lambda_0(\omega)}{\pi\rho_0(\omega)V^2}
\right) \label{eq:ezeroSIAMtdlSE}\\
&=& -\frac{4V^2}{\pi\sqrt{1-16V^2}}\ln\left[ 
\frac{1-8V^2+\sqrt{1-16V^2}}{1-8V^2-\sqrt{1-16V^2}}\right]\; .\nonumber 
\end{eqnarray}
The result agrees with the small-$V$ expression derived 
as eqs.~(E.1) and~(E.5) in Ref.~[\onlinecite{Linneweberetal2017}].
Using {\sc Mathematica}~\cite{Mathematica10} 
we find for the semi-elliptic density of states
using $d_0=\rho_0(0)=4/\pi$ and $\ln(e)=1$
\begin{equation}
C=\frac{e}{2} \exp\left( \int_{-1/2}^0{\rm d} \epsilon 
\frac{d_0-\rho_0(\epsilon)}{d_0\epsilon}
\right) = 1
\end{equation}
so that 
\begin{equation}
e_0^{\rm small}(V)=2V^2d_0 \ln\left(\frac{\pi V^2d_0}{C}\right)
= \frac{8V^2}{\pi}\ln(4V^2) \; ,
\end{equation}
with corrections of the order $V^4\ln(V^2)$.

\subsubsection{Hybridization matrix element}

For the semi-elliptic density of states, 
the hybridization function can be calculated
analytically. For $0<V<1/4$ and $0<\epsilon<1/2$ we find
\begin{eqnarray}
H(\epsilon)&=&  \int_{-1/2}^0 {\rm d}\omega
\frac{1}{\omega-\epsilon}
\frac{\rho_0(\omega)V^2}{[\omega(1-8V^2)]^2+(\pi \rho_0(\omega)V^2)^2}
\nonumber \\
&=& \frac{4V^2}{\pi} 
\int_{-1/2}^0 {\rm d}\omega \frac{1}{\omega-\epsilon}
\frac{\sqrt{1-4\omega^2}}{16V^4+\omega^2(1-16V^2)}\nonumber \\
&=&  -\frac{1}{2} \frac{\epsilon(1-8V^2)}{16V^4+(1-16V^2)\epsilon^2}
\nonumber \\
&& + \frac{4V^2(1-8V^2)}{\pi\sqrt{1-16V^2}}
\frac{\arccsch(8V^2/\sqrt{1-16V^2})}{16V^4+(1-16V^2)\epsilon^2}
\nonumber \\
&& -\frac{4V^2\sqrt{1-4\epsilon^2}}{\pi(16V^4+(1-16V^2)\epsilon^2)}
\ln\biggl(
\frac{1}{2\epsilon}+\sqrt{\frac{1}{4\epsilon^2}-1}\biggr) ,\nonumber \\
\end{eqnarray}
where $\arccsch(x)$ is the inverse hyperbolic cosecant function.

The small-$V$ expansion reads
\begin{eqnarray}
H(\epsilon)&\approx & 
-\frac{1}{2} \frac{\epsilon}{16V^4+(1-16V^2)\epsilon^2} \nonumber \\
&& +\frac{4V^2}{\pi(16V^4+(1-16V^2)\epsilon^2)} \nonumber \\
&&  \times \biggl[ \pi\epsilon+\sqrt{1-4\epsilon^2}
\ln\left(\frac{2\epsilon}{1+\sqrt{1-4\epsilon^2}}\right)\nonumber \\
&& \hphantom{\times\biggl[ } -\ln(4V^2)-8V^2\biggr] \;. 
\label{eq:SEHapprox}
\end{eqnarray}
Corrections to this expression are of the order
$V^6\ln(V^2)/$ $(16V^4+(1-16V^2)\epsilon^2)$.
The low-energy limit of this expression becomes
\begin{eqnarray}
H(\epsilon\to 0^+,V\to 0) &\approx& H^{\rm low}(\epsilon)\; ,\nonumber \\
H^{\rm low}(\epsilon)&=& -\frac{1}{2} \frac{\epsilon}{16V^4+\epsilon^2}
+\frac{4V^2}{\pi} \frac{\ln[\epsilon/(4V^2)]}{16V^4+\epsilon^2} 
\; ,\nonumber \\
\label{eq:HlowSE}
\end{eqnarray}
compare eq.~(\ref{eq:Hlowonedim}) for the one-dimensional density of states.

\subsection{Limit of small hybridizations 
for a general density of states}

Here, we collect results for any density of states in the in finite 
band-width limit
where we approximate $\rho(\epsilon)\approx d_0$ and $\Lambda_0(\epsilon)
\approx 0$.
Thus, $\Delta^{\rm ret}(\omega)\approx -\rmi \pi d_0V^2$.

\subsubsection{Ground-state energy}

For the ground-state energy we obtain 
\begin{eqnarray}
e_0(V)&=&\frac{2}{\pi}\int_{-1/2}^0{\rm d}\omega 
\cot^{-1}\left( 
\frac{\omega-V^2\Lambda_0(\omega)}{\pi\rho_0(\omega)V^2}
\right) \nonumber \\
& \approx & 2d_0V^2\ln\left[ 
\frac{2\pi d_0V^2}{C}\right]\; , 
\label{eq:ezeroSIAMtdlgeneral}
\end{eqnarray}
where $C$ is used to fit the unknown contribution to order $V^2$. In general,
\begin{equation}
C=\frac{e}{2} \exp\left( \int_{-1/2}^0{\rm d} \epsilon 
\frac{d_0-\rho_0(\epsilon)}{d_0\epsilon}
\right)
\end{equation}
for a given density of states $\rho_0(\epsilon)$.~\cite{Linneweberetal2017}

\subsubsection{Hybridization matrix element}

For $H(\epsilon)$ we obtain
\begin{eqnarray}
H(\epsilon)&\approx &  \int_{-1/2}^0 {\rm d}\omega
\frac{1}{\omega-\epsilon}
\frac{d_0V^2}{\omega^2+(\pi d_0V^2)^2}\\
&=& \frac{-2\epsilon\,\arccot(2\gamma)
+\gamma\ln[(1+4\gamma^2)\epsilon^2/(\gamma^2(1+2\epsilon)^2)]}{
2\pi (\gamma^2+\epsilon^2)}\nonumber 
\end{eqnarray}
with $\gamma=\pi d_0V^2$ and $\arccot(x)$ denotes
the inverse cotangent function.
The low-energy limit for small~$V$ becomes
\begin{eqnarray}
H(\epsilon\to 0^+,V\to 0) &\approx& H^{\rm low}(\epsilon)\; ,\nonumber \\
H^{\rm low}(\epsilon)&=& -\frac{1}{2} \frac{\epsilon}{(\pi d_0V^2)^2+\epsilon^2}
\nonumber \\
&& 
+d_0V^2\frac{\ln[\epsilon/(\pi d_0V^2)]}{(\pi d_0V^2)^2+\epsilon^2}\; .
\label{eq:Hlowgeneral}
\end{eqnarray}
The results includes 
eq.~(\ref{eq:Hlowonedim}) for the one-dimensional density of states 
($d_0=2/\pi$)
and eq.~(\ref{eq:HlowSE}) for the semi-elliptic density of states 
($d_0=4/\pi$).
Therefore, eq.~(\ref{eq:Hlowgeneral}) provides the generic low-energy
limit of the function $H(\epsilon)$.

\subsection{Limit of high dimensions}

We address the limit of high dimensions. To this end
we focus on a $d$-dimensional hyper-cubic lattice with nearest-neighbor
dispersion relation [$\veck=(k_1,k_2,\ldots,k_d)]$
\begin{equation}
\epsilon(\veck) = -\frac{2}{\sqrt{2d}} \sum_{l=1}^d \cos(k_l) \; .
\label{eq:dispersionsc}
\end{equation} 
We restrict ourselves to the case of half band-filling and
particle-hole symmetry.

\subsubsection{Bulk spin correlation function}

The spin-spin correlation function for free electrons is 
obtained from 
\begin{eqnarray}
C^{SS}(r) &=& \frac{1}{L} \sum_{\vecl} 
\langle \hat{S}^z(\vecl+\vecr) \hat{S}^z(\vecl) \rangle
\nonumber \\
&=& -\frac{1}{2} \frac{1}{L} \sum_{\vecl} 
\left|\langle \hat{c}_{\vecl+\vecr,\sigma}^+ 
\hat{c}_{\vecl,\sigma}^{\vphantom{+}} \rangle\right|^2 
\nonumber \\
&=& -\frac{1}{2} \left|P_{\sigma}^0(\vecr)\right|^2 
\label{eq:defCSS}
\end{eqnarray}
with the single-particle density matrix
\begin{eqnarray}
P_{\sigma}^0(\vecr) &=& 
\langle \hat{c}_{\vecl+\vecr,\sigma}^+ 
\hat{c}_{\vecl,\sigma}^{\vphantom{+}} \rangle
\nonumber \\
&=& \frac{1}{L} \sum_{\veck} e^{-\rmi \veck\vecr}
\langle \hat{c}_{\veck,\sigma}^+ 
\hat{c}_{\veck,\sigma}^{\vphantom{+}} \rangle \nonumber \\
&=& \int_{-\infty}^0 \rmd \epsilon \!\int_{-\infty}^{\infty} 
\frac{\rmd \eta}{2\pi} e^{\rmi \eta \epsilon}
\prod_{l=1}^d \int_{-\pi}^{\pi}
\frac{\rmd k}{2\pi} e^{2\rmi\eta\cos(k)/\sqrt{2d}-\rmi k r_l}\nonumber\\
&=& \int_{-\infty}^0 \rmd \epsilon \!\int_{-\infty}^{\infty} 
\frac{\rmd \eta}{2\pi} e^{\rmi \eta \epsilon}
\prod_{l=1}^d \rmi^{r_l} J_{r_l}(2\eta/\sqrt{2d}) \; ,
\end{eqnarray}
where $J_n(x)=(-1)^nJ_n(x)$ is the Bessel function of integer order~$n$.
We introduce the distance to the $R$th neighbor shell,
\begin{equation}
R=\sum_{l=1}^d |r_l|
\end{equation}
and use $d\gg 1$ to approximate
\begin{equation}
\prod_{l=1}^d  
J_{r_l}(2\eta/\sqrt{2d})\approx
e^{-\eta^2/2} \left(\frac{\eta}{\sqrt{2d}}\right)^R
\prod_{l=1}^d  (r_l!)^{-1}
 \; ,
\end{equation}
where we used $J_0(x\ll 1) \approx 1-x^2/4$ and $J_{n\geq 1}(x\ll 1)
\approx (x/2)^n/n!$.
Therefore,
\begin{equation}
P_{\sigma}^0(\vecr) \approx  
\left(\frac{\rmi }{\sqrt{2d}}\right)^R \prod_{l} \frac{1}{r_l!}
\int_{-\infty}^0 \rmd \epsilon \!\int_{-\infty}^{\infty} 
\frac{\rmd \eta}{2\pi} e^{\rmi \eta \epsilon} \eta^R e^{-\eta^2/2} \, .
\end{equation}
When $R=2m$ is even, we find
$P_{\sigma}^0(\vecr)= \delta_{m,0}/2$, as also follows from
particle-hole symmetry.

When $R=2m+1$ ($m\geq 0$) we note that the dominant contribution in the
$R$th neighbor shell comes from those vectors where $r_l=\pm 1$.
Their number is given by
\begin{equation}
  N(d,R)=2^R \left( \begin{array}{@{}c@{}}
    d \\ R
    \end{array}
    \right)
\approx (2d)^R \frac{1}{R!}
\label{eq:NdR}
\end{equation}
for $d\gg R$ where we used Stirling's formula,
$\ln(n!)\approx n\ln n-n$.

We write
\begin{eqnarray}
P_{\sigma}^0(\vecr) &\approx&  P_{\sigma}^0(R) \; ,\nonumber \\
P_{\sigma}^0(R)&=& 
\left(\frac{1}{\sqrt{2d}}\right)^R 
\int_{-\infty}^0 \frac{\rmd \epsilon }{\sqrt{2\pi}}
\left( \frac{\partial}{\partial\epsilon}\right)^R
\left[e^{-\epsilon^2/2}\right]
\nonumber \\
&=& 
\left(\frac{1}{\sqrt{2d}}\right)^R 
\frac{(-1)^{R-1}}{\sqrt{2\pi}}{\rm He}_{R-1}(0) \nonumber \\
&=& \left(\frac{1}{\sqrt{2d}}\right)^R 
\frac{1}{\sqrt{2\pi}} \left(-\frac{1}{2}\right)^{(R-1)/2} 
\frac{(R-1)!}{((R-1)/2)!}
\; ,\nonumber \\
\end{eqnarray}
where ${\rm He}(x)$ is a Hermite polynomial. The 
contribution of the $R$th  neighbor shell
to the spin-spin correlation function ($R=2m+1$) is
\begin{eqnarray}
C^{SS}(R)&=& -\frac{1}{2}\sum_{|\vecr|=R} \left|P_{\sigma}^0(\vecr)\right|^2
\nonumber \\
& \approx & -\frac{1}{2} \frac{1}{(2m+1)!}
\frac{1}{2\pi} \left(\frac{1}{2}\right)^{2m} 
\left(\frac{(2m)!}{m!}\right)^2 \nonumber \\
&=& -\frac{1}{4\pi} \frac{1}{2m+1} \left(\frac{1}{2}\right)^{2m} 
\left( \begin{array}{@{}c@{}}
    2m \\ m
    \end{array}
    \right)
\; .
\end{eqnarray}
This approximation becomes exact in the limit $d\to \infty$.

For large $R$ we use Stirling's formula, 
$\ln(n!)\approx n\ln n-n +\ln(2\pi n)/2$, to
find the asymptotic behavior ($R$ odd)
\begin{equation}
C^{SS}(R\gg 1)\approx 
-\left(\frac{1}{2\pi R}\right)^{3/2} \; .
\end{equation}

\subsubsection{Spin correlation function}

Along the same lines we can express the matrix element
for the spin correlation function between the impurity and the
host electrons at distance~$R=2m$ as
\begin{equation}
M_{R} =
\frac{2V}{\sqrt{2\pi}} 2^{-R/2} 
\left( \frac{-1}{\sqrt{2d}}\right)^R
\int_0^{\infty} \!\!\rmd \epsilon H(\epsilon )
e^{-\epsilon^2/2} {\rm H}_R(\epsilon/\sqrt{2}) \, ,
\end{equation}
where ${\rm H}_R(x)$ is a Hermite polynomial.
As in one dimension, correlations to odd sites 
vanish due to particle-hole symmetry.

Averaged over all sites at distance~$R\ll d$, 
the spin correlation function becomes ($\Gamma=\pi d_0 V^2$, 
$\Gamma=\sqrt{\pi/2}V^2$ for an infinite-dimensional hyper-cubic lattice)
\begin{eqnarray}
C_{dc}^S(R) &=& -\frac{1}{2} \left(\frac{2V}{\sqrt{2\pi}}\right)^2 
\left[g(R,\Gamma)\right]^2
\; , \nonumber \\
g(R,\Gamma)&=& \frac{2^{-R/2} }{\sqrt{R!}}  
\int_0^{\infty}\rmd y \tilde{H}(y)
e^{-\Gamma^2 y^2/2} {\rm H}_R(\Gamma y/\sqrt{2}) \; ,\nonumber \\
\end{eqnarray}
where we used eq.~(\ref{eq:NdR}) for the number of sites in the $R$th neighbor shell.
Moreover, for small~$V$ we may use the low-energy limit of $H(\epsilon)$,
\begin{equation}
\tilde{H}(y) \approx -\frac{1}{2} \frac{y}{1+y^2} +\frac{1}{\pi} \frac{\ln(y)}{1+y^2}
\; .
\end{equation}

To derive an asymptotic formula, we employ the approximation
\begin{equation}
{\rm H}_{R\gg 1}(x) \approx e^{x^2/2} (-2)^{R/2}(R-1)!! \cos(\sqrt{2R}x)
\label{eq:trouble}
\end{equation}
for $R\gg 1$ and $|x| <\sqrt{2R}$. Since the integral contains
another factor $\exp(-x^2/2)$, we may safely ignore the constraint
on~$x$.
Thus,  for small hybridizations, $V\ll 1$,
we may approximate
\begin{equation}
\left|g(R,\Gamma)\right| \approx \frac{(R-1)!!}{\sqrt{R!}}
\int_0^{\infty} \rmd y \tilde{H}(y)
e^{-\Gamma^2 y^2/4} \cos(\Gamma \sqrt{R}y) \; .
\end{equation}
For $\Gamma\ll 1$ we may safely ignore the exponential term in
the definition of $g(R,\Gamma)$ because the cosine is a vastly oscillating function
for $y\gtrsim 1/\Gamma$ that effectively restricts the integration to $y\lesssim 
1/(\Gamma\sqrt{R})$.  Then, the integral can be done analytically,
\begin{equation}
g(R,\Gamma) \approx (-1)^{R/2}\frac{(R-1)!!}{\sqrt{R!}}
\left[\frac{1}{2}e^{\Gamma \sqrt{R}} \Ei(-\Gamma\sqrt{R}) \right]\;,
\end{equation}
where $\Ei(x)$ is the exponential integral.
As in one dimension, the approximation works very well for all $R\geq 2$.
Using Stirling's formula $\ln(n!)\approx n\ln n-n +\ln(2\pi n)/2$, we find
($R=2m$)
\begin{equation}
\frac{((R-1)!!)^2}{R!} = \frac{(2m)!(2m)!}{2^{2m}m!m!(2m)!}\approx 
\sqrt{\frac{2}{\pi R}}
\; .
\end{equation}
Consequently, the spin correlation function becomes
\begin{equation}
C_{dc}^S(R) \approx 
-\frac{1}{2} \left(\frac{2V}{\sqrt{2\pi}}\right)^2 
\sqrt{\frac{2}{\pi R}}
\left[\frac{1}{2}e^{\Gamma \sqrt{R}} \Ei(-\Gamma\sqrt{R})\right]^2 \; .
\end{equation}
The formula is applicable for all $R\geq 2$.

The asymptotic region is reached for $\Gamma\sqrt{R}\gg 1$, i.e.,
for $R\gg 1/\Gamma^2=2/(\pi V^4)$. Even for $V=0.1$, the asymptotic
region is starts around $R_a={\cal O}(V^4)=10^4$, 
in contrast to the one-dimensional case where
the asymptotic region starts at $R_a={\cal O}(V^2)=10^2$.
In the asymptotic regime,
\begin{equation}
C_{dc}^S(R\gg 1/\Gamma^2) \approx  
-\frac{1}{2\pi^2 \Gamma} \left(\frac{1}{R}\right)^{3/2}\; ,
\end{equation}
so that the unscreened spin is given by
\begin{equation}
{\cal S}(R\gg 1/\Gamma^2)= \frac{1}{2\pi^2 \Gamma\sqrt{R}} \; .
\end{equation}
In $d\to \infty$ dimensions, the unscreened spin decays proportional to $1/\sqrt{R}$ 
whereas, in $d=1$ dimension, it decays proportional to $1/R$.

\section{Interacting SIAM}

In this section, we address the interacting single-impurity Anderson model.
First, we calculate the second-order coefficient in~$U$ for the 
ground-state energy. Next, we derive the ground-state energy for
the magnetic Hartree-Fock solution.

\subsection{Second-order coefficient for the ground-state energy}

\subsubsection{Br\"uckner-Goldstone perturbation theory}

Using Br\"uckner-Goldstone perturbation
theory,~\cite{Goldstone267,GebhardRuhl2006}
the second-order coefficient to the ground-state energy reads
\begin{equation}
-\frac{e^{(2)}(V)}{\pi \Gamma } 
=  \langle \Phi_0 | \hat{H}_{\rm int} 
\frac{1}{E_0(V)-\hat{H}_0}  \hat{H}_{\rm int} |\Phi_0\rangle \; ,
\label{eq:BGdiagam}
\end{equation}
where we used the definition of $e^{(2)}(V)$ in the main text
and $\hat{H}_{\rm int}= (\hat{n}_{d,\uparrow}-1/2)
(\hat{n}_{d,\downarrow}-1/2)$. Since we subtracted the
Hartree terms in $\hat{H}_{\rm int}$, we only sum over connected
diagrams in eq.~(\ref{eq:BGdiagam}).
With $\eta=0^+$ we can write
\begin{eqnarray}
-\frac{e^{(2)}(V)}{\pi \Gamma } 
&=&  
{\rm Re}\biggl[(-\rmi)  
\int_0^{\infty} 
\rmd t e^{-\eta t} \nonumber \\
&& \hphantom{{\rm Re}\biggl[(-\rmi)\int_0^{\infty}   }
\langle \Phi_0 | \hat{H}_{\rm int} e^{\rmi(E_0(V)-\hat{H}_0)t}  
\hat{H}_{\rm int} |\Phi_0\rangle \biggr]\nonumber \\
&=&  \int_0^{\infty} \rmd t e^{-\eta t}
{\rm Im}\left[f(t)^2\right] \; .
\label{eq:BGdiagamagain}
\end{eqnarray}
Here, we introduced
\begin{eqnarray}
f(t) &=& \langle  (\hat{n}_{d,\sigma}-1/2) 
e^{\rmi(E_0(V)-\hat{H}_0)t}  
(\hat{n}_{d,\sigma}-1/2) \rangle \nonumber \\
&=& 
\langle  \hat{d}_{\sigma}^+(t) \hat{d}_{\sigma}^{\vphantom{+}}
\rangle
\langle  \hat{d}_{\sigma}^{\vphantom{+}}(t)
\hat{d}_{\sigma}^+\rangle\; ,
\label{eq:ffunctiondef}
\end{eqnarray}
where we used Wick's theorem.
Particle-hole symmetry shows that the two factors are identical.
Using the causal Green function for the $d$-electrons,
we can write
\begin{equation}
f(t) =\Theta(t) \left[\rmi G_{d,d}^{\rm c}(t)\right]^2
\end{equation}
with 
\begin{equation}
G_{d,d}^{\rm c}(t\ge 0)=\int_{-\infty}^{\infty} \frac{\rmd \omega}{2\pi}
e^{-\rmi \omega t}
\tilde{G}_{d,d}^{\rm c}(\omega)  \; .
\label{eq:calcGtcausal}
\end{equation}
We insert the spectral representation~(\ref{eq:lehmann})
and extend the integral over $\omega$ into a contour integral in the lower complex
half-plane because we have $t>0$. The $\omega$-integral 
gives $(-\rmi) \Theta(\omega')\exp(-\rmi \omega' t)$ because
there is a pole in the lower complex half-plane only if $\omega'>0$.
Thus, 
\begin{equation}
\rmi G_{d,d}^{\rm c}(t^+)=\int_0^{\infty} \rmd \omega'
e^{-\rmi \omega' t} D_{d,d}(\omega')  \; .
\label{eq:calcGtcausalagain}
\end{equation}
We insert this result into eq.~(\ref{eq:BGdiagamagain}) and perform
the integration to find 
\begin{eqnarray}
e^{(2)}(V)
&=&  
\pi \Gamma \int_0^{\infty}\rmd \omega_1 D_{d,d}(\omega_1)
\ldots \int_0^{\infty}\rmd \omega_4 D_{d,d}(\omega_4) \nonumber \\
&& \hphantom{\pi \Gamma \int_0^{\infty}}
\times \frac{1}{\omega_1+\omega_2+\omega_3+\omega_4}
\nonumber \\
&=& \pi \int_0^{\infty} \rmd \mu [F(V,\mu)]^4 
\; ,\nonumber \\
F(V,\mu) &=& \int_0^{\infty}\rmd \omega D_{d,d}(\omega) e^{-\mu \omega/\Gamma}
\; .
\label{eq:thatworks}
\end{eqnarray}
This form is numerically more advantageous than inserting 
eq.~(\ref{eq:calcGtcausalagain}) into eq.~(\ref{eq:BGdiagamagain})
and taking the imaginary part. Using the Lehmann representation and
the (symmetric) density of states 
for the $d$-electrons, one can obtain eq.~(\ref{eq:thatworks}) from
eq.~(\ref{eq:BGdiagam}) directly.

For the one-dimensional density of states we have 
\begin{eqnarray}
F(V,\mu) &=&Z(V) e^{-v_+ \mu/(2\Gamma)} \nonumber \\
&& +\int_{0}^{1/(2\Gamma)}
\frac{\rmd u}{\pi} e^{-u \mu} 
\frac{\sqrt{1-4 u^2 \Gamma^2}}{1+u^2(1-4 u^2 \Gamma^2)} \; .\nonumber \\
\end{eqnarray}
A numerical integration in eq.~(\ref{eq:thatworks})
gives the coefficients 
$e^{(2)}(V=0.05)=0.0369271$,
$e^{(2)}(V=0.1)=0.0374447$, and 
$e^{(2)}(V=0.2)=0.0406307$.

\subsubsection{Limit of small hybridizations}

In the limit of small hybridizations, we let $\Gamma\to 0$
and ignore the pole contribution of order $V^4$.
With $\Gamma=\pi d_0 V^2$ we find that $F(V,\mu)$ becomes independent
of $V$, $F(V,\mu)\equiv F(\mu)$, with
\begin{equation}
F(\mu)= \frac{1}{\pi}
\left(\Ci(\mu) \sin(\mu) + \cos(\mu)
 \left(\frac{\pi}{2}-  \Si(\mu)\right)\right)
\, ,
\end{equation}
where $\Ci(x)$ [$\Si(x)$] is the cosine [sine] integral.

When we insert this result in eq.~(\ref{eq:thatworks}),
we obtain the numerical value $\tilde{e}^{(2)}=0.0368608$, in perfect 
agreement with Yamada's
analytical result,~\cite{Yamada1975} $\tilde{e}^{(2)}=1/4 - 7\zeta(3)/(4\pi^2)$,
and in very good agreement with the numerically obtained value for $V\leq 0.2$.
The deviations are about ten percent at $V=0.2$, about one percent at $V=0.1$,
and two per mill at $V=0.05$.

\subsection{Magnetic Hartree-Fock solution}

In the magnetic Hartree-Fock approach, the Hubbard interaction on
the impurity is replaced by
\begin{eqnarray}
H_{\rm int}^{\rm HF} 
&=& -Um \left(\hat{n}_{d,\uparrow}-1/2\right)
+Um \left(\hat{n}_{d,\downarrow}-1/2\right)+Um^2 \; , \nonumber \\
\label{eq:HFint} \\
m&=&\frac{\langle \hat{n}_{d,\uparrow}-\hat{n}_{d,\downarrow}\rangle}{2} \; .
\label{eq:selfconsHF}
\end{eqnarray}
Here, $m$ is the local polarization of the impurity.
To be definite, we restrict ourselves to $0\leq m\leq  1/2$.
Apparently, the Hartree-Fock Hamiltonian $\hat{H}^{\rm HF}=
\hat{H}_0 + H_{\rm int}^{\rm HF}$ corresponds to a non-interacting
SIAM with spin-dependent local potentials,
$E_{d,\uparrow}=Um$ and $E_{d,\downarrow}=-Um$.

\begin{figure}[t]
\includegraphics[width=8.5cm]{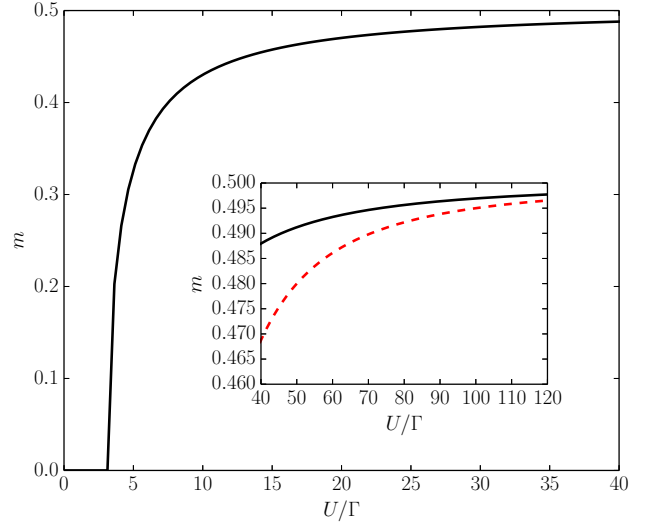}
\caption{Magnetization in Hartree-Fock theory as a function of
$U/\Gamma$ for $V=0.1$ ($\Gamma=2 V^2$ for $d_0=2/\pi$).
The onset is at $U_{\rm c}\approx 0.06283=3.131\Gamma$.
Inset: comparison of the analytic strong-coupling result~(\ref{mHFlargeU}) 
(dashed line) with the result of the numerical root for 
large~$U/\Gamma$.\label{fig:HFmagnetization}}
\end{figure}

\subsubsection{Self-consistency equation}

When we repeat the steps in Sect.~\ref{sec:impocc} we arrive at
\begin{eqnarray}
  \langle \hat{n}_{d,\sigma}\rangle &=& Z[v_{\rm b}(V,\sigma_n Um)]
  \label{eq:HFdoccupation}\\
&&+\int_{-1/2}^0 \rmd \omega
\frac{\rho_0(\omega)V^2}{(\omega+\sigma_n Um)^2+(\pi \rho_0(\omega)V^2)^2}
\nonumber 
\end{eqnarray}
with $\sigma_n=1$ ($\sigma_n=-1$) for $\sigma=\uparrow$ ($\sigma=\downarrow$).
Therefore, the self-consistency equation~(\ref{eq:selfconsHF})
becomes
\begin{eqnarray}
2m&=&Z[v_{\rm b}(V,Um)]-Z[v_{\rm b}(V,-Um)] \nonumber \\
&& +\sum_{\sigma_n=\pm 1} \int_{-1/(2\Gamma)}^0
\frac{\rmd x}{\pi} 
\frac{\sigma_n \rho_0(\Gamma x)/d_0}{
(x+\sigma_n \tilde{u})^2+ (\rho_0(\Gamma x)/d_0)^2}
\nonumber \\
\label{eq:selfconsHFfull}
\end{eqnarray}
with $\tilde{u}=Um/\Gamma$ ($\Gamma=\pi d_0 V^2$, $d_0=\rho_0(0)$).
For the one-dimensional density of states, the self-consistency equation
explicitly reads
\begin{eqnarray}
2m&=&Z[v_{\rm b}(V,Um)]-Z[v_{\rm b}(V,-Um)] \nonumber \\
&& +\sum_{\sigma_n=\pm 1} \int_{-1/(2\Gamma)}^0
\frac{\rmd x}{\pi} 
\frac{\sigma_n \sqrt{1-4\Gamma^2x^2}}{
(x+\sigma_n \tilde{u})^2(1-4\Gamma^2x^2)+1} \; .
\nonumber \\
\label{eq:selfconsHFfulloned}
\end{eqnarray}
In the limit of small hybridization and $U\ll 1$, the contribution of the
bound states and the influence of the finite bandwidth
can be ignored and we can simplify
\begin{eqnarray}
m&\approx & \sum_{\sigma_n=\pm 1} \int_{-\infty}^0
\frac{\rmd x}{2\pi} 
\frac{\sigma_n}{(x+\sigma_n \tilde{u})^2+1} 
\nonumber \\
&=& \frac{1}{\pi}\arctan(Um/\Gamma) \; , \nonumber \\
Um &\approx & \Gamma \tan(\pi m) \;.  
\label{eq:selfconsHFfullonedsmallV}
\end{eqnarray}
This shows that there is a critical interaction strength above which the magnetic
Hartree-Fock solution is energetically favorable over the paramagnetic solution.
For $V\to 0$ we find from eq.~(\ref{eq:selfconsHFfullonedsmallV}) that
$U_{\rm c}^{\rm HF,approx}(V)
=\pi \Gamma=\pi^2 d_0 V^2=2\pi V^2$. 
The numerical value for $V=0.1$, 
$U_{\rm c}^{\rm HF}(V=0.1)=0.06262$
agrees very well with this approximation,
$U_{\rm c}^{\rm HF,approx}(0.1)=0.06283$.

Eq.~(\ref{eq:selfconsHFfullonedsmallV}) also shows that
$m\to 1/2$ for $U\gg \Gamma$.
However, it incorrectly 
suggests that $m(U)=1/2- 2\Gamma/(\pi U)$ in the large-$U$ limit.
Instead, we verified numerically that
\begin{equation}
m(U) \approx \frac{1}{2} -\frac{2V^2}{U^2} + {\cal O}(1/U^3) \; .
\label{mHFlargeU}
\end{equation}
We show the result for $V=0.1$ in Fig.~\ref{fig:HFmagnetization}.

\subsubsection{Ground-state energy}

When we repeat the steps in Sect.~\ref{sec:gsenergy}, we find
\begin{eqnarray}
\Delta E_0^{\rm HF}(V,U) &=& \frac{U}{4} +Um^2 +
v_{\rm b}(V,Um)+ v_{\rm b}(V,-Um) \nonumber \\
&& + \sum_{\sigma_n=\pm 1}\int_{-1/2}^0 \frac{\rmd \omega}{\pi} 
\Cot^{-1}\left( \frac{\omega+\sigma_n Um}{\pi \rho_0(\omega)V^2}
\right)  \, , \nonumber \\
\end{eqnarray}
where we took into account the constant term~$U/4$ in the definition
of $\Delta E_0(U,V)$ and the term $Um^2$ in eq.~(\ref{eq:HFint}). Note that
$m\equiv m(U)$ is determined from the solution of the
self-consistency equation~(\ref{eq:selfconsHFfulloned}) for given~$U$.

In the large-$U$ limit, the band contribution is negligible and
the pole at $v_{\rm b}(V,Um)=-Um-V^2/(Um)$ dominates.
Together with eq.~(\ref{mHFlargeU}) we find
\begin{equation}
\Delta E_0^{\rm HF}(V,U\gg W) = - \frac{2V^2}{U} +{\cal O}(V^3/U^2) \; .
\label{eq:HFlargeU}
\end{equation}
This result is readily understood because, in the large-$U$ limit, 
the host electrons act like a single bath site
to which the impurity is coupled. The energy of this two-site model
is readily calculated and leads to eq.~(\ref{eq:HFlargeU}) in the large-$U$ limit.

\begin{figure}[ht]
\includegraphics[width=8.5cm]{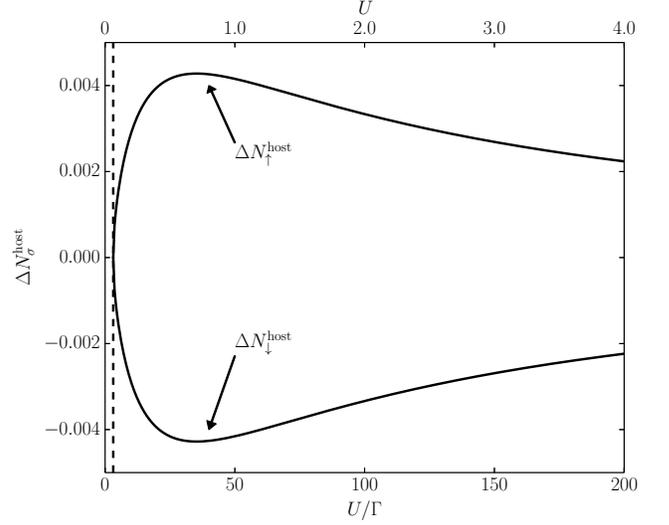}
\caption{Host electron particle number $\Delta N_{\sigma}^{\rm host}
=N_{\sigma}^{\rm host}-L/2$, eq.~(\ref{eq:hostpol})
 for $\sigma=\uparrow,\downarrow$
as a function  of the magnetization in Hartree-Fock theory for
$V=0.1$.\label{fig:Nsigmahost}}
\end{figure}

\subsubsection{Host electron polarization}

Particle-hole symmetry gives 
$\langle \hat{c}_{r,\uparrow}^+ \hat{c}_{r,\uparrow}^{\vphantom{+}}+
\hat{c}_{r,\downarrow}^+ \hat{c}_{r,\downarrow}^{\vphantom{+}}\rangle=1$ 
because $E_{d,\uparrow}=-E_{d,\downarrow}$.
However, the Hartree-Fock magnetic moment polarizes the host electrons.
In general, the particle numbers are given by
\begin{equation}
N_{\sigma} =\sum_m 
\langle \hat{a}_{m,\sigma}^+ \hat{a}_{m,\sigma}^{\vphantom{+}}\rangle
=\int_{-\infty}^0 \rmd \omega D_{\sigma}(\omega)  \; ,
\end{equation}
see Sect.~\ref{sec:DOSprops}. Using eqs.~(\ref{eq:EOMimpurityGF1}),
(\ref{eq:totalDOSforHF}), 
and~(\ref{eq:DOSfromGFagain}), we arrive at 
($\Delta N_{\sigma}=N_{\sigma}-L/2$)
\begin{equation}
\Delta N_{\sigma}= -\frac{1}{\pi}
\int_{-\infty}^0  \rmd \omega \frac{\partial }{\partial \omega}
\Cot^{-1}\left( 
\frac{\omega+E_{d,\sigma}-V^2\Lambda_0(\omega)}{\pi\rho_0(\omega)V^2}
\right)\; .
\end{equation}
With $\Cot^{-1}(-\infty)=-\pi$ we find
\begin{equation}
N_{\sigma}= \frac{L}{2} +1 -\frac{1}{\pi}\Cot^{-1}(\sigma_nUm/\Gamma)
\end{equation}
with $\Gamma=\pi \rho_0(0)V^2=2V^2$.
By definition, $N_{\sigma}=N_{\sigma}^{\rm host}+n_{d,\sigma}$ with
$n_{d,\sigma}=\langle \hat{n}_{d,\sigma}\rangle$ from eq.~(\ref{eq:HFdoccupation}).
Therefore, the host electron particle number is given by
\begin{equation}
N_{\sigma}^{\rm host}=\frac{L}{2} +1 -\frac{1}{\pi}\Cot^{-1}(\sigma_nUm/\Gamma)
-n_{d,\sigma} \; .
\label{eq:hostpol}
\end{equation}
For the symmetric Anderson model, $m=0$ and $n_{d,\sigma}=1/2$ so that
$N_{\sigma}^{\rm host}=L/2$ as it should.

The host-electron particle numbers $\Delta N_{\sigma}=N_{\sigma}-L/2$
as a function of the
magnetization are shown in Fig.~\ref{fig:Nsigmahost}.
It is seen that the impurity barely polarizes the host electrons, 
even for large magnetization where $n_{d,\uparrow}\approx 1$,
$n_{d,\downarrow}\approx 0$.
Note that a positive magnetic moment on the impurity
also leads to an excess of $\uparrow$-spins in the host electrons.

\subsubsection{Spin correlation function}

In Hartree-Fock we have $C_{dd}^S=1/8+m^2/2$. 
The result interpolates between the limiting cases 
$C_{dd}^S(U=0)=1/8$ for $m=0$ and $C_{dd}^S(U\to 0)=1/4$
for $m=1/2$.
For the spin correlation between
the impurity and the host electrons at site~$r$, Wick's theorem gives
\begin{equation}
C_{dc}^S(r)= 
\frac{m}{2} \left[ 
\langle \hat{c}_{r,\uparrow}^+ \hat{c}_{r,\uparrow}^{\vphantom{+}}\rangle
-
\langle \hat{c}_{r,\downarrow}^+ \hat{c}_{r,\downarrow}^{\vphantom{+}}\rangle
\right] 
-\frac{1}{4} \sum_{\sigma} 
\left| \langle \hat{d}_{\sigma}^+ \hat{c}_{r,\sigma}^{\vphantom{+}}\rangle \right|^2
\label{eq:CSdHFtwocontribs} \; .
\end{equation}
The first term appears because 
the Hartree-Fock magnetic moment
slightly polarizes the host electrons.
Using eq.~(\ref{eq:EOMimpurityGF1}), the site occupancies are given by
\begin{eqnarray}
\langle \hat{c}_{r,\sigma}^+ \hat{c}_{r,\sigma}^{\vphantom{+}}\rangle
&=& \frac{1}{2}-\frac{V^2}{\pi} \int_{-\infty}^0 \!\rmd \omega
{\rm Im} \biggl[
\frac{[Q_r(\omega)]^2}{\omega+\sigma_nUm -\Delta^{\rm ret}(\omega)}
\biggr]\nonumber \\
&=& \frac{1}{2} + \Delta N_{r,\sigma}^{\rm b}
+ \Delta N_{r,\sigma}^{\rm band}\; ,
\end{eqnarray}
where
\begin{eqnarray}
\Delta N_{r,\sigma}^{\rm b}
&=& V^2 Z_{\rm b}(v_{\rm b}(V,\sigma_nUm)) \left[ Q_r(v_{\rm b}(V,\sigma_nUm))\right]^2
\; ,
\nonumber \\
 \Delta N_{r,\sigma}^{\rm band}
&=& \frac{4V^2}{\pi}\int_{-\pi/2}^0 
\frac{(-1)^rA_r(p)\rmd p}{16V^4 +\cos^2p(\sin p+2\sigma_nUm)^2}\nonumber \\
\end{eqnarray}
with the bound-state weight $Z_{\rm b}(V)$ and bound-state energy
$v_{\rm b}(V,\sigma_nUm)$ from eq.~(\ref{eq:Zvbcalcl}),
$Q_r(\omega)$ from eq.~(\ref{eq:Qromega}) and
\begin{equation}
A_r(p)=-4V^2\cos(2pr) +\cos p \sin(2pr) (\sin p+2\sigma_nUm) \; .
\end{equation}
As expected for a bound state, its contribution $\Delta N_{r,\sigma}^{\rm b}$
decays exponentially 
for large distances~$r\gg 1/(8V^2)$. For $Um\neq 1/2$, 
the band contribution $\Delta N_{r,\sigma}^{\rm band}$
contains an exponentially decaying part and 
a term that decays oscillatory, proportional to $(-1)^r/r$.

The second term in eq.~(\ref{eq:CSdHFtwocontribs}) is evaluated 
as outlined in Sect.~\ref{sec:hybridizationandMr},
\begin{eqnarray}
\langle \hat{d}_{\sigma}^+ \hat{c}_{r,\sigma}^{\vphantom{+}}\rangle 
&=& 
-\frac{V}{\pi} \int_{-\infty}^0 \!\rmd \omega
{\rm Im} \biggl[
\frac{Q_r(\omega)}{\omega+\sigma_nUm -\Delta^{\rm ret}(\omega)}
\biggr]\nonumber \\
&=& M_{r,\sigma}^{\rm b} + M_{r,\sigma}^{\rm band}\nonumber \; ,\\ 
M_{r,\sigma}^{\rm b}&=& V Z_{\rm b}(v_{\rm b}(V,\sigma_nUm))Q_r(v_{\rm b}(V,\sigma_nUm))
\; , \nonumber \\
M_{2n,\sigma}^{\rm band} &=& 
\frac{2V}{\pi} \int_{-\pi/2}^0 
\frac{(-1)^n\cos p\, B_{2n}(p)\rmd p}{16V^4 +\cos^2p(\sin p+2\sigma_nUm)^2} 
\nonumber \; ,\\
B_r(p)&=& 4V^2\sin(rp)+\cos p (\sin p +2\sigma_n Um)\cos(rp) \nonumber \; ,\\
M_{2n+1,\sigma}^{\rm band} &=& 
\frac{2V}{\pi} \int_{-\pi/2}^0 
\frac{(-1)^{n+1} \cos p\,  C_{2n+1}(p)\rmd p}{
16V^4 +\cos^2p(\sin p+2\sigma_nUm)^2} 
\nonumber \; ,\\
C_r(p)&=& -4V^2\cos(rp)+\cos p (\sin p +2\sigma_n Um)\sin(rp) \nonumber \\
\end{eqnarray}
for $n\geq 0$. For $Um>1/2$, the matrix elements are small, and the 
contribution 
of the second term to
the screening are negligible. The results are discussed in the main text.

\subsection{Impurity magnetization from Bethe Ansatz}

The Bethe Ansatz results of Ref.~[\onlinecite{0022-3719-16-12-017}]
employ $H=2B$ so that the external magnetic field is given by
${\cal H}=2b \Gamma/(g\mu_{\rm B})$.

\subsubsection{Limit of vanishing interactions}
\label{subsubsect:uzero}

Since $b_0(u)\sim \sqrt{u}\to 0$, we address region~II only.
We rewrite the magnetic field in the from
\begin{equation}
\Delta b_{II}(p,u)=
\sqrt{\frac{1}{8\pi}}
\int_0^{\infty}\frac{\rmd \lambda}{\lambda^{3/2}}
\frac{\left(1-e^{-2\pi s^2 \lambda}\right)}{
\Gamma\left(\frac{1}{2}+2u \lambda\right)}
\left(\frac{2u \lambda}{e}\right)^{2u\lambda}
\label{eq:bregionIIsupp}
\end{equation}
where $\Delta b_{II}(p,u)=b_{II}(p,u)-b_0(u)$ and
we set $p=s^2/(2u)$.
Taking the limit $u\to 0$ gives ($\Gamma(1/2)=\sqrt{\pi}$)
\begin{equation}
b_{II}(s,0)=\sqrt{\frac{1}{8\pi^2}}
\int_0^{\infty}\frac{\rmd \lambda}{\lambda^{3/2}}
\left(1-e^{-2\pi s^2 \lambda}\right)=s \; ,
\label{eq:bregionIIsuppuzero}
\end{equation}
i.e., we have $p=b^2/(2u)\to \infty$ for $u\to 0$.

The magnetization reads
\begin{eqnarray}
m_{II}(s,u)&=& \frac{1}{2} -\frac{1}{2}\int_0^{\infty}
\frac{\rmd \lambda}{\sqrt{\pi}\lambda} 
\frac{e^{-2\pi s^2\lambda}}
{\Gamma\left(\frac{1}{2}+2u\lambda\right)}
\left(\frac{2u\lambda}{e}\right)^{2u\lambda} \nonumber \\
&& \hphantom{\frac{1}{2} -\frac{1}{2}\int_0^{\infty}}\times F\left(2\pi\lambda,u\right)
\label{eq:mregionIIsupp}
\end{eqnarray}
so that we obtain 
\begin{equation}
m(b)= \frac{1}{2} -\frac{1}{2}
\int_0^{\infty}
\frac{\rmd \lambda}{\sqrt{\pi}\lambda} 
\frac{e^{-2\pi b^2\lambda}}
{\sqrt{\pi}}
F\left(2\pi\lambda,0\right)
\label{eq:mregionIIsupp2}
\end{equation}
in the limit $u\to 0$. We differentiate $m(b)$ with respect to $b^2$
and insert the definition of $F(a,0)$,
\begin{eqnarray}
\frac{\rmd m(b)}{\rmd b^2}&=& 
\dashint_{-\infty}^{\infty}\frac{\rmd y}{\pi} \frac{1}{1-y^2}
\int_0^{\infty}
\frac{\rmd \mu}{2\pi} 
e^{-(b^2+y^2)\mu} \nonumber \\
&=& \frac{1}{2\pi} \frac{1}{b(1+b^2)} \; .
\label{eq:mdiff}
\end{eqnarray}
Integrating this expression with respect to $b^2$ and using $m(0)=0$ we find
\begin{equation}
m(b)=\frac{1}{\pi} \arctan(b)  \; ,
\end{equation}
as derived for the impurity magnetization of the non-interacting SIAM
in the main text.

\subsubsection{Limit of large magnetic fields}

For large~$b$, we again address region~II only. Following the lines of
the Sect.~\ref{subsubsect:uzero}, we can express the magnetic field
in the form ($p=s^2/(2u)$)
\begin{eqnarray}
b(p,u) &=& b_0(u) +s \nonumber \\
&& +\int_0^{\infty}
\frac{\rmd \lambda}{\lambda^{3/2}}
\frac{1-e^{-2\pi s^2 \lambda}}{\sqrt{8\pi}}
\left[ \frac{
\left( 2u \lambda/e\right)^{2u\lambda}}{
\Gamma\left(\frac{1}{2}+2u \lambda\right)} -\frac{1}{\sqrt{\pi}}\right]
\nonumber \\
&=& s-\sqrt{2u}Q\left(\pi s^2/u\right) \; ,\nonumber \\
Q(y)&=& \frac{1}{\sqrt{8\pi}}
\int_0^{\infty}
\frac{\rmd x}{x^{3/2}}e^{-yx}
\left[ \frac{
\left( x/e\right)^{x}}{
\Gamma\left(\frac{1}{2}+x\right)} -\frac{1}{\sqrt{\pi}}\right] \; .
\label{eq:bIIwithQ}
\end{eqnarray}
For the impurity magnetization we find
\begin{eqnarray}
m(s,u)&=&  \frac{1}{\pi} \arctan(s+u/2)+ Z(s,u) \; , \nonumber \\
Z(s,u)&=& -\frac{1}{2\sqrt{\pi}} \int_0^{\infty} \frac{\rmd \lambda}{\lambda}
e^{-2\pi s^2\lambda} F(2\pi\lambda,u) \nonumber \\
&& \hphantom{-\frac{1}{2\sqrt{\pi}} \int_0^{\infty} } 
\!\!\!\times
\left[ \frac{
\left( 2u \lambda/e\right)^{2u\lambda}}{
\Gamma\left(\frac{1}{2}+2u \lambda\right)} -\frac{1}{\sqrt{\pi}}\right].
\label{eq:magIIlargeb}
\end{eqnarray}
These formulae are valid for all magnetic fields in region~II.
In particular, they include the limit $u\to 0$.

For large fields, we see from~(\ref{eq:bIIwithQ}) that $s\approx b\gg 1$
and that only small~$x$ contribute to the integrand of $Q(y)$ for
$y=\pi  s^2/u \gg 1$. Using the small-$x$ expansion for the integrand
leads to~\cite{Mathematica10}
\begin{equation}
Q(y\gg 1) \approx - \frac{1+\ln(y)}{\sqrt{8\pi y}} \; .
\end{equation}
We use $s\approx b$ in this term to find 
\begin{equation}
s= b-\frac{u}{2\pi b}\left[ 1 + \ln\left(\pi b^2/u\right)\right]
\end{equation}
for $b^2\gg u/\pi$ from eq.~(\ref{eq:bIIwithQ}).
The same line of arguments gives
\begin{equation}
Z(s,u) \approx -\frac{u}{\pi^2} \left(\frac{1-\ln(\pi s^2/u)}{2s^3}\right)\; ,
\end{equation}
where we used $F(a\to 0,u)\approx 2\sqrt{a/\pi}$.
We replace $s$ by $b$ for large fields, insert this result into 
eq.~(\ref{eq:magIIlargeb}), and find
\begin{equation}
m_{II}(b\gg \sqrt{u/\pi})\approx \frac{1}{2}-\frac{1}{\pi b} 
+\frac{u}{2\pi b^2} +\frac{4\pi-12u-3\pi u^2}{12 \pi^2 b^3}
\label{eq:strongfieldssupp} 
\end{equation}
up to and including third order in $1/b$. 
There are no logarithmic corrections to this order.

In fact, there are no logarithmic terms to all orders of the $1/b$ expansion
because, for $b\gg 1$, both $b(p,u)$ and $m_{II}(p,u)$ can be expressed 
in terms of a series with odd powers in the parameter~$1/\sqrt{z}$ 
where $z$ obeys $p=z-\ln(2\pi e z)/(2\pi)$.~\cite{0022-3719-16-12-017}
Therefore, at large values of the external field, the impurity
magnetization does not show any signs of the logarithmic
Doniach-\v{S}unji\'c-Hamann tails in the impurity spectral 
function.~\cite{Logan1998,Logan2000,Logan2009,0022-3719-3-2-010}



\begin{thebibliography}{44}%
\makeatletter
\providecommand \@ifxundefined [1]{%
 \@ifx{#1\undefined}
}%
\providecommand \@ifnum [1]{%
 \ifnum #1\expandafter \@firstoftwo
 \else \expandafter \@secondoftwo
 \fi
}%
\providecommand \@ifx [1]{%
 \ifx #1\expandafter \@firstoftwo
 \else \expandafter \@secondoftwo
 \fi
}%
\providecommand \natexlab [1]{#1}%
\providecommand \enquote  [1]{``#1''}%
\providecommand \bibnamefont  [1]{#1}%
\providecommand \bibfnamefont [1]{#1}%
\providecommand \citenamefont [1]{#1}%
\providecommand \href@noop [0]{\@secondoftwo}%
\providecommand \href [0]{\begingroup \@sanitize@url \@href}%
\providecommand \@href[1]{\@@startlink{#1}\@@href}%
\providecommand \@@href[1]{\endgroup#1\@@endlink}%
\providecommand \@sanitize@url [0]{\catcode `\\12\catcode `\$12\catcode
  `\&12\catcode `\#12\catcode `\^12\catcode `\_12\catcode `\%12\relax}%
\providecommand \@@startlink[1]{}%
\providecommand \@@endlink[0]{}%
\providecommand \url  [0]{\begingroup\@sanitize@url \@url }%
\providecommand \@url [1]{\endgroup\@href {#1}{\urlprefix }}%
\providecommand \urlprefix  [0]{URL }%
\providecommand \Eprint [0]{\href }%
\providecommand \doibase [0]{http://dx.doi.org/}%
\providecommand \selectlanguage [0]{\@gobble}%
\providecommand \bibinfo  [0]{\@secondoftwo}%
\providecommand \bibfield  [0]{\@secondoftwo}%
\providecommand \translation [1]{[#1]}%
\providecommand \BibitemOpen [0]{}%
\providecommand \bibitemStop [0]{}%
\providecommand \bibitemNoStop [0]{.\EOS\space}%
\providecommand \EOS [0]{\spacefactor3000\relax}%
\providecommand \BibitemShut  [1]{\csname bibitem#1\endcsname}%
\let\auto@bib@innerbib\@empty
\bibitem [{\citenamefont {Anderson}(1961)}]{PhysRev.124.41}%
  \BibitemOpen
  \bibfield  {author} {\bibinfo {author} {\bibfnamefont {P.~W.}\ \bibnamefont
  {Anderson}},\ }\href@noop {} {\bibfield  {journal} {\bibinfo  {journal}
  {Phys. Rev.}\ }\textbf {\bibinfo {volume} {124}},\ \bibinfo {pages} {41}
  (\bibinfo {year} {1961})}\BibitemShut {NoStop}%
\bibitem [{\citenamefont {Anderson}(1978)}]{RevModPhys.50.191}%
  \BibitemOpen
  \bibfield  {author} {\bibinfo {author} {\bibfnamefont {P.~W.}\ \bibnamefont
  {Anderson}},\ }\href@noop {} {\bibfield  {journal} {\bibinfo  {journal} {Rev.
  Mod. Phys.}\ }\textbf {\bibinfo {volume} {50}},\ \bibinfo {pages} {191}
  (\bibinfo {year} {1978})}\BibitemShut {NoStop}%
\bibitem [{\citenamefont {Hewson}(1993)}]{Hewson}%
  \BibitemOpen
  \bibfield  {author} {\bibinfo {author} {\bibfnamefont {A.~C.}\ \bibnamefont
  {Hewson}},\ }\href@noop {} {\emph {\bibinfo {title} {{\sl The Kondo Problem
  to Heavy Fermions}}}}\ (\bibinfo  {publisher} {Cambridge University Press},\
  \bibinfo {address} {Cambridge},\ \bibinfo {year} {1993})\BibitemShut
  {NoStop}%
\bibitem [{\citenamefont {Jakobs}\ \emph {et~al.}(2010)\citenamefont {Jakobs},
  \citenamefont {Pletyukhov},\ and\ \citenamefont
  {Schoeller}}]{PhysRevB.81.195109}%
  \BibitemOpen
  \bibfield  {author} {\bibinfo {author} {\bibfnamefont {S.~G.}\ \bibnamefont
  {Jakobs}}, \bibinfo {author} {\bibfnamefont {M.}~\bibnamefont {Pletyukhov}},
  \ and\ \bibinfo {author} {\bibfnamefont {H.}~\bibnamefont {Schoeller}},\
  }\href@noop {} {\bibfield  {journal} {\bibinfo  {journal} {Phys. Rev. B}\
  }\textbf {\bibinfo {volume} {81}},\ \bibinfo {pages} {195109} (\bibinfo
  {year} {2010})}\BibitemShut {NoStop}%
\bibitem [{\citenamefont {Kinza}\ \emph {et~al.}(2013)\citenamefont {Kinza},
  \citenamefont {Ortloff}, \citenamefont {Bauer},\ and\ \citenamefont
  {Honerkamp}}]{PhysRevB.87.035111}%
  \BibitemOpen
  \bibfield  {author} {\bibinfo {author} {\bibfnamefont {M.}~\bibnamefont
  {Kinza}}, \bibinfo {author} {\bibfnamefont {J.}~\bibnamefont {Ortloff}},
  \bibinfo {author} {\bibfnamefont {J.}~\bibnamefont {Bauer}}, \ and\ \bibinfo
  {author} {\bibfnamefont {C.}~\bibnamefont {Honerkamp}},\ }\href@noop {}
  {\bibfield  {journal} {\bibinfo  {journal} {Phys. Rev. B}\ }\textbf {\bibinfo
  {volume} {87}},\ \bibinfo {pages} {035111} (\bibinfo {year}
  {2013})}\BibitemShut {NoStop}%
\bibitem [{\citenamefont {Rentrop}\ \emph {et~al.}(2016)\citenamefont
  {Rentrop}, \citenamefont {Meden},\ and\ \citenamefont
  {Jakobs}}]{PhysRevB.93.195160}%
  \BibitemOpen
  \bibfield  {author} {\bibinfo {author} {\bibfnamefont {J.~F.}\ \bibnamefont
  {Rentrop}}, \bibinfo {author} {\bibfnamefont {V.}~\bibnamefont {Meden}}, \
  and\ \bibinfo {author} {\bibfnamefont {S.~G.}\ \bibnamefont {Jakobs}},\
  }\href@noop {} {\bibfield  {journal} {\bibinfo  {journal} {Phys. Rev. B}\
  }\textbf {\bibinfo {volume} {93}},\ \bibinfo {pages} {195160} (\bibinfo
  {year} {2016})}\BibitemShut {NoStop}%
\bibitem [{\citenamefont {Zener}(1951)}]{PhysRev.81.440}%
  \BibitemOpen
  \bibfield  {author} {\bibinfo {author} {\bibfnamefont {C.}~\bibnamefont
  {Zener}},\ }\href@noop {} {\bibfield  {journal} {\bibinfo  {journal} {Phys.
  Rev.}\ }\textbf {\bibinfo {volume} {81}},\ \bibinfo {pages} {440} (\bibinfo
  {year} {1951})}\BibitemShut {NoStop}%
\bibitem [{\citenamefont {Kondo}(1964)}]{KondoPTHeorPhys}%
  \BibitemOpen
  \bibfield  {author} {\bibinfo {author} {\bibfnamefont {J.}~\bibnamefont
  {Kondo}},\ }\href@noop {} {\bibfield  {journal} {\bibinfo  {journal}
  {Progress of Theoretical Physics}\ }\textbf {\bibinfo {volume} {32}},\
  \bibinfo {pages} {37} (\bibinfo {year} {1964})}\BibitemShut {NoStop}%
\bibitem [{\citenamefont {Bulla}\ \emph {et~al.}(2008)\citenamefont {Bulla},
  \citenamefont {Costi},\ and\ \citenamefont {Pruschke}}]{RevModPhys.80.395}%
  \BibitemOpen
  \bibfield  {author} {\bibinfo {author} {\bibfnamefont {R.}~\bibnamefont
  {Bulla}}, \bibinfo {author} {\bibfnamefont {T.}~\bibnamefont {Costi}}, \ and\
  \bibinfo {author} {\bibfnamefont {T.}~\bibnamefont {Pruschke}},\ }\href@noop
  {} {\bibfield  {journal} {\bibinfo  {journal} {Rev. Mod. Phys.}\ }\textbf
  {\bibinfo {volume} {80}},\ \bibinfo {pages} {395} (\bibinfo {year}
  {2008})}\BibitemShut {NoStop}%
\bibitem [{\citenamefont {Logan}\ \emph {et~al.}(1998)\citenamefont {Logan},
  \citenamefont {Eastwood},\ and\ \citenamefont {Tusch}}]{Logan1998}%
  \BibitemOpen
  \bibfield  {author} {\bibinfo {author} {\bibfnamefont {D.~E.}\ \bibnamefont
  {Logan}}, \bibinfo {author} {\bibfnamefont {M.~P.}\ \bibnamefont {Eastwood}},
  \ and\ \bibinfo {author} {\bibfnamefont {M.~A.}\ \bibnamefont {Tusch}},\
  }\href@noop {} {\bibfield  {journal} {\bibinfo  {journal} {Journal of
  Physics: Condensed Matter}\ }\textbf {\bibinfo {volume} {10}},\ \bibinfo
  {pages} {2673} (\bibinfo {year} {1998})}\BibitemShut {NoStop}%
\bibitem [{\citenamefont {Logan}\ and\ \citenamefont
  {Glossop}(2000)}]{Logan2000}%
  \BibitemOpen
  \bibfield  {author} {\bibinfo {author} {\bibfnamefont {D.~E.}\ \bibnamefont
  {Logan}}\ and\ \bibinfo {author} {\bibfnamefont {M.~T.}\ \bibnamefont
  {Glossop}},\ }\href@noop {} {\bibfield  {journal} {\bibinfo  {journal}
  {Journal of Physics: Condensed Matter}\ }\textbf {\bibinfo {volume} {12}},\
  \bibinfo {pages} {985} (\bibinfo {year} {2000})}\BibitemShut {NoStop}%
\bibitem [{\citenamefont {Galpin}\ \emph {et~al.}(2009)\citenamefont {Galpin},
  \citenamefont {Gilbert},\ and\ \citenamefont {Logan}}]{Logan2009}%
  \BibitemOpen
  \bibfield  {author} {\bibinfo {author} {\bibfnamefont {M.~R.}\ \bibnamefont
  {Galpin}}, \bibinfo {author} {\bibfnamefont {A.~B.}\ \bibnamefont {Gilbert}},
  \ and\ \bibinfo {author} {\bibfnamefont {D.~E.}\ \bibnamefont {Logan}},\
  }\href@noop {} {\bibfield  {journal} {\bibinfo  {journal} {Journal of
  Physics: Condensed Matter}\ }\textbf {\bibinfo {volume} {21}},\ \bibinfo
  {pages} {375602} (\bibinfo {year} {2009})}\BibitemShut {NoStop}%
\bibitem [{\citenamefont {Borda}(2007)}]{PhysRevB.75.041307}%
  \BibitemOpen
  \bibfield  {author} {\bibinfo {author} {\bibfnamefont {L.}~\bibnamefont
  {Borda}},\ }\href@noop {} {\bibfield  {journal} {\bibinfo  {journal} {Phys.
  Rev. B}\ }\textbf {\bibinfo {volume} {75}},\ \bibinfo {pages} {041307}
  (\bibinfo {year} {2007})}\BibitemShut {NoStop}%
\bibitem [{\citenamefont {Mitchell}\ \emph {et~al.}(2011)\citenamefont
  {Mitchell}, \citenamefont {Becker},\ and\ \citenamefont
  {Bulla}}]{PhysRevB.84.115120}%
  \BibitemOpen
  \bibfield  {author} {\bibinfo {author} {\bibfnamefont {A.~K.}\ \bibnamefont
  {Mitchell}}, \bibinfo {author} {\bibfnamefont {M.}~\bibnamefont {Becker}}, \
  and\ \bibinfo {author} {\bibfnamefont {R.}~\bibnamefont {Bulla}},\
  }\href@noop {} {\bibfield  {journal} {\bibinfo  {journal} {Phys. Rev. B}\
  }\textbf {\bibinfo {volume} {84}},\ \bibinfo {pages} {115120} (\bibinfo
  {year} {2011})}\BibitemShut {NoStop}%
\bibitem [{\citenamefont {Lechtenberg}\ and\ \citenamefont
  {Anders}(2014)}]{PhysRevB.90.045117}%
  \BibitemOpen
  \bibfield  {author} {\bibinfo {author} {\bibfnamefont {B.}~\bibnamefont
  {Lechtenberg}}\ and\ \bibinfo {author} {\bibfnamefont {F.~B.}\ \bibnamefont
  {Anders}},\ }\href@noop {} {\bibfield  {journal} {\bibinfo  {journal} {Phys.
  Rev. B}\ }\textbf {\bibinfo {volume} {90}},\ \bibinfo {pages} {045117}
  (\bibinfo {year} {2014})}\BibitemShut {NoStop}%
\bibitem [{\citenamefont {Florens}\ and\ \citenamefont
  {Snyman}(2015)}]{PhysRevB.92.195106}%
  \BibitemOpen
  \bibfield  {author} {\bibinfo {author} {\bibfnamefont {S.}~\bibnamefont
  {Florens}}\ and\ \bibinfo {author} {\bibfnamefont {I.}~\bibnamefont
  {Snyman}},\ }\href@noop {} {\bibfield  {journal} {\bibinfo  {journal} {Phys.
  Rev. B}\ }\textbf {\bibinfo {volume} {92}},\ \bibinfo {pages} {195106}
  (\bibinfo {year} {2015})}\BibitemShut {NoStop}%
\bibitem [{\citenamefont {Ghosh}\ \emph {et~al.}(2014)\citenamefont {Ghosh},
  \citenamefont {Ribeiro},\ and\ \citenamefont
  {Haque}}]{1742-5468-2014-4-P04011}%
  \BibitemOpen
  \bibfield  {author} {\bibinfo {author} {\bibfnamefont {S.}~\bibnamefont
  {Ghosh}}, \bibinfo {author} {\bibfnamefont {P.}~\bibnamefont {Ribeiro}}, \
  and\ \bibinfo {author} {\bibfnamefont {M.}~\bibnamefont {Haque}},\
  }\href@noop {} {\bibfield  {journal} {\bibinfo  {journal} {Journal of
  Statistical Mechanics: Theory and Experiment}\ }\textbf {\bibinfo {volume}
  {2014}},\ \bibinfo {pages} {P04011} (\bibinfo {year} {2014})}\BibitemShut
  {NoStop}%
\bibitem [{\citenamefont {Holzner}\ \emph {et~al.}(2009)\citenamefont
  {Holzner}, \citenamefont {McCulloch}, \citenamefont {Schollw\"ock},
  \citenamefont {von Delft},\ and\ \citenamefont
  {Heidrich-Meisner}}]{PhysRevB.80.205114}%
  \BibitemOpen
  \bibfield  {author} {\bibinfo {author} {\bibfnamefont {A.}~\bibnamefont
  {Holzner}}, \bibinfo {author} {\bibfnamefont {I.~P.}\ \bibnamefont
  {McCulloch}}, \bibinfo {author} {\bibfnamefont {U.}~\bibnamefont
  {Schollw\"ock}}, \bibinfo {author} {\bibfnamefont {J.}~\bibnamefont {von
  Delft}}, \ and\ \bibinfo {author} {\bibfnamefont {F.}~\bibnamefont
  {Heidrich-Meisner}},\ }\href@noop {} {\bibfield  {journal} {\bibinfo
  {journal} {Phys. Rev. B}\ }\textbf {\bibinfo {volume} {80}},\ \bibinfo
  {pages} {205114} (\bibinfo {year} {2009})}\BibitemShut {NoStop}%
\bibitem [{\citenamefont {Andrei}(1980)}]{PhysRevLett.45.379}%
  \BibitemOpen
  \bibfield  {author} {\bibinfo {author} {\bibfnamefont {N.}~\bibnamefont
  {Andrei}},\ }\href@noop {} {\bibfield  {journal} {\bibinfo  {journal} {Phys.
  Rev. Lett.}\ }\textbf {\bibinfo {volume} {45}} (\bibinfo {year}
  {1980})}\BibitemShut {NoStop}%
\bibitem [{\citenamefont {Wiegmann}(1981)}]{0022-3719-14-10-014}%
  \BibitemOpen
  \bibfield  {author} {\bibinfo {author} {\bibfnamefont {P.~B.}\ \bibnamefont
  {Wiegmann}},\ }\href@noop {} {\bibfield  {journal} {\bibinfo  {journal}
  {Journal of Physics C: Solid State Physics}\ }\textbf {\bibinfo {volume}
  {14}},\ \bibinfo {pages} {1463} (\bibinfo {year} {1981})}\BibitemShut
  {NoStop}%
\bibitem [{\citenamefont {Wiegmann}\ and\ \citenamefont
  {Tsvelick}(1983)}]{0022-3719-16-12-017}%
  \BibitemOpen
  \bibfield  {author} {\bibinfo {author} {\bibfnamefont {P.~B.}\ \bibnamefont
  {Wiegmann}}\ and\ \bibinfo {author} {\bibfnamefont {A.~M.}\ \bibnamefont
  {Tsvelick}},\ }\href@noop {} {\bibfield  {journal} {\bibinfo  {journal}
  {Journal of Physics C: Solid State Physics}\ }\textbf {\bibinfo {volume}
  {16}},\ \bibinfo {pages} {2281} (\bibinfo {year} {1983})}\BibitemShut
  {NoStop}%
\bibitem [{\citenamefont {Tsvelick}\ and\ \citenamefont
  {Wiegmann}(1983)}]{0022-3719-16-12-018}%
  \BibitemOpen
  \bibfield  {author} {\bibinfo {author} {\bibfnamefont {A.~M.}\ \bibnamefont
  {Tsvelick}}\ and\ \bibinfo {author} {\bibfnamefont {P.~B.}\ \bibnamefont
  {Wiegmann}},\ }\href@noop {} {\bibfield  {journal} {\bibinfo  {journal}
  {Journal of Physics C: Solid State Physics}\ }\textbf {\bibinfo {volume}
  {16}},\ \bibinfo {pages} {2321} (\bibinfo {year} {1983})}\BibitemShut
  {NoStop}%
\bibitem [{\citenamefont {Sch\"{o}nhammer}(1990)}]{Schoenhammer1990}%
  \BibitemOpen
  \bibfield  {author} {\bibinfo {author} {\bibfnamefont {K.}~\bibnamefont
  {Sch\"{o}nhammer}},\ }\href@noop {} {\bibfield  {journal} {\bibinfo
  {journal} {Phys. Rev. B}\ }\textbf {\bibinfo {volume} {42}},\ \bibinfo
  {pages} {2591} (\bibinfo {year} {1990})}\BibitemShut {NoStop}%
\bibitem [{\citenamefont {Gebhard}(1990)}]{PhysRevB.41.9452}%
  \BibitemOpen
  \bibfield  {author} {\bibinfo {author} {\bibfnamefont {F.}~\bibnamefont
  {Gebhard}},\ }\href@noop {} {\bibfield  {journal} {\bibinfo  {journal} {Phys.
  Rev. B}\ }\textbf {\bibinfo {volume} {41}},\ \bibinfo {pages} {9452}
  (\bibinfo {year} {1990})}\BibitemShut {NoStop}%
\bibitem [{\citenamefont {B\"usser}\ \emph {et~al.}(2013)\citenamefont
  {B\"usser}, \citenamefont {Martins},\ and\ \citenamefont
  {Feiguin}}]{PhysRevB.88.245113}%
  \BibitemOpen
  \bibfield  {author} {\bibinfo {author} {\bibfnamefont {C.~A.}\ \bibnamefont
  {B\"usser}}, \bibinfo {author} {\bibfnamefont {G.~B.}\ \bibnamefont
  {Martins}}, \ and\ \bibinfo {author} {\bibfnamefont {A.~E.}\ \bibnamefont
  {Feiguin}},\ }\href@noop {} {\bibfield  {journal} {\bibinfo  {journal} {Phys.
  Rev. B}\ }\textbf {\bibinfo {volume} {88}},\ \bibinfo {pages} {245113}
  (\bibinfo {year} {2013})}\BibitemShut {NoStop}%
\bibitem [{\citenamefont {Schwabe}\ \emph {et~al.}(2015)\citenamefont
  {Schwabe}, \citenamefont {H\"ansel}, \citenamefont {Potthoff},\ and\
  \citenamefont {Mitchell}}]{PhysRevB.92.155104}%
  \BibitemOpen
  \bibfield  {author} {\bibinfo {author} {\bibfnamefont {A.}~\bibnamefont
  {Schwabe}}, \bibinfo {author} {\bibfnamefont {M.}~\bibnamefont {H\"ansel}},
  \bibinfo {author} {\bibfnamefont {M.}~\bibnamefont {Potthoff}}, \ and\
  \bibinfo {author} {\bibfnamefont {A.~K.}\ \bibnamefont {Mitchell}},\
  }\href@noop {} {\bibfield  {journal} {\bibinfo  {journal} {Phys. Rev. B}\
  }\textbf {\bibinfo {volume} {92}},\ \bibinfo {pages} {155104} (\bibinfo
  {year} {2015})}\BibitemShut {NoStop}%
\bibitem [{\citenamefont {Legeza}\ \emph {et~al.}(2006)\citenamefont {Legeza},
  \citenamefont {Gebhard},\ and\ \citenamefont {Rissler}}]{PhysRevB.74.195112}%
  \BibitemOpen
  \bibfield  {author} {\bibinfo {author} {\bibfnamefont {O.}~\bibnamefont
  {Legeza}}, \bibinfo {author} {\bibfnamefont {F.}~\bibnamefont {Gebhard}}, \
  and\ \bibinfo {author} {\bibfnamefont {J.}~\bibnamefont {Rissler}},\
  }\href@noop {} {\bibfield  {journal} {\bibinfo  {journal} {Phys. Rev. B}\
  }\textbf {\bibinfo {volume} {74}},\ \bibinfo {pages} {195112} (\bibinfo
  {year} {2006})}\BibitemShut {NoStop}%
\bibitem [{\citenamefont {Feiguin}\ and\ \citenamefont
  {B\"usser}(2011)}]{PhysRevB.84.115403}%
  \BibitemOpen
  \bibfield  {author} {\bibinfo {author} {\bibfnamefont {A.~E.}\ \bibnamefont
  {Feiguin}}\ and\ \bibinfo {author} {\bibfnamefont {C.~A.}\ \bibnamefont
  {B\"usser}},\ }\href@noop {} {\bibfield  {journal} {\bibinfo  {journal}
  {Phys. Rev. B}\ }\textbf {\bibinfo {volume} {84}},\ \bibinfo {pages} {115403}
  (\bibinfo {year} {2011})}\BibitemShut {NoStop}%
\bibitem [{\citenamefont {Schrieffer}\ and\ \citenamefont
  {Wolff}(1966)}]{PhysRev.149.491}%
  \BibitemOpen
  \bibfield  {author} {\bibinfo {author} {\bibfnamefont {J.~R.}\ \bibnamefont
  {Schrieffer}}\ and\ \bibinfo {author} {\bibfnamefont {P.~A.}\ \bibnamefont
  {Wolff}},\ }\href@noop {} {\bibfield  {journal} {\bibinfo  {journal} {Phys.
  Rev.}\ }\textbf {\bibinfo {volume} {149}},\ \bibinfo {pages} {491} (\bibinfo
  {year} {1966})}\BibitemShut {NoStop}%
\bibitem [{\citenamefont {Mahmoud}\ and\ \citenamefont
  {Gebhard}(2015)}]{Annalenpaper}%
  \BibitemOpen
  \bibfield  {author} {\bibinfo {author} {\bibfnamefont {Z.~M.~M.}\
  \bibnamefont {Mahmoud}}\ and\ \bibinfo {author} {\bibfnamefont
  {F.}~\bibnamefont {Gebhard}},\ }\href@noop {} {\bibfield  {journal} {\bibinfo
   {journal} {Ann.\ Phys. (Berlin)}\ }\textbf {\bibinfo {volume} {527}},\
  \bibinfo {pages} {794} (\bibinfo {year} {2015})}\BibitemShut {NoStop}%
\bibitem [{\citenamefont {Kawakami}\ and\ \citenamefont
  {Okiji}(1981)}]{KAWAKAMI1981483}%
  \BibitemOpen
  \bibfield  {author} {\bibinfo {author} {\bibfnamefont {N.}~\bibnamefont
  {Kawakami}}\ and\ \bibinfo {author} {\bibfnamefont {A.}~\bibnamefont
  {Okiji}},\ }\href@noop {} {\bibfield  {journal} {\bibinfo  {journal} {Physics
  Letters A}\ }\textbf {\bibinfo {volume} {86}},\ \bibinfo {pages} {483}
  (\bibinfo {year} {1981})}\BibitemShut {NoStop}%
\bibitem [{\citenamefont {White}(1992)}]{PhysRevLett.69.2863}%
  \BibitemOpen
  \bibfield  {author} {\bibinfo {author} {\bibfnamefont {S.~R.}\ \bibnamefont
  {White}},\ }\href@noop {} {\bibfield  {journal} {\bibinfo  {journal} {Phys.
  Rev. Lett.}\ }\textbf {\bibinfo {volume} {69}},\ \bibinfo {pages} {2863}
  (\bibinfo {year} {1992})}\BibitemShut {NoStop}%
\bibitem [{\citenamefont {White}(1993)}]{PhysRevB.48.10345}%
  \BibitemOpen
  \bibfield  {author} {\bibinfo {author} {\bibfnamefont {S.~R.}\ \bibnamefont
  {White}},\ }\href@noop {} {\bibfield  {journal} {\bibinfo  {journal} {Phys.
  Rev. B}\ }\textbf {\bibinfo {volume} {48}},\ \bibinfo {pages} {10345}
  (\bibinfo {year} {1993})}\BibitemShut {NoStop}%
\bibitem [{\citenamefont {Legeza}\ \emph {et~al.}(2003)\citenamefont {Legeza},
  \citenamefont {R\"oder},\ and\ \citenamefont {Hess}}]{PhysRevB.67.125114}%
  \BibitemOpen
  \bibfield  {author} {\bibinfo {author} {\bibfnamefont {O.}~\bibnamefont
  {Legeza}}, \bibinfo {author} {\bibfnamefont {J.}~\bibnamefont {R\"oder}}, \
  and\ \bibinfo {author} {\bibfnamefont {B.~A.}\ \bibnamefont {Hess}},\
  }\href@noop {} {\bibfield  {journal} {\bibinfo  {journal} {Phys. Rev. B}\
  }\textbf {\bibinfo {volume} {67}},\ \bibinfo {pages} {125114} (\bibinfo
  {year} {2003})}\BibitemShut {NoStop}%
\bibitem [{\citenamefont {Legeza}\ and\ \citenamefont
  {S\'olyom}(2004)}]{PhysRevB.70.205118}%
  \BibitemOpen
  \bibfield  {author} {\bibinfo {author} {\bibfnamefont {O.}~\bibnamefont
  {Legeza}}\ and\ \bibinfo {author} {\bibfnamefont {J.}~\bibnamefont
  {S\'olyom}},\ }\href@noop {} {\bibfield  {journal} {\bibinfo  {journal}
  {Phys. Rev. B}\ }\textbf {\bibinfo {volume} {70}},\ \bibinfo {pages} {205118}
  (\bibinfo {year} {2004})}\BibitemShut {NoStop}%
\bibitem [{\citenamefont {Yamada}(1975)}]{Yamada1975}%
  \BibitemOpen
  \bibfield  {author} {\bibinfo {author} {\bibfnamefont {K.}~\bibnamefont
  {Yamada}},\ }\href@noop {} {\bibfield  {journal} {\bibinfo  {journal}
  {Progress of Theoretical Physics}\ }\textbf {\bibinfo {volume} {53}},\
  \bibinfo {pages} {970} (\bibinfo {year} {1975})}\BibitemShut {NoStop}%
\bibitem [{cor()}]{correcttypo}%
  \BibitemOpen
  \href@noop {} {}\bibinfo {note} {We correct a typo in the expression in
  Ref.~[\onlinecite{0022-3719-16-12-017}] by replacing $\pi^{1/2}$ by
  $\pi^{3/2}$.}\BibitemShut {Stop}%
\bibitem [{\citenamefont {Doniach}\ and\ \citenamefont
  {\v{S}unji\'c}(1970)}]{0022-3719-3-2-010}%
  \BibitemOpen
  \bibfield  {author} {\bibinfo {author} {\bibfnamefont {S.}~\bibnamefont
  {Doniach}}\ and\ \bibinfo {author} {\bibfnamefont {M.}~\bibnamefont
  {\v{S}unji\'c}},\ }\href@noop {} {\bibfield  {journal} {\bibinfo  {journal}
  {Journal of Physics C: Solid State Physics}\ }\textbf {\bibinfo {volume}
  {3}},\ \bibinfo {pages} {285} (\bibinfo {year} {1970})}\BibitemShut {NoStop}%
\bibitem [{\citenamefont {Okiji}\ and\ \citenamefont
  {Kawakami}(1982)}]{Kawakamisusceptibility}%
  \BibitemOpen
  \bibfield  {author} {\bibinfo {author} {\bibfnamefont {A.}~\bibnamefont
  {Okiji}}\ and\ \bibinfo {author} {\bibfnamefont {N.}~\bibnamefont
  {Kawakami}},\ }\href@noop {} {\bibfield  {journal} {\bibinfo  {journal}
  {Solid State Communications}\ }\textbf {\bibinfo {volume} {43}},\ \bibinfo
  {pages} {365} (\bibinfo {year} {1982})}\BibitemShut {NoStop}%
\bibitem [{\citenamefont {Fritz}\ and\ \citenamefont
  {Vojta}(2013)}]{0034-4885-76-3-032501}%
  \BibitemOpen
  \bibfield  {author} {\bibinfo {author} {\bibfnamefont {L.}~\bibnamefont
  {Fritz}}\ and\ \bibinfo {author} {\bibfnamefont {M.}~\bibnamefont {Vojta}},\
  }\href@noop {} {\bibfield  {journal} {\bibinfo  {journal} {Reports on
  Progress in Physics}\ }\textbf {\bibinfo {volume} {76}},\ \bibinfo {pages}
  {032501} (\bibinfo {year} {2013})}\BibitemShut {NoStop}%
\bibitem [{\citenamefont {Linneweber}\ \emph {et~al.}(2017)\citenamefont
  {Linneweber}, \citenamefont {B\"unemann}, \citenamefont {Mahmoud},\ and\
  \citenamefont {Gebhard}}]{Linneweberetal2017}%
  \BibitemOpen
  \bibfield  {author} {\bibinfo {author} {\bibfnamefont {T.}~\bibnamefont
  {Linneweber}}, \bibinfo {author} {\bibfnamefont {J.}~\bibnamefont
  {B\"unemann}}, \bibinfo {author} {\bibfnamefont {Z.~M.~M.}\ \bibnamefont
  {Mahmoud}}, \ and\ \bibinfo {author} {\bibfnamefont {F.}~\bibnamefont
  {Gebhard}},\ }\href@noop {} {\bibfield  {journal} {\bibinfo  {journal}
  {Journal of Physics: Condensed Matter}\ }\textbf {\bibinfo {volume} {29}},\
  \bibinfo {pages} {445603} (\bibinfo {year} {2017})}\BibitemShut {NoStop}%
\bibitem [{\citenamefont {{Wolfram Research{,} Inc.}}(2015)}]{Mathematica10}%
  \BibitemOpen
  \bibfield  {author} {\bibinfo {author} {\bibnamefont {{Wolfram Research{,}
  Inc.}}},\ }\href@noop {} {\emph {\bibinfo {title} {Mathematica, {V}ersion
  10}}}\ (\bibinfo {address} {Champaign, IL},\ \bibinfo {year}
  {2015})\BibitemShut {NoStop}%
\bibitem [{\citenamefont {Goldstone}(1957)}]{Goldstone267}%
  \BibitemOpen
  \bibfield  {author} {\bibinfo {author} {\bibfnamefont {J.}~\bibnamefont
  {Goldstone}},\ }\href@noop {} {\bibfield  {journal} {\bibinfo  {journal}
  {Proceedings of the Royal Society of London A: Mathematical, Physical and
  Engineering Sciences}\ }\textbf {\bibinfo {volume} {239}},\ \bibinfo {pages}
  {267} (\bibinfo {year} {1957})}\BibitemShut {NoStop}%
\bibitem [{\citenamefont {Ruhl}\ and\ \citenamefont
  {Gebhard}(2006)}]{GebhardRuhl2006}%
  \BibitemOpen
  \bibfield  {author} {\bibinfo {author} {\bibfnamefont {D.}~\bibnamefont
  {Ruhl}}\ and\ \bibinfo {author} {\bibfnamefont {F.}~\bibnamefont {Gebhard}},\
  }\href@noop {} {\bibfield  {journal} {\bibinfo  {journal} {Journal of
  Statistical Mechanics: Theory and Experiment}\ }\textbf {\bibinfo {volume}
  {2006}},\ \bibinfo {pages} {P03015} (\bibinfo {year} {2006})}\BibitemShut
  {NoStop}%
\end{thebibliography}

%

\end{document}